\documentclass[galaxies,review,accept,moreauthors,pdftex]{Definitions/mdpi} 

\firstpage{1} 
\pubvolume{10}
\issuenum{5}
\articlenumber{102}
\pubyear{2022}
\copyrightyear{2022}
\externaleditor{{Academic Editor: Dario Gasparrini}
} 
\datereceived{31 August 2022} 
\dateaccepted{18 October 2022} 
\datepublished{20 October 2022} 
\hreflink{https://doi.org/10.3390/\linebreak galaxies10050102} 

\Title{Polarization Observations of AGN Jets: Past and Future}

\TitleCitation{Polarization Observations of AGN Jets: Past and Future}

\newcommand{\degr}{$^{\circ}$}

\Author{{Jongho Park} 
 $^{1,2,}${*}
\orcidA{} and Juan Carlos Algaba $^{3}$\orcidB{}}

\AuthorNames{Jongho Park, Juan Carlos Algaba}

\AuthorCitation{Park, J.; Algaba, J.C.}

\address{%
$^{1}$ \quad Korea Astronomy and Space Science Institute, Yuseong-gu, Daejeon 34055, Korea\\
$^{2}$ \quad Institute of Astronomy \& Astrophysics, Academia Sinica, 11F of Astronomy-Mathematics Building, \mbox{AS/NTU No. 1,} Taipei 10617, Taiwan\\
$^{3}$ \quad Department of Physics, Faculty of Science, Universiti Malaya, Kuala Lumpur 50603, Malaysi
}

\corres{Correspondence: jpark@kasi.re.kr}

\abstract{The magnetic field is believed to play a critical role in the bulk acceleration and propagation of jets produced in active galactic nuclei (AGN). Polarization observations of AGN jets provide valuable information about their magnetic fields. As a result of radiative transfer, jet structure, and stratification, among other factors, it is not always straightforward to determine the magnetic field structures from observed polarization. We review these effects and their impact on polarization emission at a variety of wavelengths, including radio, optical, and ultraviolet wavelengths in this paper. It is also possible to study the magnetic field in the launching and acceleration regions of AGN jets by using very long baseline interferometry (VLBI), which occurs on a small physical scale. Due to the weak polarization of the jets in these regions, probing the magnetic field is generally difficult. However, recent VLBI observations have detected significant polarization and Faraday rotation in some nearby sources. We present the results of these observations as well as prospects for future observations. Additionally, we briefly discuss recently developed polarization calibration and imaging techniques for VLBI data, which enable more in-depth analysis of the magnetic field structure around supermassive black holes and in AGN jets.}


\usepackage{natbib}
\newcommand\apj{ApJ}%
\newcommand\apjs{ApJS}%
\newcommand\apjl{ApJL}%
\newcommand\aj{AJ}%
\newcommand\mnras{Mon. Not. R. Astron. Soc.}%
\newcommand\araa{ARA\&A}%
\newcommand\aap{A\&A}%

\newcommand\nat{Nature}
\newcommand\pasj{Publ. Astron. Soc. Jpn.}
\newcommand\apss{Ap\&SS}
\newcommand{\jp}[1]{{#1}}

\begin{document}

\section{Introduction}
\label{sec:introduction}

In some active galactic nuclei (AGN), collimated outflows are observed, known as jets (e.g., \citep{Blandford2019}). They often move at relativistic speeds (e.g., \citep{Jorstad2017, Lister2018}) and contribute to the evolution of the interstellar and intergalactic medium by transferring momentum and energy (e.g.,~\citep{Fabian2012, Yuan2018, Yoon2018}). As one of the most energetic phenomena in the universe, AGN jets also emit high-energy photons at X-ray and $\gamma$-ray wavelengths (e.g., \citep{Kataoka2006, Abdo2010, Park2019c}) and even at very high energies (e.g.,~\citep{Aleksic2011a, Aleksic2011b, Aleksic2014}). 

It is believed that magnetic fields are important in the formation of relativistic jets in AGN. They are twisted as a result of the frame-dragging effect of spinning black holes or differential rotation of the accretion disk, which may produce jets (e.g., \citep{BZ1977, BP1982, Semenov2004, VK2004, NQ2005, McKinney2006, Komissarov2007, Komissarov2009, Tchekhovskoy2008, Tchekhovskoy2009}). The magnetic field is also crucial for the acceleration of AGN jets to relativistic speeds. Magnetohydrodynamic (MHD) models predict that jet acceleration can efficiently occur as a result of the Poynting flux to kinetic energy flux conversion through the magnetic nozzle effect (e.g., \citep{Li1992, BL1994, VK2004, Komissarov2007, Tchekhovskoy2008, Lyubarsky2009, PT2020}). Furthermore, magnetic fields, through, for example, magnetic reconnection, may contribute to strong flares that are often observed in AGN across a wide range of the electromagnetic spectrum (e.g., \citep{SS2014, Ripperda2020, Ripperda2022}).


Because these phenomena generally occur at very small spatial scales, polarization observations using very long baseline interferometry (VLBI) {are} well suited to study the magnetic fields. A good example is the linear polarization image taken with the event horizon telescope (EHT) of the ring-like structure surrounding the dark shadow of the supermassive black hole in M87 \cite{EHT2021a}. Due to the very high angular resolution of the new millimeter VLBI arrays, it is now possible to directly observe the magnetic field in the vicinity of supermassive black holes. A number of polarization observations of blazars and nearby radio galaxies using centimeter VLBI arrays have also provided valuable insights into magnetic field structures and their role in jet collimation, bulk acceleration, and \mbox{particle acceleration.}

In this review, we briefly describe some of the results of recent polarization observations of AGN jets, particularly the VLBI observations of nearby AGN jets. Readers are referred to recent reviews by Blandford et al. \citep{Blandford2019}, Boccardi et al. \citep{Boccardi2017}, and Hada et al. \citep{Hada2019} for the latest progress of observational studies of AGN jets, including results from total intensity imaging, multiwavelength studies, theoretical modeling efforts, etc. In Section~\ref{sec:blazars}, we summarize early attempts to correlate observed polarization properties across the electromagnetic spectrum, in particular between optical and radio bands, and explain these observations in terms of emission and propagation effects. Our Section~\ref{sec:nearby} briefly summarizes recent polarimetric studies using VLBI for nearby AGN jets, M87, 3C 84, and 3C 273. In Section~\ref{sec:developments}, we review recent progress in developing new algorithms for instrumental polarization calibration and polarimetric imaging of VLBI data, which could be crucial for future observations of VLBI polarization. Section~\ref{sec:conclusions} summarizes and concludes the paper.

\section{Polarization Studies in Blazar Jets}

\label{sec:blazars}



The electromagnetic spectrum, in particular from radio to optical and ultraviolet bands, of most radio-loud AGN is dominated by synchrotron radiation, arising from presumably the same electron population or, at least, populations with similar physical properties. It would thus be na\"ive to expect that observations would indicate similarities and a close connection in the properties at the different bands.



In this section, we review efforts to correlate optical and radio polarization observables in blazars, such as total ($m$) and fractional ($p$) polarization, or the electric vector position angle (EVPA, $\chi$), as well as their variability, and how understanding where the differences arise has helped to understand both the dynamics as well as the radiative processes of blazar jets.

\subsection{Early Optical--Radio Polarization Correlations}


Kinman et al. \cite{Kinman1974} studied the OJ~287 optical and radio polarization and found that the source was strongly polarized at optical wavelengths, falling to only one or two percent at short centimeter wavelengths and then rising to nearly 5\% at 11~cm. Furthermore, the strong variability found did not appear to show any correlations among the bands. They pointed out that even when considering Faraday and depolarization effects, the observations could only be explained if the source was inhomogeneous, and the emission at 11~cm was arising at a different region with a more ordered different magnetic field configurations, or with a much lower thermal electron density.  Such an inhomogeneous model was supported by their analysis of centimeter wavelength spectral indices. 


The study of Rudnick et al. \cite{Rudnick1978} included quasi-simultaneous observations at cm, mm, infrared, and optical for OJ~287, BL Lac, and 0735+17. 
They observed a very small or negligible Faraday rotation on radio bands, and a general increase of the degree of polarization toward shorter wavelengths. For OJ~287, the EVPA appear to be very similar on all observing bands, including optical with the exception of 11~cm, which was ascribed to an optical depth transition at that wavelength. These suggest that, contrary to previous observations, there is a strong relationship among the emitting region from radio to optical in this source. The case of BL~Lac seemed quite similar, with EVPA $\sim$$0$\degr\,from optical to cm wavelengths, whereas at longer wavelengths, the EVPA change could not be simply explained by opacity. On 0735+17, on the other hand, a difference of $\sim$$60$\degr\,was found between optical and radio EVPA. 



\subsection{Resolving the Polarization Structure}

Based on the observations described above, it seemed that correlation between optical and radio polarization properties occurred only in certain cases. These correlations would not only be subject to the source considered, but would also be time dependent. The initial consideration would be that such correlations would be observed only while the source would be homogeneous, the emission co-spatial, and opacity effects properly accounted for.
In the most general case, however, different physical properties of the regions, such as magnetic field strength, morphology or entanglement, size of the emitting region, or opacity, or other frequency-dependent radiative transfer effects, would generally lead to different optical and radio emission properties. Therefore, 
no correlations with optical and radio variability and polarization properties may be found a priori in all sources. 
The obvious question is, of course, what is the mechanism behind these special cases where a clear correlation can be found? Is the correlation particular to some sources with very particular or special characteristics, or is it rather a common feature to most AGNs, only hidden by external optical, geometric, and/or observational effects?

In some theoretical models, it is indeed possible that, even considering general inhomogeneous synchrotron source models, the observed radio and UV optical infrared (UVOIR) emission can be co-spatial. Ghisellini et al. \cite{Ghisellini1985} considered the emission from an inhomogeneous axially symmetric region with general assumptions on geometry and radial dependences of physical quantities. In this model, the total intensity at a given frequency is dominated by the contribution from all the regions that are optically thin. It is seen that the factor governing whether the inner or the outer regions dominate the emission is purely geometrical (see their {Figure 2}). For example, for a jet with conical geometry, the emission above the self-absorption spectral frequency $\nu_{br}$ will be dominated by the inner regions, whereas emission below $\nu_{br}$ will have contributions of larger radii with lower frequency. On the contrary, for a jet with parabolic geometry, the whole UVOIR spectrum may be dominated by the outer regions. This is remarkably interesting because recent findings \cite{AN2012,Tseng2016,Akiyama2018,Algaba2017,Hada2018,Kovalev2020,Park2021b} suggest that many sources may have a jet geometry break, with the upstream regions of the jet with a parabolic and the downstream with a conical shape. 

This suggests that a powerful tool to investigate if UVOIR emission is co-spatial would be to investigate the polarization on resolved structures, so that we can have better chances of separating the emission from different regions. Because many of the sources appear resolved only on milli-arcsecond scales, it was only possible to study the co-spatiality of the emission with the emergence of the VLBI technique. In general, the milli-arcsecond structure of most radio-loud AGN jets will be that of a compact and typically optically thick ``core'' and an optically thin jet appearing as a series of components or knots. Because the polarization of the core and/or these components may differ both in $m$ and $\chi$, they may contribute differently to the integrated polarization properties, and when comparing with the optical properties we may be able to identify one (or more) of these components to which the optical polarization is associated.

Works in this direction were first performed by Gabuzda et al. \cite{Gabuzda1994,Gabuzda1996} who compared the 6~cm VLBI and optical EVPA of a sample of sources. They found that $\chi_{opt}$ and $\chi_{VLBI}$ of the core or (in some cases) a prominently polarized jet component near the core were either aligned or perpendicular for the BL Lac objects 0735+178, 1147+245, 1219+285, 1418+546, and 1749+096, OJ~287, and BL~Lac, and the QSO  3C~279. They found the probability of this happening by chance to be very small small.  They therefore argued for a connection between the optical and the VLBI EVPAs. 

To explain this connection, they indicated that polarization of BL~Lacs at centimeter wavelengths appear to be dominated by newly emerging VLBI components. In this case, the observed bimodal distribution could arise due to the presence of an unresolved emerging VLBI component. The VLBI core radio EVPA $\chi_{VLBI}$ would then be initially perpendicular to optical $\chi_{opt}$ due to opacity effects, as long as it appears blended within an optically thick region. As the component evolves in time and reaches an optically think regime, the $\chi_{VLBI}$ and $\chi_{opt}$ would become aligned with $\chi_{opt}$. 
{
It is worth noting, however, that the flip in the EVPA would only happen in regions with very large opacity ($\tau\gtrsim5$ \cite{Wardle18}), and in general, contributions from the most optically thin regions would dominate the radio emission, and therefore an opacity-based 90\degr\,EVPA flip would occur only in rare cases.
}

\begin{figure}[H]
\includegraphics[width = 0.7\textwidth]{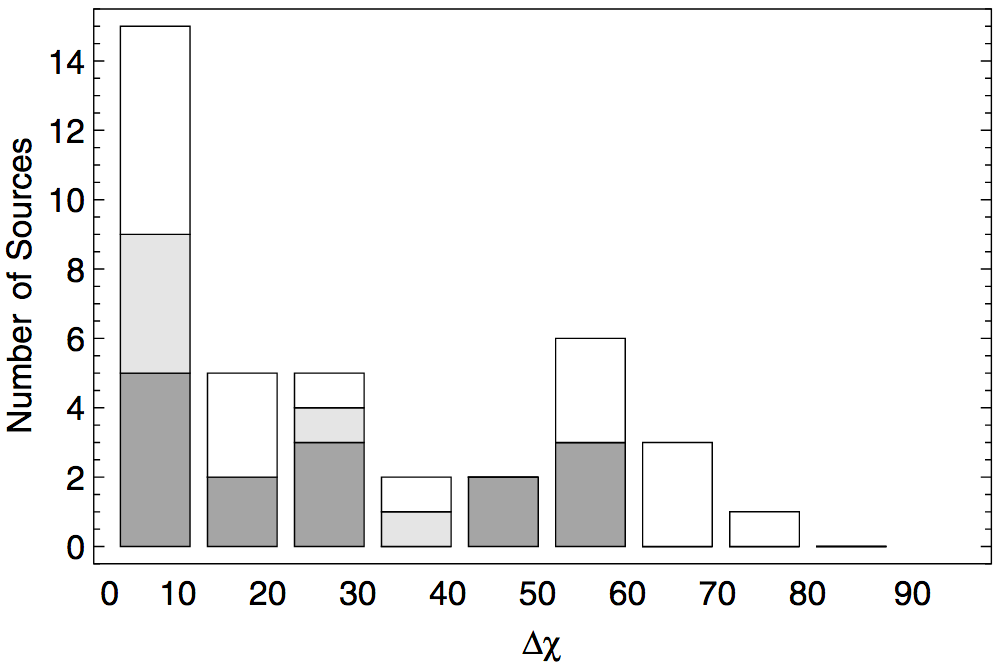}
\caption{Histogram of the difference between the radio-core and optical polarization angles $\Delta \chi$, including BL Lac objects (white), HPQs (dark grey), and LPQs (light grey). Reprinted with permission from Algaba et al. (2010). Copyright Journal compilation 2010 RAS. \label{fig:HistogramDeltaChi}}
\end{figure}

This work was later expanded upon by Algaba et al. \cite{Algaba2011,Algaba2012}. They included corrections for Faraday rotation effects and a larger sample of sources, including not only BL~Lac objects, but also low and high polarized radio quasars. These studies put a much tighter constraint on the probability of the optical and the radio VLBI EVPAs to be either parallel or perpendicular to each other in BL~Lacs. They also found that this relationship did not appear to be that simple in the case of quasars. For the latter, no obvious correlation with the degree of polarization, depolarization between optical and radio bands or magnetic field strength was found (see Figure \ref{fig:HistogramDeltaChi}). This suggested that either BL~Lacs were a particular class of objects with a peculiar emission geometry, or there were other additional factors that had not been taken into account.

The core RM may nonetheless depend on the observing frequency as $RM_{core,\nu}\propto \nu^a$. This is due to the frequency-dependent shift of the core location owing to opacity effects together with a gradient of electron density and magnetic field from the central engine~\cite{Jorstad2007}. 
Park et al. \cite{Park2018} studied the rotation measure (RM) trend in more detail with the Korean VLBI Network (KVN) by using simultaneous observations at 22, 43, and 86~GHz, hereby obtaining resolved VLBI RM analyses at one of the highest frequencies to date. Their results showed that the suggested systematic RM increase and its frequency dependence continued at larger radio frequencies. Interestingly, when comparing with optical bands, they found indications of a saturation frequency of a few hundreds GHz for the majority of sources, where the RM value is expected to reach a plateau and remain constant over higher frequencies. This suggests that we could pinpoint to an actual structure, such as a recollimation shock, causing the emission. Therefore, once we observe at large frequencies, or correct for such frequency-dependent RM up to the saturation frequency, we should be able to connect the radio, optical and high-energy polarization properties.

\subsection{A Blazar Jet Polarization Model}

Another hint in the phenomenology of the connection in the UVOIR polarization arises from D'Arcangelo et al. \cite{DArcangelo2007}, who studied the variation in the polarization characteristics of the quasar 0420--014 during an 11-day monitoring campaign in late 2005. They found a rotation of the 43~GHz VLBI EVPA in the VLBI core by an amount of more than 80\degr, a trend that was also followed by the optical EVPA. Furthermore, once the Faraday correction was considered, the 43~GHz and the optical EVPAs agreed well. Their observations matched well with a model where the bulk of the polarized emission could be associated with a conical shock, whereas the rest of the emission would be associated with mostly unpolarized~regions.



\begin{figure}[H]
\includegraphics[width = 0.6\textwidth]{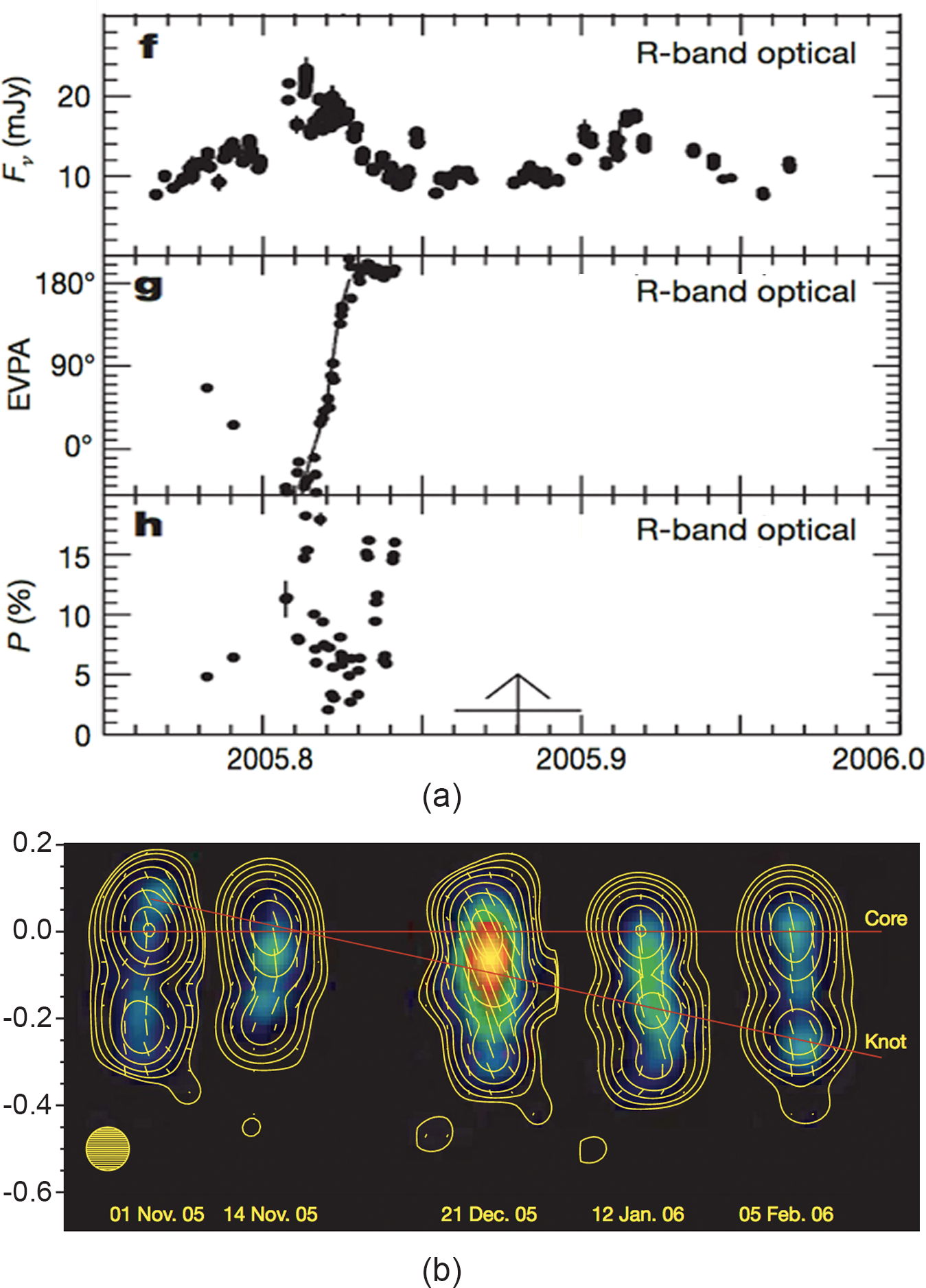}
\caption{{Time} 
 evolution of BL~Lac. (\textbf{a}) Optical flux, electric vector position angle and polarization. \mbox{(\textbf{b}) A} 43~GHz VLBI component. A clear flare together with a polarization decrease and a rotation of about 240\degr ~of the optical EVPA is seen together with the emergence from the core of a new VLBI component. (Reprinted by permission from Springer Nature, {2008} 
 \citep{Marscher2010} with permission of \mbox{the authors.)} \label{fig:EVPArotateswithflare}}
\end{figure}

Observations that captured a large EVPA rotation together with superluminal knots passing through the core strengthen the idea that polarization is dominated by shock components in the jets.  In Marscher et al. \cite{Marscher2008}, a rotation of the EVPA by about 240\degr\,in a five-day interval was observed for BL~Lac in the optical R band, whereas the degree of polarization dropped to a minimum in the middle of the rotation. Interestingly, simultaneous 7-mm VLBI observations show the appearance of a new superluminal component emerging from the core moving downward from the jet with a position angle of about 190\degr, parallel to the jet and in good agreement with the final value that the optical EVPA reached after the rotation (see Figure \ref{fig:EVPArotateswithflare}).


In a similar study of the quasar PKS~1510--089, Marscher et al. \citep{Marscher2010} found a rotation of the optical polarization vector by 720\degr\,during a five-day period in the first half of 2009 encompassing six gamma-ray flares. VLBA analysis indicated that this happened as a bright knot of emission was propagating downstream from the jet and passed through a stationary feature of the jet identified as the 43~GHz core. 

\begin{figure}[H]
\includegraphics[width = 0.9\textwidth]{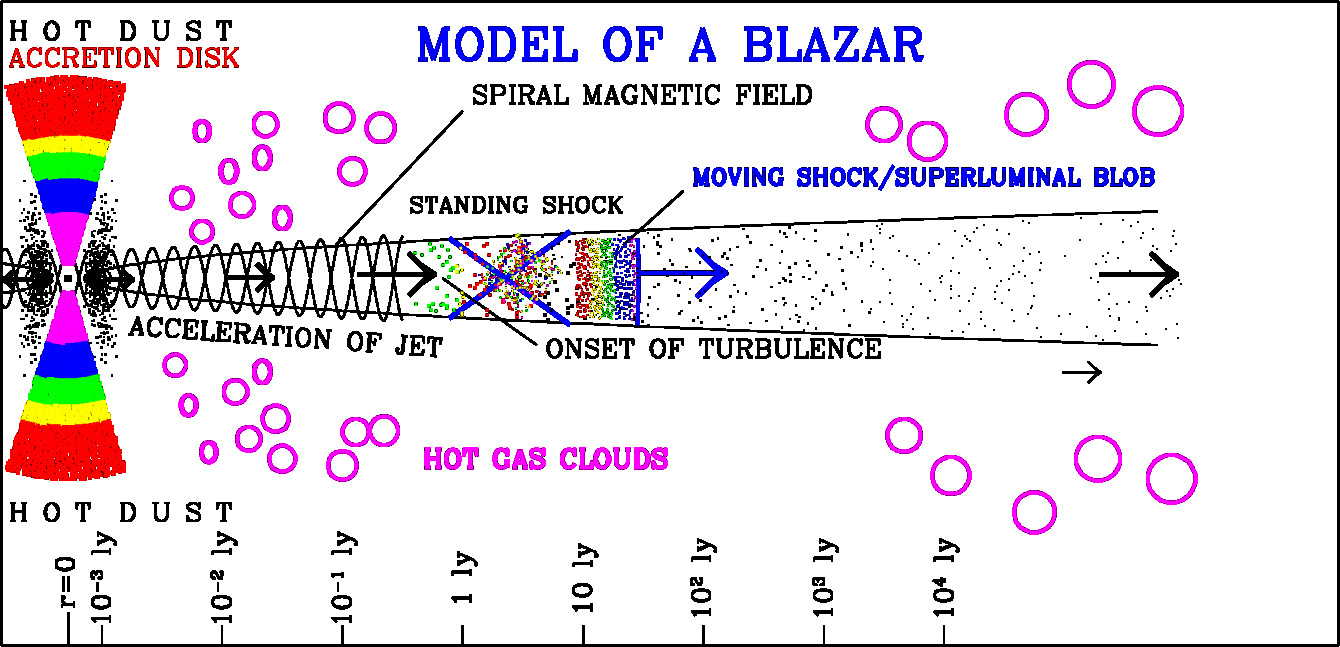}
\caption{{Blazar} model proposed by Marscher. High-energy emission may arise from the acceleration and collimation zone, from a standing shock, moving shock, turbulent region or all of these. The horizontal scale is logaritmic. Copyright 2018 by A. Marscher. \label{fig:MarscherModel}}
\end{figure}

Although a stochastic model, with a turbulent magnetic field (e.g., the turbulent extreme multi-zone model (TEMZ; \cite{Marscher2014})) can produce such rotation as a result of a random walk of the EVPA along the turbulent cells, Monte Carlo simulations show that the probability of this happening by chance together with the flares is minimal, of the order of $10^{-5}$; see e.g., \cite{Blinov17}. Instead, a more plausible model considers the moving emission feature to be following the spiral path of a helical magnetic field along the acceleration and collimation zone (ACZ). In this model, the feature covers most of the jet, to produce the flaring, and cancels most of the polarization as most of the magnetic field orientations are averaged across the feature. Only a fraction of the polarization does not cancel out, leading to the residual EVPA that systematically rotates with the propagation and is observed. A sketch of the model for PKS~1510--089 is shown in Figure \ref{fig:MarscherModel}.

A similar phenomenon has been found in other sources. For example, \mbox{Sasada et al. \citep{Sasada2010}} observed a rotation in the Stokes Q-U in 3C~454.3 during a flare in 2007. \mbox{Liodakis et al. \cite{Liodakis2020}} also found strong variations in optical polarization together with a 230\degr~EVPA rotation in 2013--2014. 
Morozova et al. \citep{Morozova2014} observed that, in S4~0954+658, the position angle of the optical polarization rotated smoothly over more than 300\degr~during an optical flare whereas a new superluminal knot appeared with an apparent speed of $19.0\pm0.3$c in 2011. 

Observations of 0420--014 during 2008--2012 showed the quasar undergoing a series of optical flares accompanied by rapid rotation of the polarization angle, an increase of activity in $\gamma$--rays, an appearance of new superluminal knots in the parsec-scale jet \cite{Troitskiy2013}. Between December 2015 and January 2017, OJ~287 showed flaring activity with its radio EVPA showing a large rotation by 340\degr~with a mean rate of $-$1.04\degr/day and optical EVPA showing a similar rotation of about $-$1.1\degr/day \citep{Myserlis2018}. These suggest that this phenomenology may be common in AGN jets.



\section{VLBI Polarization Observations of Nearby AGN Jets}
\label{sec:nearby}


Due to their close proximity, nearby sources provide a detailed view of the magnetic field structure of AGN jets when resolved by using VLBI. Specifically, the magnetic field in the jet launching region (at scales of $\lesssim10 R_S$ from the black hole; e.g., \cite{EHT2021a}) and the jet collimation and acceleration zone (at scales $\lesssim 10^4-10^6 R_S$; \cite{Marscher2008}) can be constrained. The magnetic field near a black hole is believed to play a crucial role in the launching of AGN jets, e.g., \citep{BZ1977, BP1982, NQ2005}. \jp{The relationship between jet launching and magnetically arrested disks (MADs) is of particular importance (e.g., \citep{Narayan2003, Tchekhovskoy2011}). MAD occurs when the magnetic pressure of the poloidal magnetic fields is counterbalanced by the ram pressure of the accreting gas. The fields can influence the dynamics of the accreting gas.} In recent theoretical studies, it has been demonstrated that spinning black holes can efficiently launch powerful AGN jets when their accretion disks are in a MAD state via the Blandford--Znajek process (e.g.,~\mbox{\citep{BZ1977, Tchekhovskoy2011, McKinney2012, Nakamura2018, Narayan2022}}). It is possible for the jet power to exceed the accretion power ($\dot{M_{\rm BH}}c^2$, where $\dot{M_{\rm BH}}$ is the black hole mass accretion rate and $c$ is the light speed), in this case as the jet is powered by the black hole's rotational energy extracted by the strong magnetic fields. In fact, the core-shift analysis (e.g., \citep{Zamaninasab2014}) and the SED modeling (e.g., \citep{Ghisellini2014}) indicate that the majority of radio-loud AGNs may be in a MAD state. 

Magnetic fields are also believed to play an important role in accelerating AGN jets. Theoretically, it is predicted that the transition from poloidal to toroidal-dominated magnetic fields occurs at the light cylinder radius \cite{VK2004, Komissarov2007, MB2009, CruzOsorio2022}, where $R\Omega$ equals the speed of light with $R$ being the cylindrical radius and $\Omega$ the angular frequency of field lines. For large-scale magnetic field lines threading the event horizon of the black hole, $\Omega$ is related to the black hole spin \cite{BZ1977, Tchekhovskoy2008, Tchekhovskoy2009, PT2020}. 




In spite of its importance, it has been difficult to observe the magnetic field in regions of jet launching and acceleration. One of the principal challenges is the very weak polarization of AGN jets in these regions (e.g., \citep{Lister2018}). However, recent observations have begun to detect linear polarization in the jet launching and acceleration regions of a few nearby radio-loud AGNs. The results of these observations are briefly discussed in this section.

\subsection{M87}

M87 is a giant elliptical galaxy located in the Virgo galaxy cluster. It has a powerful jet that extends out to a scale of kiloparsecs (e.g., \citep{Owen1989, Biretta1999, Perlman2001, Marshall2002, Cheung2007}). The kpc-scale jet of M87 is generally highly polarized. The first linear polarization images of the kpc-scale jet were presented by Schmidt et al. \cite{Schmidt1978}, who discovered EVPA rotations of $\sim$$75^\circ$ between optical and 6\,cm wavelengths. This result was confirmed by Dennison et al. \cite{Dennison1980}. At 15 GHz, fractional polarization tends to be higher at the edge of kpc-scale jets \citep{Owen1989}. At both radio and optical wavelengths, the projected magnetic field is largely parallel to the jet, although the field becomes perpendicular to the jet at the upstream ends of bright knots in the jet only at optical wavelengths \citep{Perlman1999}. This difference was interpreted as a result of shocks occurring in the jet interior that are the dominant sources for the optical jet emission, while the radio jet emission is primarily due to shear layers forming near the jet surface.  

A detailed study of the Faraday rotation measure distribution in the kpc-scale jet has been conducted. Previous observations with the very large array (VLA) at cm wavelengths observed a complex distribution of RM. RM magnitudes were generally low in the inner jet region at $\sim$$200\ \rm rad\ m^{-2}$, but very high positive values were found in the lobe region up to $\rm RM\sim8000$ rad\ m$^{-2}$. A further study by Algaba et al. \cite{Algaba2016} suggested that the observed RM distribution in the jet may be due to an ordered magnetic field in the Faraday screen. In a very recent study, Jansky VLA observations were used to obtain detailed linear polarization images of the kpc-scale jet over a wide frequency range between 4 and 18 GHz \citep{Pasetto2021}. An analysis of high-fidelity polarization images revealed a transverse gradient in the Faraday depth with opposite signs near the jet edges between knots E and F, indicating that the jet has a helical magnetic field. EVPA and fractional polarization rapidly change with frequency in this region, indicating the presence of an internal Faraday depolarization.

Despite the rich information regarding the linear polarization of the M87 jet at kpc scales, the jet on sub-pc to pc scales appears to be very weakly polarized (e.g., \citep{Lister2018}). However, VLBA observations at 5 and 8 GHz at pc-scales revealed patchy linear polarization distributions at $\sim$$20$ mas from the core at a level of 11\% \citep{Junor2001}. This region exhibits a wide range of Faraday rotation measures between $-2000$ and $-$12,000 $\rm{rad\ m^{-2}}$. In a follow-up multifrequency VLBA observation at 8, 12, and 15 GHz, a similar patchy distribution of rotation measures was observed in the same region \citep{ZT2002}. In contrast, a small area of the jet was found to have a positive rotation measure of $\sim$$9600\ {\rm rad\ m^{-2}}$ in this region. EVPAs in this region show good $\lambda^2$ fits, which indicates that the Faraday screen is external to the jet.

\begin{figure}[H]
\includegraphics[width = 0.7\textwidth]{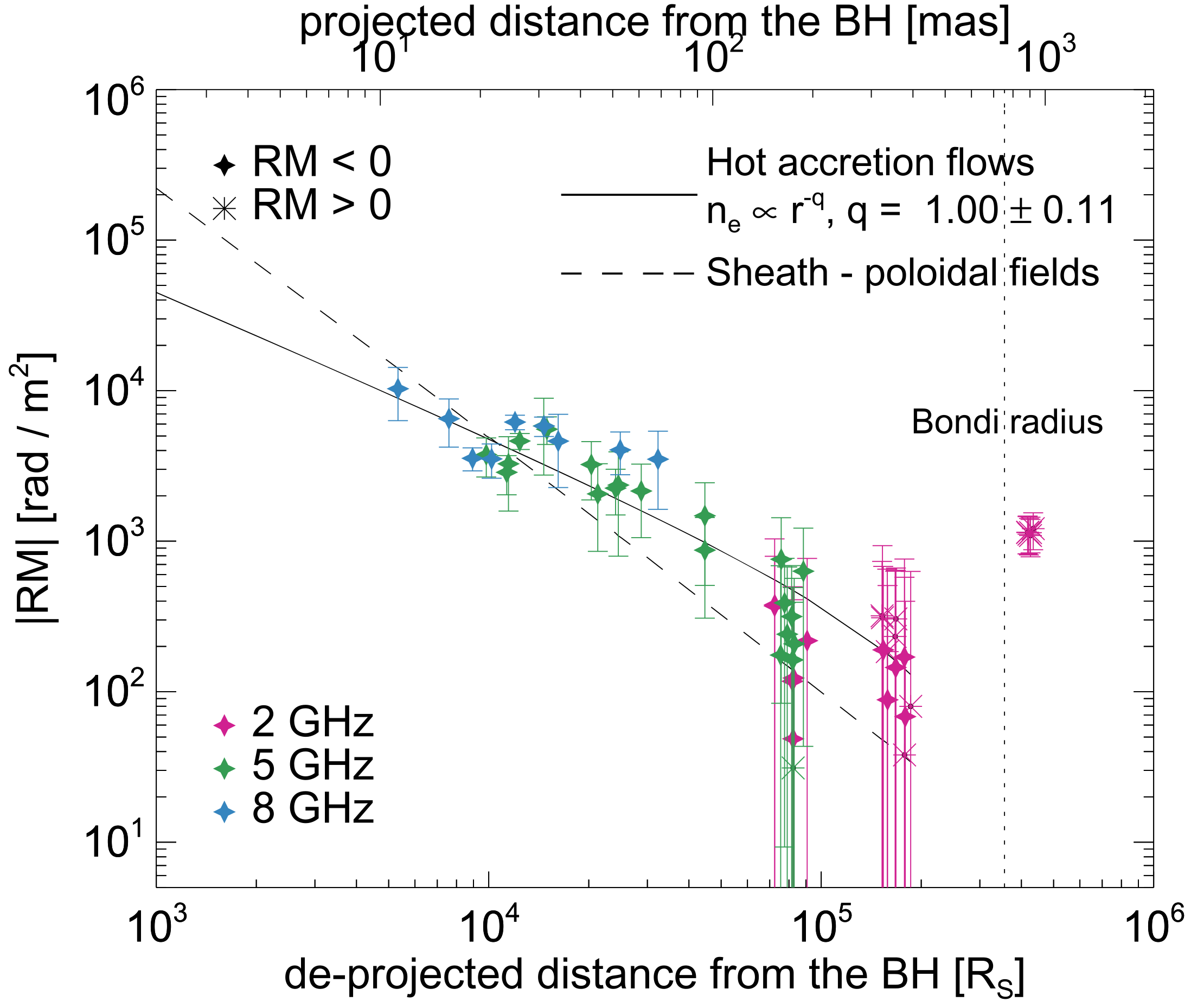}
\caption{RM magnitudes of the M87 jet as functions of de-projected jet distance from the black hole in units {of} $R_S$ \cite{Park2019a}. The data points obtained at different frequencies are shown in different colors. The diamonds and asterisks are negative and positive RMs, respectively. The vertical dotted line denotes the Bondi radius. The solid and dashed curves represent the best-fit hot accretion flows and jet sheath models, respectively. ©AAS. Reproduced with permission. \label{fig:Park2019}}
\end{figure}

Recently, Park et al. \citep{Park2019a} presented Faraday rotation measure maps of the pc-scale jet over a wide range of jet distances (between $\sim$$10$ and $\sim$$400$ mas, which corresponds to the de-projected jet distance range of $\sim$$5000$ and $\sim$$200,000\ R_S$, where $R_S$ is the Schwarzschild radius). The jet is systematically collimated (e.g., \citep{AN2012, NA2013, Hada2013, Nakamura2018}) and accelerated \mbox{(e.g., \citep{Asada2014, Mertens2016, Walker2018, Park2019b})} in this region, which is referred to as the jet acceleration and collimation zone (ACZ; \cite{Marscher2008}; see also Figure~\ref{fig:MarscherModel}). In Figure~\ref{fig:Park2019}, RM magnitudes are plotted as a function of jet distance. Interestingly, the RM magnitude decreases systematically with increasing distance from the black hole in the ACZ (inside the Bondi radius; {the radius defining the region inside of which material will be prone to be accreted toward the black hole;} refs.~\cite{AN2012, Nakamura2018}).

At various locations in the jet, the EVPAs rotate more than $45^\circ$, which is difficult to realize for realistic internal Faraday rotations (e.g., \citep{Burn1966, Homan2012, Sokoloff1998}). This result indicates that the Faraday screen is external to the jet within the Bondi radius. This is in contrast to the kpc-scale jet outside the Bondi radius, for which the JVLA data indicates that there is an internal Faraday rotation in the jet \citep{Pasetto2021}. It is expected that the pc-scale jet located within the Bondi radius is surrounded by the inflows and outflows of the black hole (e.g., \citep{YN2014, Nakamura2018}), whereas the kpc-scale jet outside the Bondi radius may be surrounded by the interstellar medium, which has a much lower electron density and weaker magnetic field. 

The observed RM data inside the Bondi radius {were} fitted with two simple analytic models. One is based on the assumption that the jet of M87 is surrounded by hot accretion flows, which are thought to exist in low-luminosity AGN such as M87 (e.g., \cite{YN2014}). Based on numerical simulation results that present toroidal magnetic fields are dominant in accretion flows, the strength of the magnetic field was assumed to be inversely proportional to the distance from the black hole (e.g., \citep{Hirose2004}). The best-fit (solid line {in Figure \ref{fig:Park2019})} indicates that the electron density of accretion flows was inversely proportional to the distance from the black hole. 

The other model is based on the assumption that the Faraday screen is a jet sheath with a similar geometry to the jet. In this model, the jet edges are assumed to become slower than the jet spine as a result of the interaction between the jet and the surrounding medium, which is referred to as the faster-spine and slower-sheath model. This structure was also indirectly inferred by the statistical analysis of blazars \cite{Clausen-Brown2013}. An assumption was made that the sheath {in M87} was dominated by a poloidal magnetic field, because otherwise transverse magnetic gradients across the jet would be detected (e.g., \citep{BM2010, Gabuzda2018}), which were not observed in the pc-scale jet.

Although the jet sheath model could not fully explain the observed data, the hot accretion flow model was able to fit the data very well. The derived density profile, $\rho\propto r^{-1}$, is consistent with the profile derived slightly inside the Bondi radius by Chandra X-ray observations \citep{Russell2015}. This profile was predicted by the black hole accretion flow model that produces substantial outflows (e.g., \citep{BB1999, BB2004, Begelman2012}). A combination of gas and magnetic pressure gradients and centrifugal force results in these outflows from the accretion flows \cite{Yuan2015}. The pressure profile inferred from the density profile by assuming an adiabatic equation of state for nonrelativistic monatomic gas is $p_{\rm gas} \propto \rho \propto r^{-5/3}$, where $\gamma = 5/3$ is the specific heat ratio. The shape of the parabolic jet is determined by the pressure profile of the external medium confining the jet; if the pressure profile is described as $p_{\rm ext} \propto r^{-\alpha}$, $\alpha \leq2$ is required for producing a parabolic jet shape (e.g., \citep{BL1994, Komissarov2009, Lyubarsky2009, Vlahakis2015}). This result, when combined with the fact that the M87 jet exhibits a parabolic shape inside the Bondi radius (e.g., \citep{AN2012, NA2013, Hada2013, Nakamura2018}), is consistent with the explanation that the M87 jet is collimated by the pressure of wind, as was reproduced in recent GRMHD simulations (e.g., \citep{Nakamura2018, Chatterjee2019}).


According to the MHD model, AGN jets are expected to accelerate to relativistic speeds when the so-called differential collimation of poloidal magnetic fields occurs, where the inner field lines closer to the jet axis are more strongly collimated than the outer field lines (e.g., \citep{Li1992, VK2003, Komissarov2007, Tchekhovskoy2009, Nakamura2018}). As recently observed by various VLBI arrays, the M87 jet is indeed gradually accelerated from non-relativistic to relativistic speeds within the Bondi radius, where it is being collimated systematically (e.g., \citep{Asada2014, Mertens2016, Walker2018, Park2019b}). Beyond the Bondi radius, the jet begins to decelerate (e.g., \citep{Biretta1995, Biretta1999, Cheung2007, Giroletti2012, Meyer2013}). Based on these results, it appears that the M87 jet may be accelerated by converting Poynting flux into kinetic energy flux.

\begin{figure}[H]
\includegraphics[width = 0.9\textwidth]{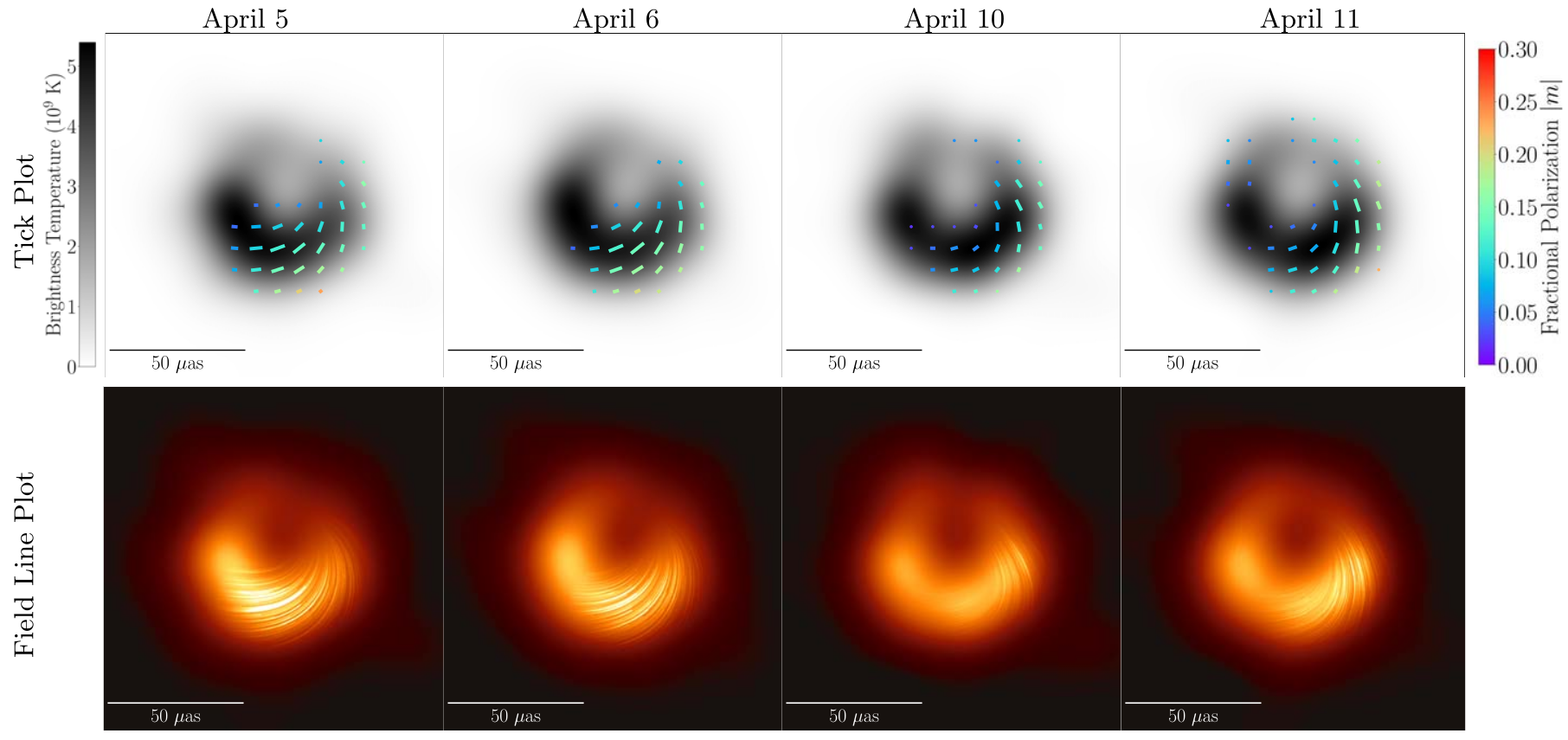}
\caption{Fiducial linear polarimetric images of the M87 black hole observed with the EHT in {2017}~\cite{EHT2021a}. Each image is an average of the five calibration/imaging methods for each day. The upper panel shows the total intensity distributions in grey color and the EVPAs with colored ticks with the colors scaling with the linearly polarized intensity distribution. The bottom panel shows the total intensity distribution in color, on top of which the black streamlines represent the observed EVPA patterns. \label{fig:ehtpol}}
\end{figure}

The EHT collaboration has recently published linear polarization images of the ring surrounding the shadow of the supermassive black hole in M87 observed at 1.3~mm (\mbox{Figure~\ref{fig:ehtpol}; \citep{EHT2021a}}). There is a significant amount of polarization in the southwestern part of the ring, reaching up to the level of 15\% there. The most surprising finding was the arrangement of EVPAs in a nearly azimuthal pattern. Due to the effects of, e.g., light bending and relativistic aberration, the observed polarization vectors for relativistically moving plasma around the black hole may not be perpendicular to the projected magnetic field vectors (see, e.g., \citep{EHT2021b, Narayan2021} for more details).

A comparison of the observed linear polarization images has been conducted by the EHT collaboration with a large library of simulated polarimetric images generated by GRMHD simulations, which include various physical effects such as light bending, relativistic aberration, Faraday rotation, etc. \citep{EHT2021b}. According to GRMHD models, there are two types of accretion flows: standard and normal evolution (SANE) models and MAD models. SANE disks have a lower magnetic flux and weak magnetic fields are sheared out by the motion of the plasma, resulting in toroidal configurations (e.g., \citep{Hirose2004, Narayan2012, Ricarte2021}). MAD models have strong fluxes, and the magnetic fields are mainly poloidal-dominated (e.g.,~\citep{BR1974, Igumenshchev2003, Narayan2003, Tchekhovskoy2011}). The magnetic pressure of these fields is balanced by the ram pressure of the accreting gas, which influences the dynamics of the disk.

Based on the observed linear polarization properties, MAD models appear to be more favorable \citep{EHT2021b}. It was difficult for SANE models to reproduce the observed moderate level of polarization as well as the twisted EVPA pattern. As a result, it appears that strong, ordered, and poloidal-dominated magnetic fields exist in the vicinity of the M87 black hole. When combined with a spinning black hole, this strong magnetic field is capable of launching relativistic jets (e.g., \citep{BZ1977, Tchekhovskoy2012}). There are in fact powerful jets in M87, with jet power estimates ranging from $\sim10^{42}$ to $\sim10^{45}\ {\rm erg\ s^{-1}}$ (e.g., \citep{BB1996, Owen2000, Allen2006, Rafferty2006, Stawarz2006, BL2009, Broderick2015}). Based on this result, the idea that AGN jets may come from MADs has been directly confirmed, which was previously suggested indirectly by both SED modeling studies (e.g., \citep{Ghisellini2014}) and core-shift studies (e.g., \citep{Zamaninasab2014}) for radio-loud AGNs.

\subsection{3C~84}

The 3C 84 is a radio source associated with the AGN NGC 1275 (also known as Perseus A), which hosts powerful jets. The proximity of 3C 84 ($z=0.017559$; \cite{Strauss1992}) and its large mass of $M_{\rm BH} = (0.8-2)\times 10^9 M_\odot$ \citep{Scharwachter2013, Giovannini2018} offer a unique opportunity to study the detailed physics of AGN jets. As an example, the receding jet in the north of the radio core exhibits a free--free absorption feature, which enables us to study the accretion flows (e.g., \citep{Walker2000}). An extremely wide jet opening angle at a short distance from the core with clear limb-brightening allows us to study the transverse jet velocity structure and the origin of the jets. It has also been suggested by recent VLBI observations that the jet is actively interacting with the interstellar medium due to a strong collision with a compact dense cloud \citep{Kino2021}, as well as a cocoon-like structure surrounding the jet \citep{Savolainen2021}.

The 3C 84 is a well-known ``unpolarized'' calibrator at centimeter wavelengths. Considering that the jet produces synchrotron emission, which is generally highly polarized, this is an intriguing phenomenon. However, VLBA observations in 2004 detected significant linearly polarized emission across the tip of the bright southern jet component, which is located $\sim$$14$ from the core \citep{Taylor2006}. A Faraday rotation measure ranging from 6500 to 7500 $\rm rad\ m^{-2}$ was determined for this polarization patch. High-frequency observations also revealed significant linear polarization in 3C 84. Observations with the combined array for research in millimeter wavelength astronomy (CARMA) and the submillimeter array (SMA) at 1.3 and 0.9 mm have detected very large Faraday rotation measures of \mbox{$(8.7\pm2.3)\times10^5\ {\rm rad\ m^{-2}}$~\cite{Plambeck2014}}. Because these arrays are connected interferometers, it was unclear where the linearly polarized emission and Faraday rotation originated. It was suggested that they originated from the nucleus of the system in the light of the comparison with the VLBA polarization observation \citep{Taylor2006} and the application of quasi-spherical radiatively inefficient accretion flow models \citep{Bower2003, Marrone2007}.

In recent VLBA observations at 43 GHz, it has been discovered that the hotspot at the southern tip of the innermost restarted jet has been significantly polarized since late 2015 \citep{Nagai2017}. The brightening in polarization may be the result of the hotspot moving from the southern end of the western limb to the southern end of the eastern limb. The RM of $(6.3\pm1.9)\times10^5\ {\rm rad\ m^{-2}}$ was derived from the hotspot. Based on a comparison with previous CARMA and SMA observations, it was concluded that the RM observed was caused by a clumpy or inhomogeneous ambient medium.

\begin{figure}[H]
\includegraphics[width = 0.75\textwidth]{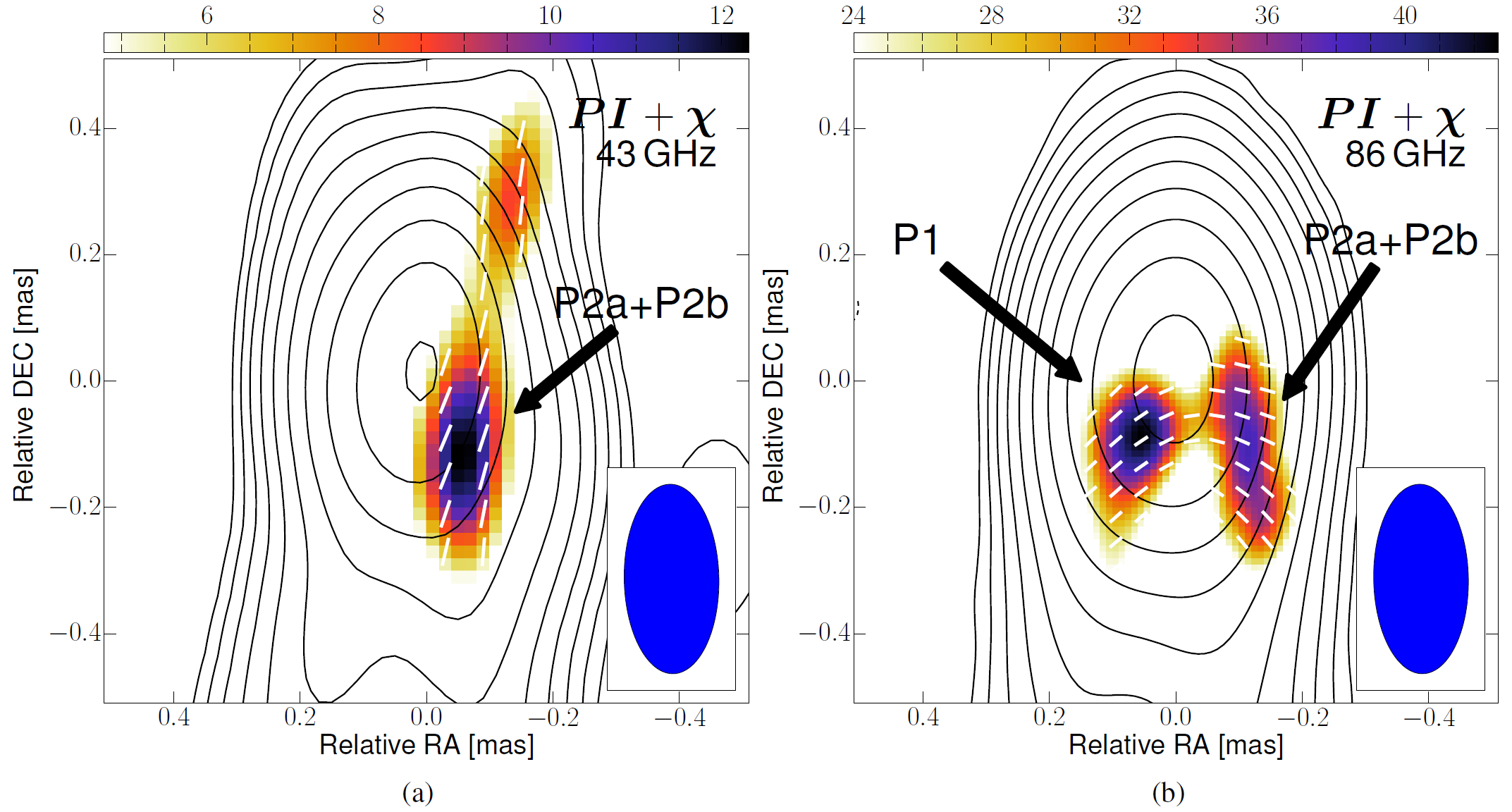}
\caption{Linear polarization images of the 3C 84 jet observed with the VLBA at 43 GHz (\textbf{left}) and with the GMVA at 86 GHz (\textbf{right}), taken from {Kim et al.} \cite{Kim2019}. The images were obtained quasi-simultaneously in May 2015 and convolved with the same beam. The contours represent the total intensity distributions. The colors show the linearly polarized intensity distributions, and the white ticks are the EVPAs. \label{fig:kim2019a}}
\end{figure}

\begin{figure}[H]
\includegraphics[width = 0.67\textwidth]{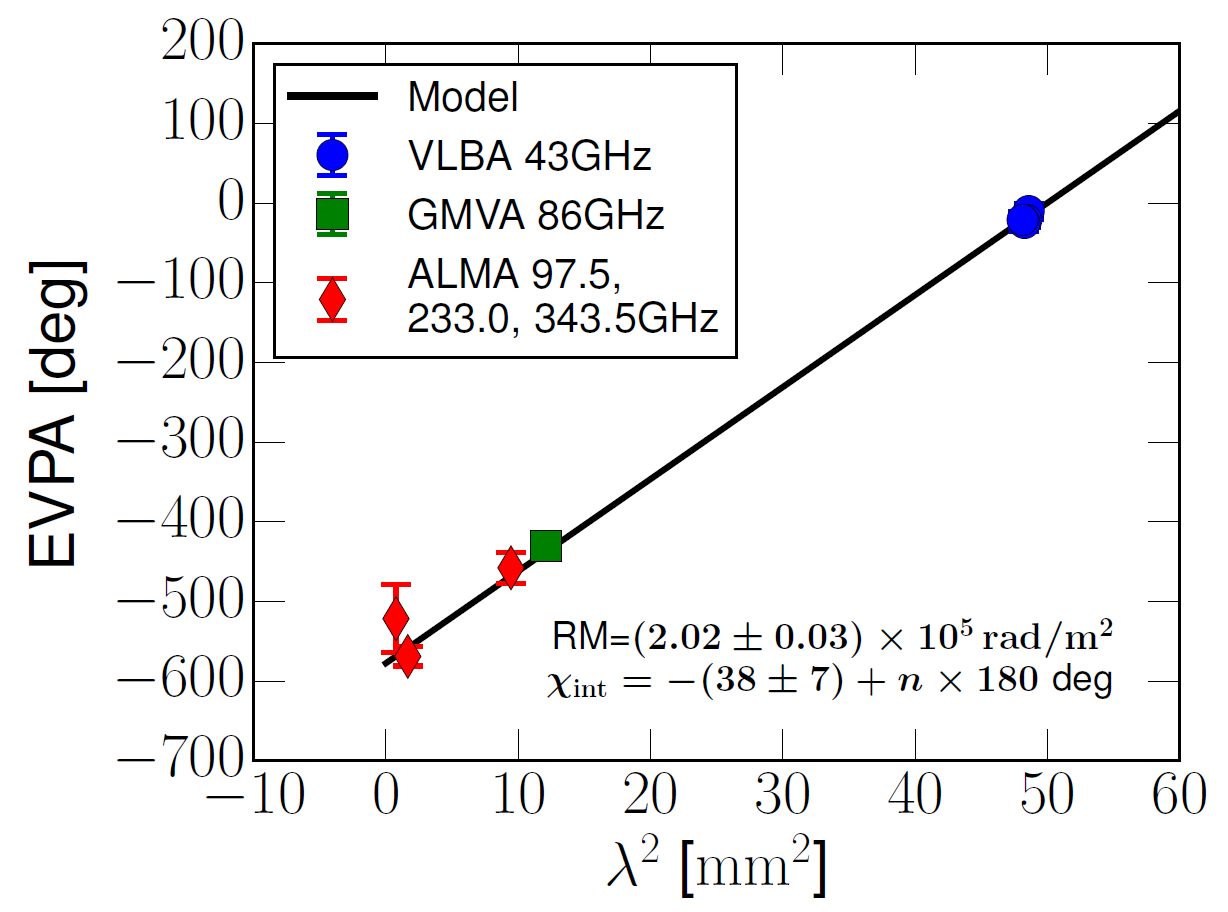}
\caption{EVPAs of 3C 84 as functions of $\lambda^2$ from contemporaneous VLBA 43 GHz (blue), GMVA 86~GHz (green), and ALMA 97.5, 233.0, and 343.5 GHz data (red). The black solid line is a linear fit to the data points, from which the best-fit $\rm RM = (2.02\pm0.03)\times10^5\ rad\ m^{-2}$ and $\chi_{\rm int} = -(38\pm7)^\circ$ are {derived} \cite{Kim2019}. \label{fig:kim2019b}}
\end{figure}

GMVA observations at 86 GHz in May 2015 detected linearly polarized emission in the jet near the core region (\cite{Kim2019}; see Figure~\ref{fig:kim2019a}). The linearly polarized regions are associated with the two limbs that emerge from the core. In combination with a contemporaneous VLBA 43 GHz image and ALMA multifrequency observations, a large RM of $\sim$$2~\times~10^5\ {\rm rad\ m^{-2}}$ was derived (Figure~\ref{fig:kim2019b}). On the basis of the large RM value, the observed depolarization at longer wavelengths, and the RM variability observed in a series of 43 GHz VLBA images, the Faraday rotation was proposed to occur in a boundary layer in a transversely stratified jet \cite{Kim2019}.

\unskip

The results of these studies indicate that AGN jets can be significantly depolarized at long wavelengths near black holes. Although the reason for the depolarization remains unclear, it may be related to turbulent black hole accretion flows. As an example, the M87 jet is substantially depolarized within the Bondi radius (e.g., \citep{ZT2002, Park2019a}), whereas it is moderately to highly polarized outside the Bondi radius (e.g., \citep{Owen1990, Perlman1999, Pasetto2021}). At short wavelengths, depolarization is expected to have a small effect regardless of its source (e.g., \citep{Burn1966, Sokoloff1998, Pasetto2018}). As a result, future high-resolution VLBI observations at mm wavelengths should allow us to investigate the magnetic field structures and the distributions of Faraday rotation measures in the launching and acceleration regions of AGN jets.

\subsection{3C 273}

The 3C 273 is a nearby flat spectrum radio quasar e.g., \citep{Paltani1998} at $z=0.158$ \citep{Strauss1992}. It hosts a powerful jet that extends to kiloparsec-scales (e.g., \citep{Meyer2016}). On parsec-scales, the jet exhibits a double helix morphology \citep{LZ2001}, where it is actively collimated \citep{Akiyama2018, Okino2021}. The jet has a rich linear polarization structure on pc scales, but the core is unpolarized (e.g.,~\mbox{\citep{Attridge2005, Hada2016}}). ALMA observations have detected a very high RM of $(5.0\pm0.3)\times10^5{\rm rad\ m^{-2}}$ at 1.3~mm~ \citep{Hovatta2019}, which is an order of magnitude greater than the RMs detected in the downstream jet~\mbox{\citep{Attridge2005, Hada2016}}. Based on this result, it appears that the core is depolarized at low frequencies but becomes polarized at mm wavelengths.

\begin{figure}[H]
\includegraphics[width = 0.45\textwidth]{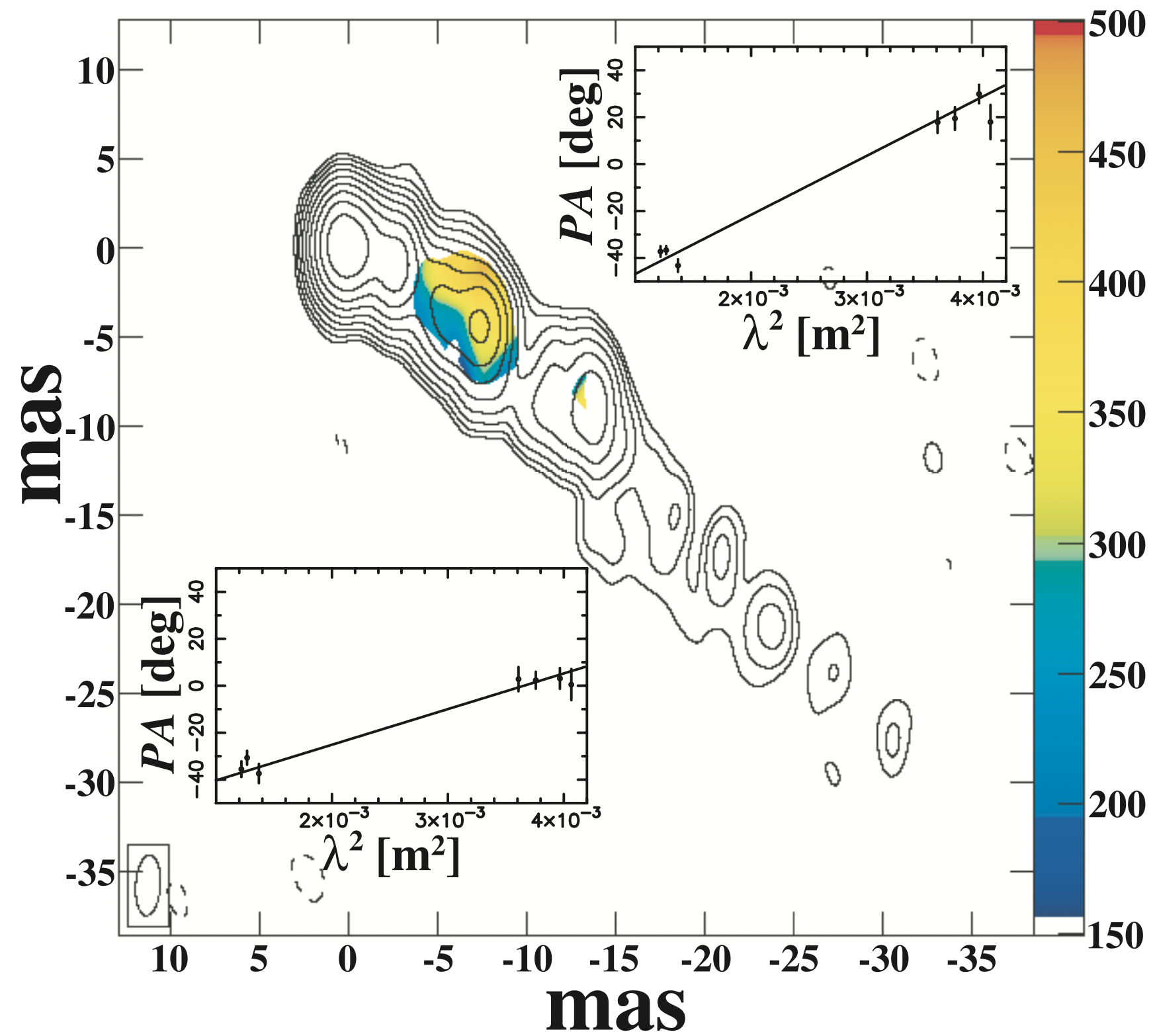}
\caption{{Colors} denote the RM distribution in the 3C 273 jet observed by Asada {et al.} \citep{Asada2002}, while the contours represent the total intensity distribution. The inlet plots represent the EVPAs as functions of $\lambda^2$ in the two sides of the jet where the transverse RM gradient is observed. \label{fig:asada2002}}
\end{figure}

Having a rich linear polarization structure on pc-scales, the wide jet of this source makes it possible to study the origin of Faraday rotation in AGN jets. Previous VLBA multifrequency observations between 5 and 8 GHz in 1995 revealed a transverse gradient in the RM across the jet (\citep{Asada2002}; Figure~\ref{fig:asada2002}). According to this result, the jet is surrounded by a helical or toroidal magnetic field. Subsequent VLBA observations confirmed this behavior, along with a temporal variation (e.g., \citep{ZT2005, Asada2008, Hovatta2012}). Several studies have investigated transverse gradients in RM across jets in other sources since this study (see \cite{Gabuzda2018} and \mbox{references therein). }

\begin{figure}[H]
\includegraphics[trim = 0mm -30mm 0mm 0mm, width = 0.42\textwidth]{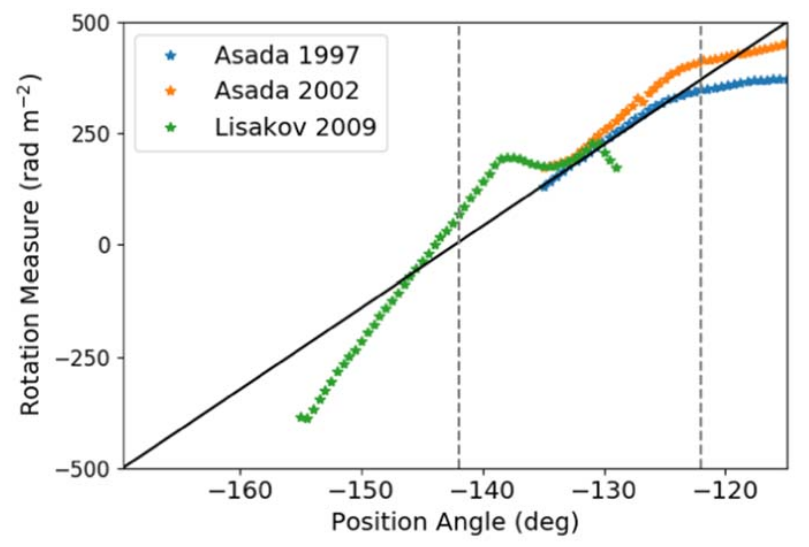}
\includegraphics[width = 0.32\textwidth]{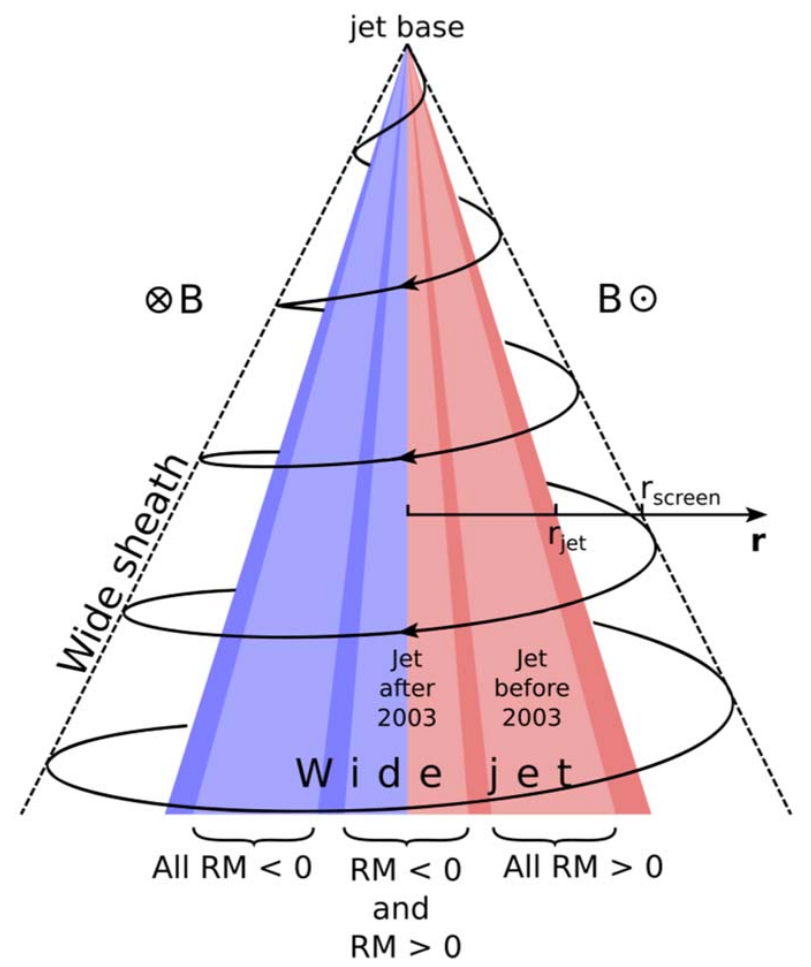}
\caption{(\textbf{Left}) The RMs of the 3C 273 jet as functions of the jet position angle at the projected distance of 9 mas from the core observed in 1997 (blue) and 2002 (orange), reported by \mbox{{Asada et al.} \citep{Asada2008}}, and in 2009 reported by Lisakov et al. \citep{Lisakov2021}. The two vertical dashed lines represent the average jet position angle before (at $-122^\circ$) and after (at $-142^\circ$) the major jet swing occurred in 2003. The black solid line denotes a linear fit to all data points. (\textbf{Right}) Illustration of the jet-sheath model developed by Lisakov et al. \citep{Lisakov2021} to explain the observed RM data. The jet illuminates different parts of the sheath that is threaded by a helical magnetic field in different epochs, resulting in the change in the transverse RM gradients. ©AAS. Reproduced with permission. \label{fig:lisakov2021}}
\end{figure}

VLBA observations in 2009 also detected transverse RM gradients in the 3C 273 jet~\citep{Lisakov2021}. Compared to Asada et al.'s earlier epochs of 1997 and 2002 \citep{Asada2008}, the jet position angle substantially changed during this epoch. The left panel of Figure~\ref{fig:lisakov2021} compares the RMs at the projected jet distance of 9 mas from the core as a function of the position angle of the jet. The RMs in different epochs are smoothly interconnected. In addition, a continuous gradient from negative to position RMs was observed with increasing position angle for the first time. In light of this result, the Faraday screen appears to be wider than the narrow relativistic jet of a single epoch, and different portions of the screen are illuminated by the jet as the jet's position angle changes over time. Over the course of more than ten years, the Faraday screen does not seem to have changed very much in its properties. By using this information, Lisakov et al. \citep{Lisakov2021} proposed a model for the jet-sheath system in 3C 273 in which the sheath is threaded by a helical magnetic field serving as a Faraday screen, as shown in the right panel of Figure~\ref{fig:lisakov2021}. According to these results, investigating the evolution of AGN jets and Faraday screens can provide valuable information regarding the nature of the Faraday rotation.

\unskip

\section{New Developments \& Future Prospects}
\label{sec:developments}

Recent developments have been made in the algorithms for instrumental polarization calibration as well as linear polarization imaging for VLBI data. Specifically, these developments were made for the purpose of directly imaging supermassive black holes in M87 and Sagittarius A* with EHT observations. Nevertheless, they will also be useful for studies of linear polarization of AGN jets by using cm-VLBI observations. We briefly describe them here and discuss their implications for future research.

Instrumental polarization signals are present in all realistic VLBI data because the antenna polarization is not perfectly circular or linear (e.g., \citep{Thompson2017}). These signals can cause an unpolarized source to appear polarized and are usually of the same order as the intrinsic polarization signals of the source. The \texttt{LPCAL} task \cite{Leppanen1995} implemented in the astronomical image processing system (AIPS; \cite{Greisen2003}) has long been widely used to calibrate the instrumental polarization of VLBI data. \texttt{LPCAL} has been successful in a great deal of studies, but it has three main limitations. The first is that it assumes that linear polarization and total intensity structures of calibrators are similar, the so-called ``similarity approximation,'' which does not stand up in all cases, especially at high frequencies. It also does not permit us to obtain instrumental polarization solutions based on data from multiple calibrators, despite the fact that instrumental polarization is generally expected to be the same for all sources. In addition, it is based on linearized instrumental polarization models, which are violated when the level of instrumental polarization becomes moderate or high.

In order to overcome these limitations, \texttt{GPCAL} \citep{Park2021a} and \texttt{PolSolve} \citep{Martividal2021}, which is an extension of \texttt{LPCAL}, have been developed. These pipelines utilize the instrumental polarization model with second-order terms included. By iterating CLEAN for Stokes $Q$ and $U$ data and solving for instrumental polarization models using the CLEAN models, they take into account the complexity of polarization structures of calibrators. Furthermore, data from multiple calibrators can be used simultaneously to improve the accuracy of calibration. \texttt{GPCAL} is based on AIPS and Difmap and utilizes ParselTongue, a Python interface to AIPS, whereas \texttt{PolSolve} is based on NRAO's common astronomy software applications (CASA).

The \texttt{eht-imaging} software library \cite{Chael2016, Chael2018} implements an imaging algorithm for VLBI data based on the regularized maximum likelihood (RML) method. In addition, it is capable of performing polarimetric image reconstruction by using RML and instrumental polarization calibration on an iterative basis. In addition, there are two methods based on Markov chain Monte Carlo (MCMC) schemes: D-term modeling code (DMC; \cite{Pesce2021}) and THEMIS \citep{Broderick2020}. The methods explore simultaneously the posterior space of the full Stokes image, the complex gains, and instrumental polarization for all stations. All of these five methods have been applied to the first linear polarization imaging of the M87 black hole and have yielded consistent results \citep{EHT2021a}.

\begin{figure}[H]
\includegraphics[width = 0.9\textwidth]{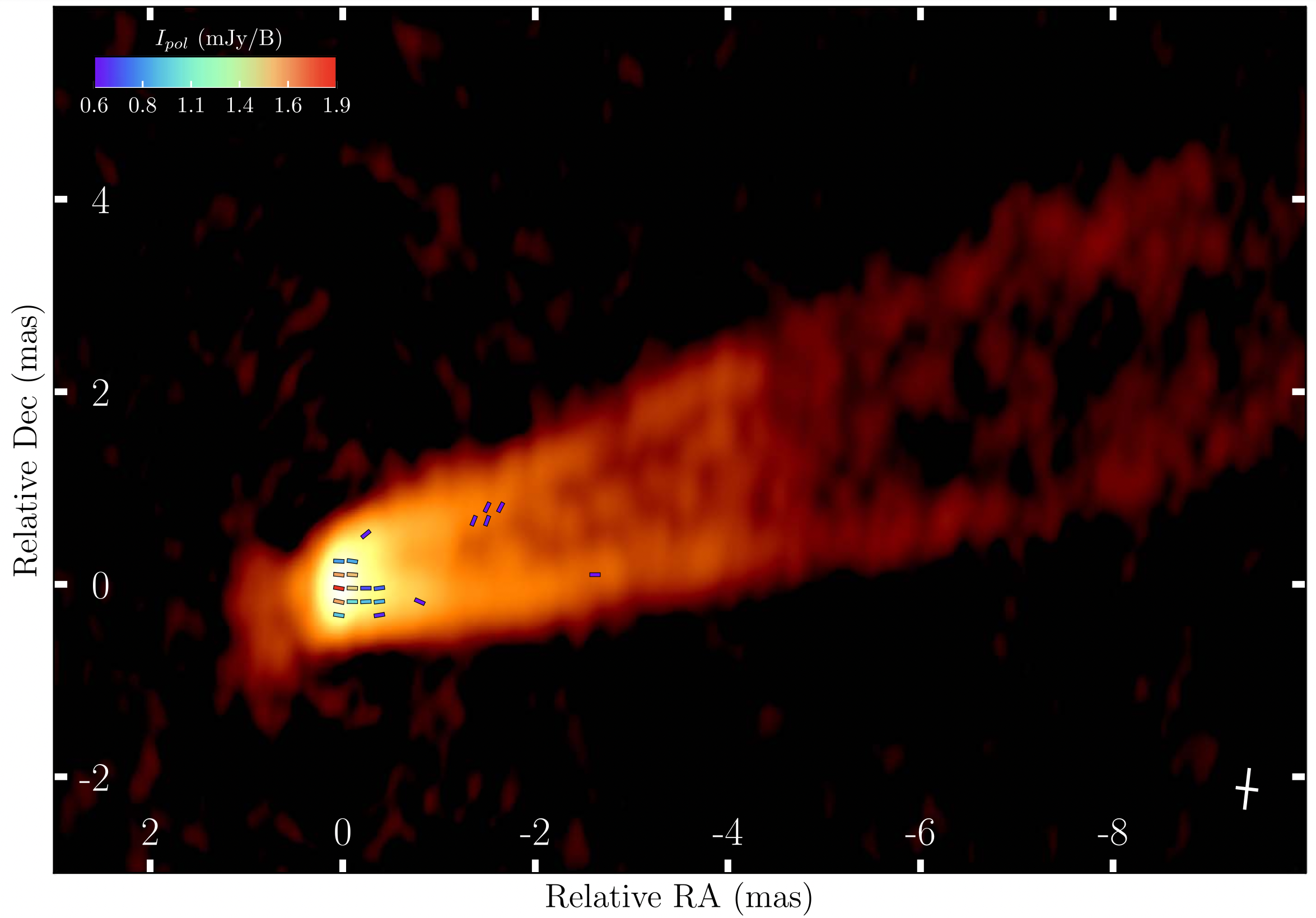}
\caption{{An} average image of the linear polarization images of the M87 jet observed with the VLBA at 43 GHz in four epochs in 2007 and {2018} \cite{Park2021c}. The colors show the total intensity emission distribution and the ticks represent the EVPAs with the colors scaling with linear polarization intensity. ©AAS. Reproduced with permission. \label{fig:park2021}}
\end{figure}

In Figure~\ref{fig:park2021}, we present the linear polarization image of the M87 jet obtained with the VLBA at 43 GHz in order to demonstrate the importance of improved instrumental polarization calibration by using the newly developed algorithms. This image is an average of four images obtained with instrumental polarization calibration by using \texttt{GPCAL} between 2007 and 2018 \citep{Park2021c}. In the image, a compact linear polarization structure is visible near the total intensity core. The same dataset has, however, been analyzed in previous studies, from which more complex linear polarization structures have been derived in the core region \citep{Walker2018, Kravchenko2020}. These studies used the calibrator 3C 279 for instrumental polarization calibration by applying \texttt{LPCAL}. Park et al. performed the tests by using synthetic data and closure traces {(}
{closure traces are quantities that are insensitive to both antenna gain and polarization leakage corruptions \citep{BP2020}}{)}  and demonstrated that the similarity approximation used in \texttt{LPCAL} did not hold up well for 3C 279 because its polarization structure is complex, resulting in artifacts in linear polarization images previously presented.

\section{Circular Polarization}

{
Thus far, we have discussed the results of linear polarization studies of AGN jets. The circular polarization (CP) of AGN jets has received relatively little attention because they usually exhibit very low levels of CP. In this section, we briefly introduce some of the CP~studies. 
}

{
Circular polarization in AGN jets are believed to be produced by two main mechanisms. A small amount of CP can be produced by synchrotron emission. In this case, the observed CP may be used to probe the magnetic field structure of the source. It is also possible for CP to occur as a result of propagation effects through Faraday conversion. In the case of electromagnetic waves propagating a medium with a magnetic field component perpendicular to the propagation direction, plasma electrons accelerated by the electric field component parallel to the field oscillate freely, whereas those accelerated by the orthogonal component experience Lorentz force. There is a delay in the parallel component with respect to the orthogonal component, which is manifested as a CP. In this case, CP observations can be used to investigate low-energy relativistic particles in plasma. Jones~\cite{Jones1988} suggested that the latter mechanism should be more efficient at producing CP than the former. 
}

{
The first reliable observations of circular polarization were made by Gilbert \& Conway~\cite{GilbertConway1970}, who found a fractional circular polarization level of $m_C\sim0.2$\% for CTA 102 and 2203--18 at 49~cm. Weiler \& de Pater \cite{WeilerdePater1983} as well as Komesaroff et al. \cite{Komesaroff1984} obtained a large number of integrated CP observations of AGNs, finding typical levels of $m_C \lesssim 0.1$\%.
}

{
It was not possible to detect reliable resolved CP in AGN jets until Homan and Wardle introduced calibration and imaging techniques that could disentangle source-intrinsic CP signals from instrumental signals. These techniques estimate the ratio of antenna gains between right- and left-hand circular polarizations (R/L) using many sources observed in the same run. They could derive the R/L gain ratio as a function of time by assuming that the sources have low CP levels and no preferred CP signs. Observing a sufficient number of sources over a long period of time is essential for the success of this method. A series of studies conducted by this group detected CP at a level of $m_C\sim0.5$\% between 5 and 22~GHz and up to $\sim3\%$ in 3C~84. According to these studies, there is no apparent correlation between linear and circular polarizations, and in the majority of the sources, the CP sign tends to remain stable over time. Although they were unable to distinguish between the dominant mechanism generating the CP, the remarkable consistency of the CP sign suggested a stable unidirectional magnetic field over very long periods of time.
}



It was discovered by O'Sullivan that there was an anti-correlation between the linear and circular polarizations in PKS~2126--158 for the first time. This behavior is expected if Faraday conversion of linear to circular polarization occurs, and this result demonstrates that Faraday conversion may be in action in AGN jets.

\section{Conclusions}
\label{sec:conclusions}

This review summarizes recent progress in polarimetric observations of AGN jets using VLBI. Our discussion focused on blazars, nearby AGN jets, and recent developments in VLBI polarization calibration and imaging techniques. Below is a brief summary of each~topic.

\begin{itemize}
    \item \textbf{{Blazars.}
} 
    It seems clear we have advanced a lot in the understanding of the polarization emission in blazars, and the connection between the properties between optical and radio polarization. Although it was clear that various radiative transfer mechanisms, such as Faraday effects, would affect the multi-band polarization observation, the understanding of their extent and characteristics has significantly improved in the recent years.
    
 The works by Gabuzda et al. \citep{Gabuzda1994,Gabuzda1996} and Algaba et al. \citep{Algaba2011,Algaba2012} aiming to resolve the location and characteristics of the specific radio polarized regions have been enhanced and complemented by more detailed analysis such as the analysis by Park et al. \citep{Park2018}, who considered (i) higher frequencies, (ii) explicit RM dependence with frequency, and (iii) a saturation point where the optical depth transitions into optically thin, and RM does not increase any more. Once these aspects are considered, the connection between optical and radio emission are better understood.
    
 On the other hand, the origins of the emission are now also better understood to be ascribed to a shock or magnetic reconnection phenomenology. The multi-band study of the time evolution of polarization together with follow-up of the VLBI structural changes has provided not only understanding of the intrinsic jet polarization properties, but also of the jet dynamics and evolution.

    \vspace{0.2cm}
    \item \textbf{{Nearby AGN jets.}} 
    Due to its proximity and very massive black hole, M87 enables detailed polarimetric studies of the supermassive black hole and its jet. In multifrequency VLBA observations, RM magnitudes were found to decrease with increasing distance from the black hole within the Bondi radius \cite{Park2019a}. An analytic black hole accretion flow model was applied to the RM data, from which the pressure profile of the medium surrounding the jet was determined. The result is in good agreement with the current understanding that the M87 jet is systematically collimated by the pressure of winds, which have actually been reproduced in recent GRMHD simulations \citep{Nakamura2018, Chatterjee2019}. Recent observations with the EHT have revealed a linear polarization image of the M87 black hole \cite{EHT2021a}. The observed EVPA of the ring follows an azimuthal pattern. After comparing the results with GRMHD models, it was concluded that M87 is in a MAD state, in which there is a strong poloidal magnetic field in the vicinity of the black hole \cite{EHT2021b}. Combined with a spinning black hole, this magnetic field can produce powerful jets, as seen in M87.
    
The nearby radio galaxy 3C 84 demonstrates many interesting features, including restarted jet activity, a strong collision between the jet and a compact dense cloud, and so on. The pc-scale jet of this source has been known for its very weak linear polarization. However, recent VLBI observations made at 43 and 86 GHz have detected significant patches of linear polarization in the jet right downstream the core \cite{Kim2019}. By using these results, combined with ALMA-only linear polarization measurements at 98, 233, and 344 GHz, a large RM of $\sim$$2\times10^5\ {\rm rad\ m^{-2}}$ was derived. The RM, depolarization, and RM variability may be explained by Faraday rotation occurring in a boundary layer of a transversely stratified jet \cite{Kim2019}. The results of this study suggest that future observations of polarization at mm wavelengths can provide valuable information on the magnetic field structure and the RM distribution in nearby AGNs, because shorter wavelengths are expected to produce less depolarization.

 The nearby flat spectrum radio quasar 3C 273 hosts a powerful jet extending to kpc-scales. Due to the rich linear polarization structure of the jet and its large width on pc scales, VLBA observations in 1997 and 2002 revealed that the jet exhibits a significant transverse gradient in RMs \cite{Asada2002, Asada2008}. This result indicates that a helical or toroidal magnetic field is wrapping around the jet. Further observations of linear polarization with the VLBA provided even more interesting results. Lisakov~et~al.~\cite{Lisakov2021} found that transverse RM gradients were also observed in the jet based on VLBA observations in 2009. As a result of the swing in jet direction in 2003, the jet position angles between 1997/2002 and 2009 were significantly different. In contrast, the RMs at the same jet distance in the different epochs are smoothly interconnected. It appears that the Faraday screen is wider than the narrow relativistic jet observed at one epoch, and that the jet illuminates different parts of the screen in different epochs. Based on this result, the jet-sheath model with a sheath threaded by a helical magnetic field serving as the Faraday screen was derived.

    \vspace{0.2cm}
    \item \textbf{{Recent Developments in VLBI polarimetry.}} A number of novel algorithms for instrumental polarization calibration and polarimetric imaging have been developed recently, including \texttt{GPCAL} \cite{Park2021a}, \texttt{PolSolve} \cite{Martividal2021}, \texttt{eht-imaging} \cite{Chael2016, Chael2018}, DMC \cite{Pesce2021}, and THEMIS \cite{Broderick2020}. These softwares and pipelines overcome some of the limitations of the existing software, \texttt{LPCAL}, which has been widely used for instrument polarization calibration for VLBI data for decades. They have been applied to the first polarimetric imaging of the M87 black hole conducted with the EHT and have been demonstrated to be effective \cite{EHT2021a}. In the region of jet launching and acceleration/collimation, AGN jets are typically weakly polarized, presumably as a result of turbulent accretion flows surrounding them. Therefore, these advanced software tools will be useful for studying the magnetic field structure of AGN jets with future polarimetric VLBI observations. In a recent study, it was demonstrated that the high calibration accuracy provided by GPCAL can significantly improve a linear polarization view of the sub-parsec core of the M87 \cite{Park2021c}.
    
    \vspace{0.2cm}
    {\item \textbf{{Circular Polarization}} Circular polarization in AGN jets is believed to be produced by either synchrotron emission or Faraday conversion. The resulting CP is typically of the order of $m_C\lesssim0.5$\% and is challenging to detect reliably. Nevertheless, a few studies, using dedicated calibration techniques, found that the CP sign is generally stable over time for several sources. This suggests that the magnetic fields responsible for the observed CP may also be stable. Some studies have attempted to model the observed CP signs in several AGN jets by assuming a helical magnetic field in the jets as well as linear polarization vectors and transverse RM gradients. Moreover, an anti-correlation between the fractional linear and circular polarization was discovered in PKS 2126--158, indicating that Faraday conversion may be in action in this source.}

\end{itemize}


\authorcontributions{{Conceptualization, J.P. and J.C.A.; Writing---Original Draft Preparation, J.P. and J.C.A.; Writing---Review \& Editing, J.P. and J.C.A.; Visualization, J.P. and J.C.A. All authors have read and agreed to the published version of the manuscript.}}

\funding{J.C.A would like to acknowledge the beneficial grant provided by the University Malaya’s Grant, GPF081-2020.}

\dataavailability{{Not applicable.}}

\acknowledgments{{J.P. acknowledges ﬁnancial support through the EACOA Fellowship awarded by the East Asia Core Observatories Association, which consists of the Academia Sinica Institute of Astronomy and Astrophysics, the National Astronomical Observatory of Japan, Center for Astronomical Mega-Science, Chinese Academy of Sciences, and the Korea Astronomy and Space Science~Institute.}}

\conflictsofinterest{{The authors declare no conflict of interest.}}

\begin{adjustwidth}{-\extralength}{0cm}
\reftitle{References}

\end{adjustwidth}


\begin{thebibliography}{999}

\bibitem[{Blandford} \em{et~al.}(2019){Blandford}, {Meier}, and
{Readhead}]{Blandford2019}
{Blandford}, R.; {Meier}, D.; {Readhead}, A.
\newblock {Relativistic Jets from Active Galactic Nuclei}.
\newblock {\em \araa} {\bf 2019}, {\em 57},~467--509.
\newblock  [\href{http://doi.org/10.1146/annurev-astro-081817-051948}{CrossRef}]

\bibitem[{Jorstad} \em{et~al.}(2017){Jorstad}, {Marscher}, {Morozova},
{Troitsky}, {Agudo}, {Casadio}, {Foord}, {G{\'o}mez}, {MacDonald}, {Molina},
{L{\"a}hteenm{\"a}ki}, {Tammi}, and {Tornikoski}]{Jorstad2017}
{Jorstad}, S.G.; {Marscher}, A.P.; {Morozova}, D.A.; {Troitsky}, I.S.; {Agudo},
I.; {Casadio}, C.; {Foord}, A.; {G{\'o}mez}, J.L.; {MacDonald}, N.R.;
{Molina}, S.N.;  et~al.
\newblock {Kinematics of Parsec-scale Jets of Gamma-Ray Blazars at 43 GHz
within the VLBA-BU-BLAZAR Program}.
\newblock {\em \apj} {\bf 2017}, {\em 846},~98.
\newblock  [\href{http://dx.doi.org/10.3847/1538-4357/aa8407}{CrossRef}]

\bibitem[{Lister} \em{et~al.}(2018){Lister}, {Aller}, {Aller}, {Hodge},
{Homan}, {Kovalev}, {Pushkarev}, and {Savolainen}]{Lister2018}
{Lister}, M.L.; {Aller}, M.F.; {Aller}, H.D.; {Hodge}, M.A.; {Homan}, D.C.;
{Kovalev}, Y.Y.; {Pushkarev}, A.B.; {Savolainen}, T.
\newblock {MOJAVE. XV. VLBA 15 GHz Total Intensity and Polarization Maps of 437
Parsec-scale AGN Jets from 1996 to 2017}.
\newblock {\em \apjs} {\bf 2018}, {\em 234},~12.
\newblock  [\href{http://dx.doi.org/10.3847/1538-4365/aa9c44}{CrossRef}]

\bibitem[{Fabian}(2012)]{Fabian2012}
{Fabian}, A.C.
\newblock {Observational Evidence of Active Galactic Nuclei Feedback}.
\newblock {\em \araa} {\bf 2012}, {\em 50},~455--489.
\newblock  [\href{http://dx.doi.org/10.1146/ annurev-astro-081811-125521}{CrossRef}]

\bibitem[{Yuan} \em{et~al.}(2018){Yuan}, {Yoon}, {Li}, {Gan}, {Ho}, and
{Guo}]{Yuan2018}
{Yuan}, F.; {Yoon}, D.; {Li}, Y.P.; {Gan}, Z.M.; {Ho}, L.C.; {Guo}, F.
\newblock {Active Galactic Nucleus Feedback in an Elliptical Galaxy with the
Most Updated AGN Physics. I. Low Angular Momentum Case}.
\newblock {\em \apj} {\bf 2018}, {\em 857},~121,
\newblock  [\href{http://dx.doi.org/10.3847/1538-4357/aab8f8}{CrossRef}]

\bibitem[{Yoon} \em{et~al.}(2018){Yoon}, {Yuan}, {Gan}, {Ostriker}, {Li}, and
{Ciotti}]{Yoon2018}
{Yoon}, D.; {Yuan}, F.; {Gan}, Z.M.; {Ostriker}, J.P.; {Li}, Y.P.; {Ciotti}, L.
\newblock {Active Galactic Nucleus Feedback in an Elliptical Galaxy with the
Most Updated AGN Physics. II. High Angular Momentum Case}.
\newblock {\em \apj} {\bf 2018}, {\em 864},~6.
\newblock  [\href{http://dx.doi.org/10.3847/1538-4357/aad37e}{CrossRef}]

\bibitem[{Kataoka} \em{et~al.}(2006){Kataoka}, {Stawarz}, {Aharonian},
{Takahara}, {Ostrowski}, and {Edwards}]{Kataoka2006}
{Kataoka}, J.; {Stawarz}, {\L}.; {Aharonian}, F.; {Takahara}, F.; {Ostrowski},
M.; {Edwards}, P.G.
\newblock {The X-Ray Jet in Centaurus A: Clues to the Jet Structure and
Particle Acceleration}.
\newblock {\em \apj} {\bf 2006}, {\em 641},~158--168.  
\newblock  [\href{http://dx.doi.org/10.1086/500407}{CrossRef}]

\bibitem[{Abdo} \em{et~al.}(2010){Abdo}, {Ackermann}, {Agudo}, {Ajello},
{Allafort}, {Aller}, {Aller}, {Antolini}, {Arkharov}, {Axelsson}, {Bach},
{Baldini}, {Ballet}, {Barbiellini}, {Bastieri}, {Bechtol}, {Bellazzini},
{Berdyugin}, {Berenji}, {Blandford}, {Blinov}, {Bloom}, {Boettcher},
{Bonamente}, {Borgland}, {Bouvier}, {Bregeon}, {Brez}, {Brigida}, {Bruel},
{Buehler}, {Buemi}, {Burnett}, {Buson}, {Caliandro}, {Cameron}, {Caraveo},
{Carosati}, {Carrigan}, {Casandjian}, {Cavazzuti}, {Cecchi}, {{\c{C}}elik},
{Chekhtman}, {Chen}, {Cheung}, {Chiang}, {Ciprini}, {Claus}, {Cohen-Tanugi},
{Conrad}, {Corbel}, {Costamante}, {Dermer}, {de Angelis}, {de Palma},
{Donato}, {Silva}, {Drell}, {Dubois}, {Dumora}, {Farnier}, {Favuzzi},
{Fegan}, {Ferrara}, {Focke}, {Forn{\'e}}, {Fortin}, {Fukazawa}, {Funk},
{Fusco}, {Gargano}, {Gasparrini}, {Gehrels}, {Germani}, {Giebels},
{Giglietto}, {Giordano}, {Giroletti}, {Glanzman}, {Godfrey}, {Grenier},
{Grove}, {Guiriec}, {Gurwell}, {Gusbar}, {G{\'o}mez}, {Hadasch},
{Hagen-Thorn}, {Hayashida}, {Hays}, {Horan}, {Hughes}, {J{\'o}hannesson},
{Johnson}, {Johnson}, {Kamae}, {Katagiri}, {Kataoka}, {Kawai}, {Kimeridze},
{Kn{\"o}dlseder}, {Konstantinova}, {Kopatskaya}, {Koptelova}, {Kovalev},
{Kurtanidze}, {Kuss}, {Lahteenmaki}, {Lande}, {Larionov}, {Larionova},
{Larionova}, {Larsson}, {Latronico}, {Lee}, {Leto}, {Lister}, {Longo},
{Loparco}, {Lott}, {Lovellette}, {Lubrano}, {Madejski}, {Makeev}, {Massaro},
{Mazziotta}, {McConville}, {McEnery}, {McHardy}, {Michelson}, {Mitthumsiri},
{Mizuno}, {Moiseev}, {Monte}, {Monzani}, {Morozova}, {Morselli},
{Moskalenko}, {Murgia}, {Naumann-Godo}, {Nikolashvili}, {Nolan}, {Norris},
{Nuss}, {Ohno}, {Ohsugi}, {Okumura}, {Omodei}, {Orlando}, {Ormes}, {Ozaki},
{Paneque}, {Panetta}, {Parent}, {Pasanen}, {Pelassa}, {Pepe},
{Pesce-Rollins}, {Piron}, {Porter}, {Pushkarev}, {Rain{\`o}}, {Raiteri},
{Rando}, {Razzano}, {Reimer}, {Reimer}, {Reinthal}, {Ripken}, {Ritz},
{Roca-Sogorb}, {Rodriguez}, {Roth}, {Roustazadeh}, {Ryde}, {Sadrozinski},
{Sander}, {Scargle}, {Sgr{\`o}}, {Sigua}, {Smith}, {Sokolovsky}, {Spandre},
{Spinelli}, {Starck}, {Strickman}, {Suson}, {Takahashi}, {Takahashi},
{Takalo}, {Tanaka}, {Taylor}, {Thayer}, {Thayer}, {Thompson}, {Tibaldo},
{Tornikoski}, {Torres}, {Tosti}, {Tramacere}, {Trigilio}, {Troitsky},
{Umana}, {Usher}, {Vandenbroucke}, {Vasileiou}, {Vilchez}, {Villata},
{Vitale}, {Waite}, {Wang}, {Winer}, {Wood}, {Yang}, {Ylinen}, and
{Ziegler}]{Abdo2010}
{Abdo}, A.A.; {Ackermann}, M.; {Agudo}, I.; {Ajello}, M.; {Allafort}, A.;
{Aller}, H.D.; {Aller}, M.F.; {Antolini}, E.; {Arkharov}, A.A.; {Axelsson},
M.;  et~al.
\newblock {Fermi Large Area Telescope and Multi-wavelength Observations of the
Flaring Activity of PKS 1510-089 between 2008 September and 2009 June}.
\newblock {\em \apj} {\bf 2010}, {\em 721},~1425--1447.
\newblock  [\href{http://dx.doi.org/10.1088/0004-637X/721/2/1425}{CrossRef}]

\bibitem[{Park} \em{et~al.}(2019){Park}, {Lee}, {Kim}, {Hodgson}, {Trippe},
{Kim}, {Algaba}, {Kino}, {Zhao}, {Lee}, and {Gurwell}]{Park2019c}
{Park}, J.; {Lee}, S.S.; {Kim}, J.Y.; {Hodgson}, J.A.; {Trippe}, S.; {Kim},
D.W.; {Algaba}, J.C.; {Kino}, M.; {Zhao}, G.Y.; {Lee}, J.W.;  et~al.
\newblock {Ejection of Double Knots from the Radio Core of PKS 1510-089 during
the Strong Gamma-Ray Flares in 2015}.
\newblock {\em \apj} {\bf 2019}, {\em 877},~106.
\newblock  [\href{http://dx.doi.org/10.3847/1538-4357/ab1b27}{CrossRef}]

\bibitem[{Aleksi{\'c}} \em{et~al.}(2011{\natexlab{a}}){Aleksi{\'c}},
{Antonelli}, {Antoranz}, {Backes}, {Barrio}, {Bastieri}, {Becerra
Gonz{\'a}lez}, {Bednarek}, {Berdyugin}, {Berger}, {Bernardini}, {Biland},
{Blanch}, {Bock}, {Boller}, {Bonnoli}, {Borla Tridon}, {Braun}, {Bretz},
{Ca{\~n}ellas}, {Carmona}, {Carosi}, {Colin}, {Colombo}, {Contreras},
{Cortina}, {Cossio}, {Covino}, {Dazzi}, {de Angelis}, {de Cea Del Pozo}, {de
Lotto}, {Delgado Mendez}, {Diago Ortega}, {Doert}, {Dom{\'\i}nguez}, {Dominis
Prester}, {Dorner}, {Doro}, {Elsaesser}, {Ferenc}, {Fonseca}, {Font},
{Fruck}, {Garc{\'\i}a L{\'o}pez}, {Garczarczyk}, {Garrido}, {Giavitto},
{Godinovi{\'c}}, {Hadasch}, {H{\"a}fner}, {Herrero}, {Hildebrand}, {Hose},
{Hrupec}, {Huber}, {Jogler}, {Klepser}, {Kr{\"a}henb{\"u}hl}, {Krause}, {La
Barbera}, {Lelas}, {Leonardo}, {Lindfors}, {Lombardi}, {L{\'o}pez}, {Lorenz},
{Majumdar}, {Makariev}, {Maneva}, {Mankuzhiyil}, {Mannheim}, {Maraschi},
{Mariotti}, {Mart{\'\i}nez}, {Mazin}, {Meucci}, {Miranda}, {Mirzoyan},
{Miyamoto}, {Mold{\'o}n}, {Moralejo}, {Nieto}, {Nilsson}, {Orito}, {Oya},
{Paoletti}, {Pardo}, {Paredes}, {Partini}, {Pasanen}, {Pauss},
{Perez-Torres}, {Persic}, {Peruzzo}, {Pilia}, {Pochon}, {Prada}, {Prada
Moroni}, {Prandini}, {Puljak}, {Reichardt}, {Reinthal}, {Rhode}, {Rib{\'o}},
{Rico}, {R{\"u}gamer}, {R{\"u}ger}, {Saggion}, {Saito}, {Saito}, {Salvati},
{Satalecka}, {Scalzotto}, {Scapin}, {Schultz}, {Schweizer}, {Shayduk},
{Shore}, {Sillanp{\"a}{\"a}}, {Sitarek}, {Sobczynska}, {Spanier}, {Spiro},
{Stamerra}, {Steinke}, {Storz}, {Strah}, {Suri{\'c}}, {Takalo}, {Tavecchio},
{Temnikov}, {Terzi{\'c}}, {Tescaro}, {Teshima}, {Thom}, {Tibolla}, {Torres},
{Treves}, {Vankov}, {Vogler}, {Wagner}, {Weitzel}, {Zabalza}, {Zandanel}, and
{Zanin}]{Aleksic2011a}
{Aleksi{\'c}}, J.; {Antonelli}, L.A.; {Antoranz}, P.; {Backes}, M.; {Barrio},
J.A.; {Bastieri}, D.; {Becerra Gonz{\'a}lez}, J.; {Bednarek}, W.;
{Berdyugin}, A.; {Berger}, K.;  et~al.
\newblock {MAGIC Observations and multiwavelength properties of the quasar 3C
279 in 2007 and 2009}.
\newblock {\em \aap} {\bf 2011}, {\em 530},~A4.
\newblock  [\href{http://dx.doi.org/10.1051/0004-6361/201116497}{CrossRef}]

\bibitem[{Aleksi{\'c}} \em{et~al.}(2011{\natexlab{b}}){Aleksi{\'c}},
{Antonelli}, {Antoranz}, {Backes}, {Barrio}, {Bastieri}, {Becerra
Gonz{\'a}lez}, {Bednarek}, {Berdyugin}, {Berger}, {Bernardini}, {Biland},
{Blanch}, {Bock}, {Boller}, {Bonnoli}, {Borla Tridon}, {Braun}, {Bretz},
{Ca{\~n}ellas}, {Carmona}, {Carosi}, {Colin}, {Colombo}, {Contreras},
{Cortina}, {Cossio}, {Covino}, {Dazzi}, {De Angelis}, {De Cea del Pozo}, {De
Lotto}, {Delgado Mendez}, {Diago Ortega}, {Doert}, {Dom{\'\i}nguez}, {Dominis
Prester}, {Dorner}, {Doro}, {Elsaesser}, {Ferenc}, {Fonseca}, {Font},
{Fruck}, {Garc{\'\i}a L{\'o}pez}, {Garczarczyk}, {Garrido}, {Giavitto},
{Godinovi{\'c}}, {Hadasch}, {H{\"a}fner}, {Herrero}, {Hildebrand},
{H{\"o}hne-M{\"o}nch}, {Hose}, {Hrupec}, {Huber}, {Jogler}, {Klepser},
{Kr{\"a}henb{\"u}hl}, {Krause}, {La Barbera}, {Lelas}, {Leonardo},
{Lindfors}, {Lombardi}, {L{\'o}pez}, {Lorenz}, {Makariev}, {Maneva},
{Mankuzhiyil}, {Mannheim}, {Maraschi}, {Mariotti}, {Mart{\'\i}nez}, {Mazin},
{Meucci}, {Miranda}, {Mirzoyan}, {Miyamoto}, {Mold{\'o}n}, {Moralejo},
{Nieto}, {Nilsson}, {Orito}, {Oya}, {Paneque}, {Paoletti}, {Pardo},
{Paredes}, {Partini}, {Pasanen}, {Pauss}, {Perez-Torres}, {Persic},
{Peruzzo}, {Pilia}, {Pochon}, {Prada}, {Prada Moroni}, {Prandini}, {Puljak},
{Reichardt}, {Reinthal}, {Rhode}, {Rib{\'o}}, {Rico}, {R{\"u}gamer},
{Saggion}, {Saito}, {Saito}, {Salvati}, {Satalecka}, {Scalzotto}, {Scapin},
{Schultz}, {Schweizer}, {Shayduk}, {Shore}, {Sillanp{\"a}{\"a}}, {Sitarek},
{Sobczynska}, {Spanier}, {Spiro}, {Stamerra}, {Steinke}, {Storz}, {Strah},
{Suri{\'c}}, {Takalo}, {Tavecchio}, {Temnikov}, {Terzi{\'c}}, {Tescaro},
{Teshima}, {Thom}, {Tibolla}, {Torres}, {Treves}, {Vankov}, {Vogler},
{Wagner}, {Weitzel}, {Zabalza}, {Zandanel}, {Zanin}, {MAGIC Collaboration},
{Tanaka}, {Wood}, and {Buson}]{Aleksic2011b}
{Aleksi{\'c}}, J.; {Antonelli}, L.A.; {Antoranz}, P.; {Backes}, M.; {Barrio},
J.A.; {Bastieri}, D.; {Becerra Gonz{\'a}lez}, J.; {Bednarek}, W.;
{Berdyugin}, A.; {Berger}, K.;  et~al.
\newblock {MAGIC Discovery of Very High Energy Emission from the FSRQ PKS
1222+21}.
\newblock {\em \apjl} {\bf 2011}, {\em 730},~L8.
\newblock  [\href{http://dx.doi.org/10.1088/2041-8205/730/1/L8}{CrossRef}]

\bibitem[{Aleksi{\'c}} \em{et~al.}(2014){Aleksi{\'c}}, {Ansoldi}, {Antonelli},
{Antoranz}, {Babic}, {Bangale}, {Barres de Almeida}, {Barrio}, {Becerra
Gonz{\'a}lez}, {Bednarek}, {Bernardini}, {Biland}, {Blanch}, {Bonnefoy},
{Bonnoli}, {Borracci}, {Bretz}, {Carmona}, {Carosi}, {Carreto Fidalgo},
{Colin}, {Colombo}, {Contreras}, {Cortina}, {Covino}, {Da Vela}, {Dazzi}, {De
Angelis}, {De Caneva}, {De Lotto}, {Delgado Mendez}, {Doert},
{Dom{\'\i}nguez}, {Dominis Prester}, {Dorner}, {Doro}, {Einecke},
{Eisenacher}, {Elsaesser}, {Farina}, {Ferenc}, {Fonseca}, {Font}, {Frantzen},
{Fruck}, {Garc{\'\i}a L{\'o}pez}, {Garczarczyk}, {Garrido Terrats}, {Gaug},
{Godinovi{\'c}}, {Gonz{\'a}lez Mu{\~n}oz}, {Gozzini}, {Hadasch}, {Hayashida},
{Herrera}, {Herrero}, {Hildebrand}, {Hose}, {Hrupec}, {Idec}, {Kadenius},
{Kellermann}, {Kodani}, {Konno}, {Krause}, {Kubo}, {Kushida}, {La Barbera},
{Lelas}, {Lewandowska}, {Lindfors}, {Lombardi}, {L{\'o}pez},
{L{\'o}pez-Coto}, {L{\'o}pez-Oramas}, {Lorenz}, {Lozano}, {Makariev},
{Mallot}, {Maneva}, {Mankuzhiyil}, {Mannheim}, {Maraschi}, {Marcote},
{Mariotti}, {Mart{\'\i}nez}, {Mazin}, {Menzel}, {Meucci}, {Miranda},
{Mirzoyan}, {Moralejo}, {Munar-Adrover}, {Nakajima}, {Niedzwiecki},
{Nilsson}, {Nishijima}, {Noda}, {Nowak}, {Orito}, {Overkemping}, {Paiano},
{Palatiello}, {Paneque}, {Paoletti}, {Paredes}, {Paredes-Fortuny}, {Partini},
{Persic}, {Prada}, {Prada Moroni}, {Prandini}, {Preziuso}, {Puljak},
{Reinthal}, {Rhode}, {Rib{\'o}}, {Rico}, {Rodriguez Garcia}, {R{\"u}gamer},
{Saggion}, {Saito}, {Saito}, {Satalecka}, {Scalzotto}, {Scapin}, {Schultz},
{Schweizer}, {Shore}, {Sillanp{\"a}{\"a}}, {Sitarek}, {Snidaric},
{Sobczynska}, {Spanier}, {Stamatescu}, {Stamerra}, {Steinbring}, {Storz},
{Strzys}, {Sun}, {Suri{\'c}}, {Takalo}, {Takami}, {Tavecchio}, {Temnikov},
{Terzi{\'c}}, {Tescaro}, {Teshima}, {Thaele}, {Tibolla}, {Torres}, {Toyama},
{Treves}, {Uellenbeck}, {Vogler}, {Wagner}, {Zandanel}, {Zanin}, {MAGIC
Collaboration}, {Lucarelli}, {Pittori}, {Vercellone}, {Verrecchia}, {AGILE
Collaboration}, {Buson}, {D'Ammando}, {Stawarz}, {Giroletti}, {Orienti},
{Fermi-LAT Collaboration}, {Mundell}, {Steele}, {Zarpudin}, {Raiteri},
{Villata}, {Sandrinelli}, {L{\"a}hteenm{\"a}ki}, {Tammi}, {Tornikoski},
{Hovatta}, {Readhead}, {Max-Moerbeck}, {Richards}, {Jorstad}, {Marscher},
{Gurwell}, {Larionov}, {Blinov}, {Konstantinova}, {Kopatskaya}, {Larionova},
{Larionova}, {Morozova}, {Troitsky}, {Mokrushina}, {Pavlova}, {Chen}, {Lin},
{Panwar}, {Agudo}, {Casadio}, {G{\'o}mez}, {Molina}, {Kurtanidze},
{Nikolashvili}, {Kurtanidze}, {Chigladze}, {Acosta-Pulido}, {Carnerero},
{Manilla-Robles}, {Ovcharov}, {Bozhilov}, {Metodieva}, {Aller}, {Aller},
{Fuhrman}, {Angelakis}, {Nestoras}, {Krichbaum}, {Zensus}, {Ungerechts}, and
{Sievers}]{Aleksic2014}
{Aleksi{\'c}}, J.; {Ansoldi}, S.; {Antonelli}, L.A.; {Antoranz}, P.; {Babic},
A.; {Bangale}, P.; {Barres de Almeida}, U.; {Barrio}, J.A.; {Becerra
Gonz{\'a}lez}, J.; {Bednarek}, W.;  et~al.
\newblock {MAGIC gamma-ray and multi-frequency observations of flat spectrum
radio quasar PKS 1510-089 in early 2012}.
\newblock {\em \aap} {\bf 2014}, {\em 569},~A46.
\newblock  [\href{http://dx.doi.org/10.1051/0004-6361/201423484}{CrossRef}]

\bibitem[{Blandford} and {Znajek}(1977)]{BZ1977}
{Blandford}, R.D.; {Znajek}, R.L.
\newblock {Electromagnetic extraction of energy from Kerr black holes.}
\newblock {\em \mnras} {\bf 1977}, {\em 179},~433--456.
\newblock  [\href{http://dx.doi.org/10.1093/mnras/179.3.433}{CrossRef}]

\bibitem[{Blandford} and {Payne}(1982)]{BP1982}
{Blandford}, R.D.; {Payne}, D.G.
\newblock {Hydromagnetic flows from accretion disks and the production of radio
jets.}
\newblock {\em \mnras} {\bf 1982}, {\em 199},~883--903.
\newblock  [\href{http://dx.doi.org/10.1093/mnras/199.4.883}{CrossRef}]

\bibitem[{Semenov} \em{et~al.}(2004){Semenov}, {Dyadechkin}, and
{Punsly}]{Semenov2004}
{Semenov}, V.; {Dyadechkin}, S.; {Punsly}, B.
\newblock {Simulations of Jets Driven by Black Hole Rotation}.
\newblock {\em Science} {\bf 2004}, {\em 305},~978--980.  [\href{http://dx.doi.org/10.1126/science.1100638}{CrossRef}] [\href{http://www.ncbi.nlm.nih.gov/pubmed/15310894}{PubMed}]

\bibitem[{Vlahakis} and {K{\"o}nigl}(2004)]{VK2004}
{Vlahakis}, N.; {K{\"o}nigl}, A.
\newblock {Magnetic Driving of Relativistic Outflows in Active Galactic Nuclei.
I. Interpretation of Parsec-Scale Accelerations}.
\newblock {\em \apj} {\bf 2004}, {\em 605},~656--661.
\newblock  [\href{http://dx.doi.org/10.1086/382670}{CrossRef}]

\bibitem[{Narayan} and {Quataert}(2005)]{NQ2005}
{Narayan}, R.; {Quataert}, E.
\newblock {Black Hole Accretion}.
\newblock {\em Science} {\bf 2005}, {\em 307},~77--80. [\href{http://dx.doi.org/10.1126/science.1105746}{CrossRef}] [\href{http://www.ncbi.nlm.nih.gov/pubmed/15637269}{PubMed}]

\bibitem[{McKinney}(2006)]{McKinney2006}
{McKinney}, J.C.
\newblock {General relativistic magnetohydrodynamic simulations of the jet
formation and large-scale propagation from black hole accretion systems}.
\newblock {\em \mnras} {\bf 2006}, {\em 368},~1561--1582.
\newblock  [\href{http://dx.doi.org/10.1111/j.1365-2966.2006.10256.x}{CrossRef}]

\bibitem[{Komissarov} \em{et~al.}(2007){Komissarov}, {Barkov}, {Vlahakis}, and
{K{\"o}nigl}]{Komissarov2007}
{Komissarov}, S.S.; {Barkov}, M.V.; {Vlahakis}, N.; {K{\"o}nigl}, A.
\newblock {Magnetic acceleration of relativistic active galactic nucleus jets}.
\newblock {\em \mnras} {\bf 2007}, {\em 380},~51--70.
\newblock  [\href{http://dx.doi.org/10.1111/j.1365-2966.2007.12050.x}{CrossRef}]

\bibitem[{Komissarov} \em{et~al.}(2009){Komissarov}, {Vlahakis}, {K{\"o}nigl},
and {Barkov}]{Komissarov2009}
{Komissarov}, S.S.; {Vlahakis}, N.; {K{\"o}nigl}, A.; {Barkov}, M.V.
\newblock {Magnetic acceleration of ultrarelativistic jets in gamma-ray burst
sources}.
\newblock {\em \mnras} {\bf 2009}, {\em 394},~1182--1212.
\newblock  [\href{http://dx.doi.org/10.1111/j.1365-2966.2009.14410.x}{CrossRef}]

\bibitem[{Tchekhovskoy} \em{et~al.}(2008){Tchekhovskoy}, {McKinney}, and
{Narayan}]{Tchekhovskoy2008}
{Tchekhovskoy}, A.; {McKinney}, J.C.; {Narayan}, R.
\newblock {Simulations of ultrarelativistic magnetodynamic jets from gamma-ray
burst engines}.
\newblock {\em \mnras} {\bf 2008}, {\em 388},~551--572.
\newblock  [\href{http://dx.doi.org/10.1111/j.1365-2966.2008.13425.x}{CrossRef}]

\bibitem[{Tchekhovskoy} \em{et~al.}(2009){Tchekhovskoy}, {McKinney}, and
{Narayan}]{Tchekhovskoy2009}
{Tchekhovskoy}, A.; {McKinney}, J.C.; {Narayan}, R.
\newblock {Efficiency of Magnetic to Kinetic Energy Conversion in a Monopole
Magnetosphere}.
\newblock {\em \apj} {\bf 2009}, {\em 699},~1789--1808.
\newblock  [\href{http://dx.doi.org/10.1088/0004-637X/699/2/1789}{CrossRef}]

\bibitem[{Li} \em{et~al.}(1992){Li}, {Chiueh}, and {Begelman}]{Li1992}
{Li}, Z.Y.; {Chiueh}, T.; {Begelman}, M.C.
\newblock {Electromagnetically Driven Relativistic Jets: A Class of
Self-similar Solutions}.
\newblock {\em \apj} {\bf 1992}, {\em 394},~459.
\newblock  [\href{http://dx.doi.org/10.1086/171597}{CrossRef}]

\bibitem[{Begelman} and {Li}(1994)]{BL1994}
{Begelman}, M.C.; {Li}, Z.Y.
\newblock {Asymptotic Domination of Cold Relativistic MHD Winds by Kinetic
Energy Flux}.
\newblock {\em \apj} {\bf 1994}, {\em 426},~269.
\newblock  [\href{http://dx.doi.org/10.1086/174061}{CrossRef}]

\bibitem[{Lyubarsky}(2009)]{Lyubarsky2009}
{Lyubarsky}, Y.
\newblock {Asymptotic Structure of Poynting-Dominated Jets}.
\newblock {\em \apj} {\bf 2009}, {\em 698},~1570--1589.
\newblock  [\href{http://dx.doi.org/10.1088/0004-637X/698/2/1570}{CrossRef}]

\bibitem[{Pu} and {Takahashi}(2020)]{PT2020}
{Pu}, H.Y.; {Takahashi}, M.
\newblock {Properties of Trans-fast Magnetosonic Jets in Black Hole
Magnetospheres}.
\newblock {\em \apj} {\bf 2020}, {\em 892},~37.
\newblock  [\href{http://dx.doi.org/10.3847/1538-4357/ab77ab}{CrossRef}]

\bibitem[{Sironi} and {Spitkovsky}(2014)]{SS2014}
{Sironi}, L.; {Spitkovsky}, A.
\newblock {Relativistic Reconnection: An Efficient Source of Non-thermal
Particles}.
\newblock {\em \apjl} {\bf 2014}, {\em 783},~L21.
\newblock  [\href{http://dx.doi.org/10.1088/2041-8205/783/1/L21}{CrossRef}]

\bibitem[{Ripperda} \em{et~al.}(2020){Ripperda}, {Bacchini}, and
{Philippov}]{Ripperda2020}
{Ripperda}, B.; {Bacchini}, F.; {Philippov}, A.A.
\newblock {Magnetic Reconnection and Hot Spot Formation in Black Hole Accretion
Disks}.
\newblock {\em \apj} {\bf 2020}, {\em 900},~100.
\newblock  [\href{http://dx.doi.org/10.3847/1538-4357/ababab}{CrossRef}]

\bibitem[{Ripperda} \em{et~al.}(2022){Ripperda}, {Liska}, {Chatterjee},
{Musoke}, {Philippov}, {Markoff}, {Tchekhovskoy}, and {Younsi}]{Ripperda2022}
{Ripperda}, B.; {Liska}, M.; {Chatterjee}, K.; {Musoke}, G.; {Philippov}, A.A.;
{Markoff}, S.B.; {Tchekhovskoy}, A.; {Younsi}, Z.
\newblock {Black Hole Flares: Ejection of Accreted Magnetic Flux through 3D
Plasmoid-mediated Reconnection}.
\newblock {\em \apjl} {\bf 2022}, {\em 924},~L32.
\newblock  [\href{http://dx.doi.org/10.3847/2041-8213/ac46a1}{CrossRef}]

\bibitem[{Event Horizon Telescope Collaboration} \em{et~al.}(2021){Event
Horizon Telescope Collaboration}, {Akiyama}, {Algaba}, {Alberdi}, {Alef},
{Anantua}, {Asada}, {Azulay}, {Baczko}, {Ball}, {Balokovi{\'c}}, {Barrett},
{Benson}, {Bintley}, {Blackburn}, {Blundell}, {Boland}, {Bouman}, {Bower},
{Boyce}, {Bremer}, {Brinkerink}, {Brissenden}, {Britzen}, {Broderick},
{Broguiere}, {Bronzwaer}, {Byun}, {Carlstrom}, {Chael}, {Chan}, {Chatterjee},
{Chatterjee}, {Chen}, {Chen}, {Chesler}, {Cho}, {Christian}, {Conway},
{Cordes}, {Crawford}, {Crew}, {Cruz-Osorio}, {Cui}, {Davelaar}, {De
Laurentis}, {Deane}, {Dempsey}, {Desvignes}, {Dexter}, {Doeleman}, {Eatough},
{Falcke}, {Farah}, {Fish}, {Fomalont}, {Ford}, {Fraga-Encinas}, {Freeman},
{Friberg}, {Fromm}, {Fuentes}, {Galison}, {Gammie}, {Garc{\'\i}a}, {Gentaz},
{Georgiev}, {Goddi}, {Gold}, {G{\'o}mez}, {G{\'o}mez-Ruiz}, {Gu}, {Gurwell},
{Hada}, {Haggard}, {Hecht}, {Hesper}, {Ho}, {Ho}, {Honma}, {Huang}, {Huang},
{Hughes}, {Ikeda}, {Inoue}, {Issaoun}, {James}, {Jannuzi}, {Janssen},
{Jeter}, {Jiang}, {Jimenez-Rosales}, {Johnson}, {Jorstad}, {Jung}, {Karami},
{Karuppusamy}, {Kawashima}, {Keating}, {Kettenis}, {Kim}, {Kim}, {Kim},
{Kim}, {Kino}, {Koay}, {Kofuji}, {Koch}, {Koyama}, {Kramer}, {Kramer},
{Krichbaum}, {Kuo}, {Lauer}, {Lee}, {Levis}, {Li}, {Li}, {Lindqvist}, {Lico},
{Lindahl}, {Liu}, {Liu}, {Liuzzo}, {Lo}, {Lobanov}, {Loinard}, {Lonsdale},
{Lu}, {MacDonald}, {Mao}, {Marchili}, {Markoff}, {Marrone}, {Marscher},
{Mart{\'\i}-Vidal}, {Matsushita}, {Matthews}, {Medeiros}, {Menten}, {Mizuno},
{Mizuno}, {Moran}, {Moriyama}, {Moscibrodzka}, {M{\"u}ller}, {Musoke},
{Mej{\'\i}as}, {Michalik}, {Nadolski}, {Nagai}, {Nagar}, {Nakamura},
{Narayan}, {Narayanan}, {Natarajan}, {Nathanail}, {Neilsen}, {Neri}, {Ni},
{Noutsos}, {Nowak}, {Okino}, {Olivares}, {Ortiz-Le{\'o}n}, {Oyama},
{{\"O}zel}, {Palumbo}, {Park}, {Patel}, {Pen}, {Pesce}, {Pi{\'e}tu},
{Plambeck}, {PopStefanija}, {Porth}, {P{\"o}tzl}, {Prather},
{Preciado-L{\'o}pez}, {Psaltis}, {Pu}, {Ramakrishnan}, {Rao}, {Rawlings},
{Raymond}, {Rezzolla}, {Ricarte}, {Ripperda}, {Roelofs}, {Rogers}, {Ros},
{Rose}, {Roshanineshat}, {Rottmann}, {Roy}, {Ruszczyk}, {Rygl},
{S{\'a}nchez}, {S{\'a}nchez-Arguelles}, {Sasada}, {Savolainen}, {Schloerb},
{Schuster}, {Shao}, {Shen}, {Small}, {Sohn}, {SooHoo}, {Sun}, {Tazaki},
{Tetarenko}, {Tiede}, {Tilanus}, {Titus}, {Toma}, {Torne}, {Trent},
{Traianou}, {Trippe}, {van Bemmel}, {van Langevelde}, {van Rossum}, {Wagner},
{Ward-Thompson}, {Wardle}, {Weintroub}, {Wex}, {Wharton}, {Wielgus}, {Wong},
{Wu}, {Yoon}, {Young}, {Young}, {Younsi}, {Yuan}, {Yuan}, {Zensus}, {Zhao},
and {Zhao}]{EHT2021a}
{Event Horizon Telescope Collaboration}.; {Akiyama}, K.; {Algaba}, J.C.;
{Alberdi}, A.; {Alef}, W.; {Anantua}, R.; {Asada}, K.; {Azulay}, R.;
{Baczko}, A.K.; {Ball}, D.;  et~al.
\newblock {First M87 Event Horizon Telescope Results. VII. Polarization of the
Ring}.
\newblock {\em \apjl} {\bf 2021}, {\em 910},~L12.
\newblock  [\href{http://dx.doi.org/10.3847/2041-8213/abe71d}{CrossRef}]

\bibitem[{Boccardi} \em{et~al.}(2017){Boccardi}, {Krichbaum}, {Ros}, and
{Zensus}]{Boccardi2017}
{Boccardi}, B.; {Krichbaum}, T.P.; {Ros}, E.; {Zensus}, J.A.
\newblock {Radio observations of active galactic nuclei with mm-VLBI}.
\newblock {\em AARv} {\bf 2017}, {\em 25},~4.
\newblock  [\href{http://dx.doi.org/10.1007/s00159-017-0105-6}{CrossRef}]

\bibitem[{Hada}(2019)]{Hada2019}
{Hada}, K.
\newblock {Relativistic Jets from AGN Viewed at Highest Angular Resolution}.
\newblock {\em Galaxies} {\bf 2019}, {\em 8},~1.  [\href{http://dx.doi.org/10.3390/galaxies8010001}{CrossRef}]

\bibitem[{Kinman} \em{et~al.}(1974){Kinman}, {Grasdalen}, and
{Rieke}]{Kinman1974}
{Kinman}, T.D.; {Grasdalen}, G.L.; {Rieke}, G.H.
\newblock {Optical and Infrared Observations of the Jet of M87}.
\newblock {\em \apjl} {\bf 1974}, {\em 194},~L1.
\newblock  [\href{http://dx.doi.org/10.1086/181656}{CrossRef}]

\bibitem[{Rudnick} \em{et~al.}(1978){Rudnick}, {Owen}, {Jones}, {Puschell}, and
{Stein}]{Rudnick1978}
{Rudnick}, L.; {Owen}, F.N.; {Jones}, T.W.; {Puschell}, J.J.; {Stein}, W.A.
\newblock {Coordinated centimeter, millimeter, infrared, and visual polarimetry
of compact nonthermal sources.}
\newblock {\em \apjl} {\bf 1978}, {\em 225},~L5--L9.
\newblock  [\href{http://dx.doi.org/10.1086/182781}{CrossRef}]

\bibitem[{Ghisellini} \em{et~al.}(1985){Ghisellini}, {Maraschi}, and
{Treves}]{Ghisellini1985}
{Ghisellini}, G.; {Maraschi}, L.; {Treves}, A.
\newblock {Inhomogeneous synchrotron-self-compton models and the problem of
relativistic beaming of BL Lac objects.}
\newblock {\em \aap} {\bf 1985}, {\em 146},~204--212.

\bibitem[{Asada} and {Nakamura}(2012)]{AN2012}
{Asada}, K.; {Nakamura}, M.
\newblock {The Structure of the M87 Jet: A Transition from Parabolic to Conical
Streamlines}.
\newblock {\em \apjl} {\bf 2012}, {\em 745},~L28.
\newblock  [\href{http://dx.doi.org/10.1088/2041-8205/745/2/L28}{CrossRef}]

\bibitem[{Tseng} \em{et~al.}(2016){Tseng}, {Asada}, {Nakamura}, {Pu}, {Algaba},
and {Lo}]{Tseng2016}
{Tseng}, C.Y.; {Asada}, K.; {Nakamura}, M.; {Pu}, H.Y.; {Algaba}, J.C.; {Lo},
W.P.
\newblock {Structural Transition in the NGC 6251 Jet: An Interplay with the
Supermassive Black Hole and Its Host Galaxy}.
\newblock {\em \apj} {\bf 2016}, {\em 833},~288.
\newblock  [\href{http://dx.doi.org/10.3847/1538-4357/833/2/288}{CrossRef}]

\bibitem[{Akiyama} \em{et~al.}(2018){Akiyama}, {Asada}, {Fish}, {Nakamura},
{Hada}, {Nagai}, and {Lonsdale}]{Akiyama2018}
{Akiyama}, K.; {Asada}, K.; {Fish}, V.; {Nakamura}, M.; {Hada}, K.; {Nagai},
H.; {Lonsdale}, C.
\newblock {The Global Jet Structure of the Archetypical Quasar 3C 273}.
\newblock {\em Galaxies} {\bf 2018}, {\em 6},~15.
\newblock  [\href{http://dx.doi.org/10.3390/galaxies6010015}{CrossRef}]

\bibitem[{Algaba} \em{et~al.}(2017){Algaba}, {Nakamura}, {Asada}, and
{Lee}]{Algaba2017}
{Algaba}, J.C.; {Nakamura}, M.; {Asada}, K.; {Lee}, S.S.
\newblock {Resolving the Geometry of the Innermost Relativistic Jets in Active
Galactic Nuclei}.
\newblock {\em \apj} {\bf 2017}, {\em 834},~65.
\newblock  [\href{http://dx.doi.org/10.3847/1538-4357/834/1/65}{CrossRef}]

\bibitem[{Hada} \em{et~al.}(2018){Hada}, {Doi}, {Wajima}, {D'Ammand o},
{Orienti}, {Giroletti}, {Giovannini}, {Nakamura}, and {Asada}]{Hada2018}
{Hada}, K.; {Doi}, A.; {Wajima}, K.; {D'Ammand o}, F.; {Orienti}, M.;
{Giroletti}, M.; {Giovannini}, G.; {Nakamura}, M.; {Asada}, K.
\newblock {Collimation, Acceleration, and Recollimation Shock in the Jet of
Gamma-Ray Emitting Radio-loud Narrow-line Seyfert 1 Galaxy 1H0323+342}.
\newblock {\em \apj} {\bf 2018}, {\em 860},~141.
\newblock  [\href{http://dx.doi.org/10.3847/1538-4357/aac49f}{CrossRef}]

\bibitem[{Kovalev} \em{et~al.}(2020){Kovalev}, {Pushkarev}, {Nokhrina},
{Plavin}, {Beskin}, {Chernoglazov}, {Lister}, and {Savolainen}]{Kovalev2020}
{Kovalev}, Y.Y.; {Pushkarev}, A.B.; {Nokhrina}, E.E.; {Plavin}, A.V.; {Beskin},
V.S.; {Chernoglazov}, A.V.; {Lister}, M.L.; {Savolainen}, T.
\newblock {A transition from parabolic to conical shape as a common effect in
nearby AGN jets}.
\newblock {\em \mnras} {\bf 2020}, {\em 495},~3576--3591.
\newblock  [\href{http://dx.doi.org/10.1093/mnras/staa1121}{CrossRef}]

\bibitem[{Park} \em{et~al.}(2021){Park}, {Hada}, {Nakamura}, {Asada}, {Zhao},
and {Kino}]{Park2021b}
{Park}, J.; {Hada}, K.; {Nakamura}, M.; {Asada}, K.; {Zhao}, G.; {Kino}, M.
\newblock {Jet Collimation and Acceleration in the Giant Radio Galaxy NGC 315}.
\newblock {\em \apj} {\bf 2021}, {\em 909},~76,
\newblock  [\href{http://dx.doi.org/10.3847/1538-4357/abd6ee}{CrossRef}]

\bibitem[{Gabuzda} and {Sitko}(1994)]{Gabuzda1994}
{Gabuzda}, D.C.; {Sitko}, M.L.
\newblock {Nearly Simultaneous Very Long Baseline Interferometry and Optical
Polarimetry of Blazars}.
\newblock {\em \aj} {\bf 1994}, {\em 107},~884.
\newblock  [\href{http://dx.doi.org/10.1086/116902}{CrossRef}]

\bibitem[{Gabuzda} \em{et~al.}(1996){Gabuzda}, {Sitko}, and
{Smith}]{Gabuzda1996}
{Gabuzda}, D.C.; {Sitko}, M.L.; {Smith}, P.S.
\newblock {Correlations Between the VLBI and Optical Polarization of BL
Lacertae Objects}.
\newblock {\em \aj} {\bf 1996}, {\em 112},~1877.
\newblock  [\href{http://dx.doi.org/10.1086/118149}{CrossRef}]

\bibitem[Wardle(2018)]{Wardle18}
Wardle, J.
\newblock The Variable Rotation Measure Distribution in 3C 273 on Parsec
Scales.
\newblock {\em Galaxies} {\bf 2018}, {\em 6},~5.  [\href{http://dx.doi.org/10.3390/galaxies6010005}{CrossRef}]

\bibitem[{Algaba} \em{et~al.}(2011){Algaba}, {Gabuzda}, and
{Smith}]{Algaba2011}
{Algaba}, J.C.; {Gabuzda}, D.C.; {Smith}, P.S.
\newblock {Search for correlations between the optical and radio polarization
of active galactic nuclei---I. VLBA polarization data at 15 + 22 + 43 GHz}.
\newblock {\em \mnras} {\bf 2011}, {\em 411},~85--101.
\newblock  [\href{http://dx.doi.org/10.1111/j.1365-2966.2010.17654.x}{CrossRef}]

\bibitem[{Algaba} \em{et~al.}(2012){Algaba}, {Gabuzda}, and
{Smith}]{Algaba2012}
{Algaba}, J.C.; {Gabuzda}, D.C.; {Smith}, P.S.
\newblock {Search for correlations between the optical and radio polarization
of active galactic nuclei---II. VLBA polarization data at 12+15+22+24+43
GHz}.
\newblock {\em \mnras} {\bf 2012}, {\em 420},~542--553.
\newblock  [\href{http://dx.doi.org/10.1111/j.1365-2966.2011.20061.x}{CrossRef}]

\bibitem[{Jorstad} \em{et~al.}(2007){Jorstad}, {Marscher}, {Stevens}, {Smith},
{Forster}, {Gear}, {Cawthorne}, {Lister}, {Stirling}, {G{\'o}mez}, {Greaves},
and {Robson}]{Jorstad2007}
{Jorstad}, S.G.; {Marscher}, A.P.; {Stevens}, J.A.; {Smith}, P.S.; {Forster},
J.R.; {Gear}, W.K.; {Cawthorne}, T.V.; {Lister}, M.L.; {Stirling}, A.M.;
{G{\'o}mez}, J.L.;  et~al.
\newblock {Multiwaveband Polarimetric Observations of 15 Active Galactic Nuclei
at High Frequencies: Correlated Polarization Behavior}.
\newblock {\em \aj} {\bf 2007}, {\em 134},~799--824.
\newblock  [\href{http://dx.doi.org/10.1086/519996}{CrossRef}]

\bibitem[{Park} \em{et~al.}(2018){Park}, {Kam}, {Trippe}, {Kang}, {Byun},
{Kim}, {Algaba}, {Lee}, {Zhao}, {Kino}, {Shin}, {Hada}, {Lee}, {Oh},
{Hodgson}, and {Sohn}]{Park2018}
{Park}, J.; {Kam}, M.; {Trippe}, S.; {Kang}, S.; {Byun}, D.Y.; {Kim}, D.W.;
{Algaba}, J.C.; {Lee}, S.S.; {Zhao}, G.Y.; {Kino}, M.;  et~al.
\newblock {Revealing the Nature of Blazar Radio Cores through Multifrequency
Polarization Observations with the Korean VLBI Network}.
\newblock {\em \apj} {\bf 2018}, {\em 860},~112.
\newblock  [\href{http://dx.doi.org/10.3847/1538-4357/aac490}{CrossRef}]

\bibitem[{D'Arcangelo} \em{et~al.}(2007){D'Arcangelo}, {Marscher}, {Jorstad},
{Smith}, {Larionov}, {Hagen-Thorn}, {Kopatskaya}, {Williams}, and
{Gear}]{DArcangelo2007}
{D'Arcangelo}, F.D.; {Marscher}, A.P.; {Jorstad}, S.G.; {Smith}, P.S.;
{Larionov}, V.M.; {Hagen-Thorn}, V.A.; {Kopatskaya}, E.N.; {Williams}, G.G.;
{Gear}, W.K.
\newblock {Rapid Multiwaveband Polarization Variability in the Quasar PKS
0420-014: Optical Emission from the Compact Radio Jet}.
\newblock {\em \apjl} {\bf 2007}, {\em 659},~L107--L110.
\newblock  [\href{http://dx.doi.org/10.1086/517525}{CrossRef}]

\bibitem[{Marscher} \em{et~al.}(2008){Marscher}, {Jorstad}, {D'Arcangelo},
{Smith}, {Williams}, {Larionov}, {Oh}, {Olmstead}, {Aller}, {Aller},
{McHardy}, {L{\"a}hteenm{\"a}ki}, {Tornikoski}, {Valtaoja}, {Hagen-Thorn},
{Kopatskaya}, {Gear}, {Tosti}, {Kurtanidze}, {Nikolashvili}, {Sigua},
{Miller}, and {Ryle}]{Marscher2008}
{Marscher}, A.P.; {Jorstad}, S.G.; {D'Arcangelo}, F.D.; {Smith}, P.S.;
{Williams}, G.G.; {Larionov}, V.M.; {Oh}, H.; {Olmstead}, A.R.; {Aller},
M.F.; {Aller}, H.D.;  et~al.
\newblock {The inner jet of an active galactic nucleus as revealed by a
radio-to-{\ensuremath{\gamma}}-ray outburst}.
\newblock {\em \nat} {\bf 2008}, {\em 452},~966--969. [\href{http://dx.doi.org/10.1038/nature06895}{CrossRef}] [\href{http://www.ncbi.nlm.nih.gov/pubmed/18432239}{PubMed}]

\bibitem[{Marscher} \em{et~al.}(2010){Marscher}, {Jorstad}, {Larionov},
{Aller}, {Aller}, {L{\"a}hteenm{\"a}ki}, {Agudo}, {Smith}, {Gurwell},
{Hagen-Thorn}, {Konstantinova}, {Larionova}, {Larionova}, {Melnichuk},
{Blinov}, {Kopatskaya}, {Troitsky}, {Tornikoski}, {Hovatta}, {Schmidt},
{D'Arcangelo}, {Bhattarai}, {Taylor}, {Olmstead}, {Manne-Nicholas},
{Roca-Sogorb}, {G{\'o}mez}, {McHardy}, {Kurtanidze}, {Nikolashvili},
{Kimeridze}, and {Sigua}]{Marscher2010}
{Marscher}, A.P.; {Jorstad}, S.G.; {Larionov}, V.M.; {Aller}, M.F.; {Aller},
H.D.; {L{\"a}hteenm{\"a}ki}, A.; {Agudo}, I.; {Smith}, P.S.; {Gurwell}, M.;
{Hagen-Thorn}, V.A.;  et~al.
\newblock {Probing the Inner Jet of the Quasar PKS 1510-089 with Multi-Waveband
Monitoring During Strong Gamma-Ray Activity}.
\newblock {\em \apjl} {\bf 2010}, {\em 710},~L126--L131.
\newblock  [\href{http://dx.doi.org/10.1088/2041-8205/710/2/L126}{CrossRef}]

\bibitem[{Marscher}(2014)]{Marscher2014}
{Marscher}, A.P.
\newblock {Turbulent, Extreme Multi-zone Model for Simulating Flux and
Polarization Variability in Blazars}.
\newblock {\em \apj} {\bf 2014}, {\em 780},~87,
\newblock  [\href{http://dx.doi.org/10.1088/0004-637X/780/1/87}{CrossRef}]

\bibitem[Blinov \em{et~al.}(2017)Blinov, Pavlidou, Papadakis, Kiehlmann,
Liodakis, Panopoulou, Angelakis, Baloković, Hovatta, King, Kus, Kylafis,
Mahabal, Maharana, Myserlis, Paleologou, Papamastorakis, Pazderski, Pearson,
Ramaprakash, Readhead, Reig, Tassis, and Zensus]{Blinov17}
Blinov, D.; Pavlidou, V.; Papadakis, I.; Kiehlmann, S.; Liodakis, I.;
Panopoulou, G.V.; Angelakis, E.; Baloković, M.; Hovatta, T.; King, O.G.;
et~al.
\newblock {RoboPol: Connection between optical polarization plane rotations and
gamma-ray flares in blazars}.
\newblock {\em Mon. Not. R. Astron. Soc.} {\bf 2017},
{\em 474},~1296--1306. Available online: \url{https://academic.oup.com/mnras/article-pdf/474/1/1296/22367606/stx2786.pdf} ({accessed on 18 October 2022}
) .
\newblock  [\href{http://dx.doi.org/10.1093/mnras/stx2786}{CrossRef}]

\bibitem[{Sasada} \em{et~al.}(2010){Sasada}, {Uemura}, {Arai}, {Fukazawa},
{Kawabata}, {Ohsugi}, {Yamashita}, {Isogai}, {Nagae}, {Uehara}, {Mizuno},
{Katagiri}, {Takahashi}, {Sato}, and {Kino}]{Sasada2010}
{Sasada}, M.; {Uemura}, M.; {Arai}, A.; {Fukazawa}, Y.; {Kawabata}, K.S.;
{Ohsugi}, T.; {Yamashita}, T.; {Isogai}, M.; {Nagae}, O.; {Uehara}, T.;
et~al.
\newblock {Multiband Photopolarimetric Monitoring of an Outburst of the Blazar
3C 454.3 in 2007}.
\newblock {\em \pasj} {\bf 2010}, {\em 62},~645.
\newblock  [\href{http://dx.doi.org/10.1093/pasj/62.3.645}{CrossRef}]

\bibitem[{Liodakis} \em{et~al.}(2020){Liodakis}, {Blinov}, {Jorstad},
{Arkharov}, {Di Paola}, {Efimova}, {Grishina}, {Kiehlmann}, {Kopatskaya},
{Larionov}, {Larionova}, {Larionova}, {Marscher}, {Morozova}, {Nikiforova},
{Pavlidou}, {Traianou}, {Troitskaya}, {Troitsky}, {Uemura}, and
{Weaver}]{Liodakis2020}
{Liodakis}, I.; {Blinov}, D.; {Jorstad}, S.G.; {Arkharov}, A.A.; {Di Paola},
A.; {Efimova}, N.V.; {Grishina}, T.S.; {Kiehlmann}, S.; {Kopatskaya}, E.N.;
{Larionov}, V.M.;  et~al.
\newblock {Two Flares with One Shock: The Interesting Case of 3C 454.3}.
\newblock {\em \apj} {\bf 2020}, {\em 902},~61.
\newblock  [\href{http://dx.doi.org/10.3847/1538-4357/abb1b8}{CrossRef}]

\bibitem[{Morozova} \em{et~al.}(2014){Morozova}, {Larionov}, {Troitsky},
{Jorstad}, {Marscher}, {G{\'o}mez}, {Blinov}, {Efimova}, {Hagen-Thorn},
{Hagen-Thorn}, {Joshi}, {Konstantinova}, {Kopatskaya}, {Larionova},
{Larionova}, {L{\"a}hteenm{\"a}ki}, {Tammi}, {Rastorgueva-Foi}, {McHardy},
{Tornikoski}, {Agudo}, {Casadio}, {Molina}, {Volvach}, and
{Volvach}]{Morozova2014}
{Morozova}, D.A.; {Larionov}, V.M.; {Troitsky}, I.S.; {Jorstad}, S.G.;
{Marscher}, A.P.; {G{\'o}mez}, J.L.; {Blinov}, D.A.; {Efimova}, N.V.;
{Hagen-Thorn}, V.A.; {Hagen-Thorn}, E.I.;  et~al.
\newblock {The Outburst of the Blazar S4 0954+658 in 2011 March-April}.
\newblock {\em \aj} {\bf 2014}, {\em 148},~42.
\newblock  [\href{http://dx.doi.org/10.1088/0004-6256/148/3/42}{CrossRef}]

\bibitem[{Troitskiy} \em{et~al.}(2013){Troitskiy}, {Morozova}, {Jorstad},
{Marscher}, {Larionov}, {Blinov}, {Agudo}, and {Smith}]{Troitskiy2013}
{Troitskiy}, I.S.; {Morozova}, D.A.; {Jorstad}, S.G.; {Marscher}, A.P.;
{Larionov}, V.M.; {Blinov}, D.A.; {Agudo}, I.; {Smith}, P.S.
\newblock {Multiwavelength polarization observations of the
{\ensuremath{\gamma}}-ray bright quasar PKS 0420-014}.
\newblock
{\em Eur. Phys. J. Web Conf.}   \textbf{2013}, \emph{61}, 07008.
\newblock  [\href{http://dx.doi.org/10.1051/epjconf/20136107008}{CrossRef}]

\bibitem[{Myserlis} \em{et~al.}(2018){Myserlis}, {Komossa}, {Angelakis},
{G{\'o}mez}, {Karamanavis}, {Krichbaum}, {Bach}, and {Grupe}]{Myserlis2018}
{Myserlis}, I.; {Komossa}, S.; {Angelakis}, E.; {G{\'o}mez}, J.L.;
{Karamanavis}, V.; {Krichbaum}, T.P.; {Bach}, U.; {Grupe}, D.
\newblock {High cadence, linear, and circular polarization monitoring of OJ
287. Helical magnetic field in a bent jet}.
\newblock {\em \aap} {\bf 2018}, {\em 619},~A88.
\newblock  [\href{http://dx.doi.org/10.1051/0004-6361/201732273}{CrossRef}]

\bibitem[{Narayan} \em{et~al.}(2003){Narayan}, {Igumenshchev}, and
{Abramowicz}]{Narayan2003}
{Narayan}, R.; {Igumenshchev}, I.V.; {Abramowicz}, M.A.
\newblock {Magnetically Arrested Disk: An Energetically Efficient Accretion
Flow}.
\newblock {\em \pasj} {\bf 2003}, {\em 55},~L69--L72.
\newblock  [\href{http://dx.doi.org/10.1093/pasj/55.6.L69}{CrossRef}]

\bibitem[{Tchekhovskoy} \em{et~al.}(2011){Tchekhovskoy}, {Narayan}, and
{McKinney}]{Tchekhovskoy2011}
{Tchekhovskoy}, A.; {Narayan}, R.; {McKinney}, J.C.
\newblock {Efficient generation of jets from magnetically arrested accretion on
a rapidly spinning black hole}.
\newblock {\em \mnras} {\bf 2011}, {\em 418},~L79--L83.
\newblock  [\href{http://dx.doi.org/10.1111/j.1745-3933.2011.01147.x}{CrossRef}]

\bibitem[{McKinney} \em{et~al.}(2012){McKinney}, {Tchekhovskoy}, and {Bland
ford}]{McKinney2012}
{McKinney}, J.C.; {Tchekhovskoy}, A.; {Bland ford}, R.D.
\newblock {General relativistic magnetohydrodynamic simulations of magnetically
choked accretion flows around black holes}.
\newblock {\em \mnras} {\bf 2012}, {\em 423},~3083--3117.
\newblock  [\href{http://dx.doi.org/10.1111/j.1365-2966.2012.21074.x}{CrossRef}]

\bibitem[{Nakamura} \em{et~al.}(2018){Nakamura}, {Asada}, {Hada}, {Pu},
{Noble}, {Tseng}, {Toma}, {Kino}, {Nagai}, {Takahashi}, {Algaba}, {Orienti},
{Akiyama}, {Doi}, {Giovannini}, {Giroletti}, {Honma}, {Koyama}, {Lico},
{Niinuma}, and {Tazaki}]{Nakamura2018}
{Nakamura}, M.; {Asada}, K.; {Hada}, K.; {Pu}, H.Y.; {Noble}, S.; {Tseng}, C.;
{Toma}, K.; {Kino}, M.; {Nagai}, H.; {Takahashi}, K.;  et~al.
\newblock {Parabolic Jets from the Spinning Black Hole in M87}.
\newblock {\em \apj} {\bf 2018}, {\em 868},~146.
\newblock  [\href{http://dx.doi.org/10.3847/1538-4357/aaeb2d}{CrossRef}]

\bibitem[{Narayan} \em{et~al.}(2022){Narayan}, {Chael}, {Chatterjee},
{Ricarte}, and {Curd}]{Narayan2022}
{Narayan}, R.; {Chael}, A.; {Chatterjee}, K.; {Ricarte}, A.; {Curd}, B.
\newblock {Jets in magnetically arrested hot accretion flows: Geometry, power,
and black hole spin-down}.
\newblock {\em \mnras} {\bf 2022}, {\em 511},~3795--3813.
\newblock  [\href{http://dx.doi.org/10.1093/mnras/stac285}{CrossRef}]

\bibitem[{Zamaninasab} \em{et~al.}(2014){Zamaninasab}, {Clausen-Brown},
{Savolainen}, and {Tchekhovskoy}]{Zamaninasab2014}
{Zamaninasab}, M.; {Clausen-Brown}, E.; {Savolainen}, T.; {Tchekhovskoy}, A.
\newblock {Dynamically important magnetic fields near accreting supermassive
black holes}.
\newblock {\em \nat} {\bf 2014}, {\em 510},~126--128.
\newblock  [\href{http://dx.doi.org/10.1038/nature13399}{CrossRef}]

\bibitem[{Ghisellini} \em{et~al.}(2014){Ghisellini}, {Tavecchio}, {Maraschi},
{Celotti}, and {Sbarrato}]{Ghisellini2014}
{Ghisellini}, G.; {Tavecchio}, F.; {Maraschi}, L.; {Celotti}, A.; {Sbarrato},
T.
\newblock {The power of relativistic jets is larger than the luminosity of
their accretion disks}.
\newblock {\em \nat} {\bf 2014}, {\em 515},~376--378.
\newblock  [\href{http://dx.doi.org/10.1038/nature13856}{CrossRef}]

\bibitem[{McKinney} and {Blandford}(2009)]{MB2009}
{McKinney}, J.C.; {Blandford}, R.D.
\newblock {Stability of relativistic jets from rotating, accreting black holes
via fully three-dimensional magnetohydrodynamic simulations}.
\newblock {\em \mnras} {\bf 2009}, {\em 394},~L126--L130.
\newblock  [\href{http://dx.doi.org/10.1111/j.1745-3933.2009.00625.x}{CrossRef}]

\bibitem[{Cruz-Osorio} \em{et~al.}(2022){Cruz-Osorio}, {Fromm}, {Mizuno},
{Nathanail}, {Younsi}, {Porth}, {Davelaar}, {Falcke}, {Kramer}, and
{Rezzolla}]{CruzOsorio2022}
{Cruz-Osorio}, A.; {Fromm}, C.M.; {Mizuno}, Y.; {Nathanail}, A.; {Younsi}, Z.;
{Porth}, O.; {Davelaar}, J.; {Falcke}, H.; {Kramer}, M.; {Rezzolla}, L.
\newblock {State-of-the-art energetic and morphological modelling of the
launching site of the M87 jet}.
\newblock {\em Nat. Astron.} {\bf 2022}, {\em 6},~103--108.
\newblock  [\href{http://dx.doi.org/10.1038/s41550-021-01506-w}{CrossRef}]

\bibitem[{Owen} \em{et~al.}(1989){Owen}, {Hardee}, and {Cornwell}]{Owen1989}
{Owen}, F.N.; {Hardee}, P.E.; {Cornwell}, T.J.
\newblock {High-Resolution, High Dynamic Range VLA Images of the M87 Jet at 2
Centimeters}.
\newblock {\em \apj} {\bf 1989}, {\em 340},~698.
\newblock  [\href{http://dx.doi.org/10.1086/167430}{CrossRef}]

\bibitem[{Biretta} \em{et~al.}(1999){Biretta}, {Sparks}, and
{Macchetto}]{Biretta1999}
{Biretta}, J.A.; {Sparks}, W.B.; {Macchetto}, F.
\newblock {Hubble Space Telescope Observations of Superluminal Motion in the
M87 Jet}.
\newblock {\em \apj} {\bf 1999}, {\em 520},~621--626.
\newblock  [\href{http://dx.doi.org/10.1086/307499}{CrossRef}]

\bibitem[{Perlman} \em{et~al.}(2001){Perlman}, {Biretta}, {Sparks},
{Macchetto}, and {Leahy}]{Perlman2001}
{Perlman}, E.S.; {Biretta}, J.A.; {Sparks}, W.B.; {Macchetto}, F.D.; {Leahy},
J.P.
\newblock {The Optical-Near-Infrared Spectrum of the M87 Jet fromHubble Space
Telescope Observations}.
\newblock {\em \apj} {\bf 2001}, {\em 551},~206--222.
\newblock  [\href{http://dx.doi.org/10.1086/320052}{CrossRef}]

\bibitem[{Marshall} \em{et~al.}(2002){Marshall}, {Canizares}, and
{Schulz}]{Marshall2002}
{Marshall}, H.L.; {Canizares}, C.R.; {Schulz}, N.S.
\newblock {The High-Resolution X-Ray Spectrum of SS 433 Using the Chandra
HETGS}.
\newblock {\em \apj} {\bf 2002}, {\em 564},~941--952.
\newblock  [\href{http://dx.doi.org/10.1086/324398}{CrossRef}]

\bibitem[{Cheung} \em{et~al.}(2007){Cheung}, {Harris}, and
{Stawarz}]{Cheung2007}
{Cheung}, C.C.; {Harris}, D.E.; {Stawarz}, {\L}.
\newblock {Superluminal Radio Features in the M87 Jet and the Site of Flaring
TeV Gamma-Ray Emission}.
\newblock {\em \apjl} {\bf 2007}, {\em 663},~L65--L68.
\newblock  [\href{http://dx.doi.org/10.1086/520510}{CrossRef}]

\bibitem[{Schmidt} \em{et~al.}(1978){Schmidt}, {Peterson}, and
{Beaver}]{Schmidt1978}
{Schmidt}, G.D.; {Peterson}, B.M.; {Beaver}, E.A.
\newblock {Imaging polarimetry of the jet of M87 and 3C 273.}
\newblock {\em \apjl} {\bf 1978}, {\em 220},~L31--L36.
\newblock  [\href{http://dx.doi.org/10.1086/182631}{CrossRef}]

\bibitem[{Dennison}(1980)]{Dennison1980}
{Dennison}, B.
\newblock {Faraday rotation in the M87 radio/X-ray halo}.
\newblock {\em \apj} {\bf 1980}, {\em 236},~761--768.
\newblock  [\href{http://dx.doi.org/10.1086/157801}{CrossRef}]

\bibitem[{Perlman} \em{et~al.}(1999){Perlman}, {Biretta}, {Zhou}, {Sparks}, and
{Macchetto}]{Perlman1999}
{Perlman}, E.S.; {Biretta}, J.A.; {Zhou}, F.; {Sparks}, W.B.; {Macchetto}, F.D.
\newblock {Optical and Radio Polarimetry of the M87 Jet at 0.2'' Resolution}.
\newblock {\em \aj} {\bf 1999}, {\em 117},~2185--2198.
\newblock  [\href{http://dx.doi.org/10.1086/300844}{CrossRef}]

\bibitem[{Algaba} \em{et~al.}(2016){Algaba}, {Asada}, and
{Nakamura}]{Algaba2016}
{Algaba}, J.C.; {Asada}, K.; {Nakamura}, M.
\newblock {Resolving the Rotation Measure of the M87 Jet on Kiloparsec Scales}.
\newblock {\em \apj} {\bf 2016}, {\em 823},~86.
\newblock  [\href{http://dx.doi.org/10.3847/0004-637X/823/2/86}{CrossRef}]

\bibitem[{Pasetto} \em{et~al.}(2021){Pasetto}, {Carrasco-Gonz{\'a}lez},
{G{\'o}mez}, {Mart{\'\i}}, {Perucho}, {O'Sullivan}, {Anderson},
{D{\'\i}az-Gonz{\'a}lez}, {Fuentes}, and {Wardle}]{Pasetto2021}
{Pasetto}, A.; {Carrasco-Gonz{\'a}lez}, C.; {G{\'o}mez}, J.L.; {Mart{\'\i}},
J.M.; {Perucho}, M.; {O'Sullivan}, S.P.; {Anderson}, C.;
{D{\'\i}az-Gonz{\'a}lez}, D.J.; {Fuentes}, A.; {Wardle}, J.
\newblock {Reading M87's DNA: A Double Helix Revealing a Large-scale Helical
Magnetic Field}.
\newblock {\em \apjl} {\bf 2021}, {\em 923},~L5.
\newblock  [\href{http://dx.doi.org/10.3847/2041-8213/ac3a88}{CrossRef}]

\bibitem[{Junor} \em{et~al.}(2001){Junor}, {Biretta}, and {Wardle}]{Junor2001}
{Junor}, W.; {Biretta}, J.A.; {Wardle}, J.F.C.
\newblock {VLBA lambda lambda 6, 4 cm polarimetry of Vir A}.
\newblock In Proceedings of the Galaxies and their Constituents at the Highest Angular Resolutions, Manchester, UK, 15--18 August   2001.

\bibitem[{Zavala} and {Taylor}(2002)]{ZT2002}
{Zavala}, R.T.; {Taylor}, G.B.
\newblock {Faraday Rotation Measures in the Parsec-Scale Jets of the Radio
Galaxies M87, 3C 111, and 3C 120}.
\newblock {\em \apjl} {\bf 2002}, {\em 566},~L9--L12.
\newblock  [\href{http://dx.doi.org/10.1086/339441}{CrossRef}]

\bibitem[{Park} \em{et~al.}(2019){Park}, {Hada}, {Kino}, {Nakamura}, {Ro}, and
{Trippe}]{Park2019a}
{Park}, J.; {Hada}, K.; {Kino}, M.; {Nakamura}, M.; {Ro}, H.; {Trippe}, S.
\newblock {Faraday Rotation in the Jet of M87 inside the Bondi Radius:
Indication of Winds from Hot Accretion Flows Confining the Relativistic Jet}.
\newblock {\em \apj} {\bf 2019}, {\em 871},~257.
\newblock  [\href{http://dx.doi.org/10.3847/1538-4357/aaf9a9}{CrossRef}]

\bibitem[{Nakamura} and {Asada}(2013)]{NA2013}
{Nakamura}, M.; {Asada}, K.
\newblock {The Parabolic Jet Structure in M87 as a Magnetohydrodynamic Nozzle}.
\newblock {\em \apj} {\bf 2013}, {\em 775},~118.
\newblock  [\href{http://dx.doi.org/10.1088/0004-637X/775/2/118}{CrossRef}]

\bibitem[{Hada} \em{et~al.}(2013){Hada}, {Doi}, {Nagai}, {Inoue}, {Honma},
{Giroletti}, and {Giovannini}]{Hada2013}
{Hada}, K.; {Doi}, A.; {Nagai}, H.; {Inoue}, M.; {Honma}, M.; {Giroletti}, M.;
{Giovannini}, G.
\newblock {Evidence for a Nuclear Radio Jet and its Structure down to lsim100
Schwarzschild Radii in the Center of the Sombrero Galaxy (M 104, NGC 4594)}.
\newblock {\em \apj} {\bf 2013}, {\em 779},~6.
\newblock  [\href{http://dx.doi.org/10.1088/0004-637X/779/1/6}{CrossRef}]

\bibitem[{Asada} \em{et~al.}(2014){Asada}, {Nakamura}, {Doi}, {Nagai}, and
{Inoue}]{Asada2014}
{Asada}, K.; {Nakamura}, M.; {Doi}, A.; {Nagai}, H.; {Inoue}, M.
\newblock {Discovery of Sub- to Superluminal Motions in the M87 Jet: An
Implication of Acceleration from Sub-relativistic to Relativistic Speeds}.
\newblock {\em \apjl} {\bf 2014}, {\em 781},~L2.
\newblock  [\href{http://dx.doi.org/10.1088/2041-8205/781/1/L2}{CrossRef}]

\bibitem[{Mertens} \em{et~al.}(2016){Mertens}, {Lobanov}, {Walker}, and
{Hardee}]{Mertens2016}
{Mertens}, F.; {Lobanov}, A.P.; {Walker}, R.C.; {Hardee}, P.E.
\newblock {Kinematics of the jet in M 87 on scales of 100--1000 Schwarzschild
radii}.
\newblock {\em \aap} {\bf 2016}, {\em 595},~A54.
\newblock  [\href{http://dx.doi.org/10.1051/0004-6361/201628829}{CrossRef}]

\bibitem[{Walker} \em{et~al.}(2018){Walker}, {Hardee}, {Davies}, {Ly}, and
{Junor}]{Walker2018}
{Walker}, R.C.; {Hardee}, P.E.; {Davies}, F.B.; {Ly}, C.; {Junor}, W.
\newblock {The Structure and Dynamics of the Subparsec Jet in M87 Based on 50
VLBA Observations over 17 Years at 43 GHz}.
\newblock {\em \apj} {\bf 2018}, {\em 855},~128.
\newblock  [\href{http://dx.doi.org/10.3847/1538-4357/aaafcc}{CrossRef}]

\bibitem[{Park} \em{et~al.}(2019){Park}, {Hada}, {Kino}, {Nakamura}, {Hodgson},
{Ro}, {Cui}, {Asada}, {Algaba}, {Sawada-Satoh}, {Lee}, {Cho}, {Shen},
{Jiang}, {Trippe}, {Niinuma}, {Sohn}, {Jung}, {Zhao}, {Wajima}, {Tazaki},
{Honma}, {An}, {Akiyama}, {Byun}, {Kim}, {Zhang}, {Cheng}, {Kobayashi},
{Shibata}, {Lee}, {Roh}, {Oh}, {Yeom}, {Jung}, {Oh}, {Kim}, {Hwang}, and
{Hagiwara}]{Park2019b}
{Park}, J.; {Hada}, K.; {Kino}, M.; {Nakamura}, M.; {Hodgson}, J.; {Ro}, H.;
{Cui}, Y.; {Asada}, K.; {Algaba}, J.C.; {Sawada-Satoh}, S.;  et~al.
\newblock {Kinematics of the M87 Jet in the Collimation Zone: Gradual
Acceleration and Velocity Stratification}.
\newblock {\em \apj} {\bf 2019}, {\em 887},~147.
\newblock  [\href{http://dx.doi.org/10.3847/1538-4357/ab5584}{CrossRef}]

\bibitem[{Burn}(1966)]{Burn1966}
{Burn}, B.J.
\newblock {On the depolarization of discrete radio sources by Faraday
dispersion}.
\newblock {\em \mnras} {\bf 1966}, {\em 133},~67.\linebreak
\newblock  [\href{http://dx.doi.org/10.1093/mnras/133.1.67}{CrossRef}]

\bibitem[{Homan}(2012)]{Homan2012}
{Homan}, D.C.
\newblock {Inverse Depolarization: A Potential Probe of Internal Faraday
Rotation and Helical Magnetic Fields in Extragalactic Radio Jets}.
\newblock {\em \apjl} {\bf 2012}, {\em 747},~L24.
\newblock  [\href{http://dx.doi.org/10.1088/2041-8205/747/2/L24}{CrossRef}]

\bibitem[{Sokoloff} \em{et~al.}(1998){Sokoloff}, {Bykov}, {Shukurov},
{Berkhuijsen}, {Beck}, and {Poezd}]{Sokoloff1998}
{Sokoloff}, D.D.; {Bykov}, A.A.; {Shukurov}, A.; {Berkhuijsen}, E.M.; {Beck},
R.; {Poezd}, A.D.
\newblock {Depolarization and Faraday effects in galaxies}.
\newblock {\em \mnras} {\bf 1998}, {\em 299},~189--206.
\newblock  [\href{http://dx.doi.org/10.1046/j.1365-8711.1998.01782.x}{CrossRef}]

\bibitem[{Yuan} and {Narayan}(2014)]{YN2014}
{Yuan}, F.; {Narayan}, R.
\newblock {Hot Accretion Flows Around Black Holes}.
\newblock {\em \araa} {\bf 2014}, {\em 52},~529--588.
\newblock  [\href{http://dx.doi.org/10.1146/annurev-astro-082812-141003}{CrossRef}]

\bibitem[{Hirose} \em{et~al.}(2004){Hirose}, {Krolik}, {De Villiers}, and
{Hawley}]{Hirose2004}
{Hirose}, S.; {Krolik}, J.H.; {De Villiers}, J.P.; {Hawley}, J.F.
\newblock {Magnetically Driven Accretion Flows in the Kerr Metric. II.
Structure of the Magnetic Field}.
\newblock {\em \apj} {\bf 2004}, {\em 606},~1083--1097.
\newblock  [\href{http://dx.doi.org/10.1086/383184}{CrossRef}]

\bibitem[{Clausen-Brown} \em{et~al.}(2013){Clausen-Brown}, {Savolainen},
{Pushkarev}, {Kovalev}, and {Zensus}]{Clausen-Brown2013}
{Clausen-Brown}, E.; {Savolainen}, T.; {Pushkarev}, A.B.; {Kovalev}, Y.Y.;
{Zensus}, J.A.
\newblock {Causal connection in parsec-scale relativistic jets: Results from
the MOJAVE VLBI survey}.
\newblock {\em \aap} {\bf 2013}, {\em 558},~A144.
\newblock  [\href{http://dx.doi.org/10.1051/0004-6361/201322203}{CrossRef}]

\bibitem[{Broderick} and {McKinney}(2010)]{BM2010}
{Broderick}, A.E.; {McKinney}, J.C.
\newblock {Parsec-scale Faraday Rotation Measures from General Relativistic
Magnetohydrodynamic Simulations of Active Galactic Nucleus Jets}.
\newblock {\em \apj} {\bf 2010}, {\em 725},~750--773.
\newblock  [\href{http://dx.doi.org/10.1088/0004-637X/725/1/750}{CrossRef}]

\bibitem[{Gabuzda}(2018)]{Gabuzda2018}
{Gabuzda}, D.
\newblock {Evidence for Helical Magnetic Fields Associated with AGN Jets and
the Action of a Cosmic Battery}.
\newblock {\em Galaxies} {\bf 2018}, {\em 7},~5.
\newblock  [\href{http://dx.doi.org/10.3390/galaxies7010005}{CrossRef}]

\bibitem[{Russell} \em{et~al.}(2015){Russell}, {Fabian}, {McNamara}, and
{Broderick}]{Russell2015}
{Russell}, H.R.; {Fabian}, A.C.; {McNamara}, B.R.; {Broderick}, A.E.
\newblock {Inside the Bondi radius of M87}.
\newblock {\em \mnras} {\bf 2015}, {\em 451},~588--600.
\newblock  [\href{http://dx.doi.org/10.1093/mnras/stv954}{CrossRef}]

\bibitem[{Blandford} and {Begelman}(1999)]{BB1999}
{Blandford}, R.D.; {Begelman}, M.C.
\newblock {On the fate of gas accreting at a low rate on to a black hole}.
\newblock {\em \mnras} {\bf 1999}, {\em 303},~L1--L5,
\newblock  [\href{http://dx.doi.org/10.1046/j.1365-8711.1999.02358.x}{CrossRef}]

\bibitem[{Blandford} and {Begelman}(2004)]{BB2004}
{Blandford}, R.D.; {Begelman}, M.C.
\newblock {Two-dimensional adiabatic flows on to a black hole---I. Fluid
accretion}.
\newblock {\em \mnras} {\bf 2004}, {\em 349},~68--86.
\newblock  [\href{http://dx.doi.org/10.1111/j.1365-2966.2004.07425.x}{CrossRef}]

\bibitem[{Begelman}(2012)]{Begelman2012}
{Begelman}, M.C.
\newblock {Radiatively inefficient accretion: Breezes, winds and
hyperaccretion}.
\newblock {\em \mnras} {\bf 2012}, {\em 420},~2912--2923.
\newblock  [\href{http://dx.doi.org/10.1111/j.1365-2966.2011.20071.x}{CrossRef}]

\bibitem[{Yuan} \em{et~al.}(2015){Yuan}, {Gan}, {Narayan}, {Sadowski}, {Bu},
and {Bai}]{Yuan2015}
{Yuan}, F.; {Gan}, Z.; {Narayan}, R.; {Sadowski}, A.; {Bu}, D.; {Bai}, X.N.
\newblock {Numerical Simulation of Hot Accretion Flows. III. Revisiting Wind
Properties Using the Trajectory Approach}.
\newblock {\em \apj} {\bf 2015}, {\em 804},~101.
\newblock  [\href{http://dx.doi.org/10.1088/0004-637X/804/2/101}{CrossRef}]

\bibitem[{Vlahakis}(2015)]{Vlahakis2015}
{Vlahakis}, N. {Theory of Relativistic Jets}.
\newblock In {\em The Formation and Disruption of Black Hole Jets};
{Contopoulos}, I., {Gabuzda}, D., {Kylafis}, N., Eds.;  {Astrophysics and Space Science Library; Spring: Cham, Switzerland}, 
2015; Volume 414, p. 177.
\newblock  [\href{http://dx.doi.org/10.1007/978-3-319-10356-3_7}{CrossRef}]

\bibitem[{Chatterjee} \em{et~al.}(2019){Chatterjee}, {Liska}, {Tchekhovskoy},
and {Markoff}]{Chatterjee2019}
{Chatterjee}, K.; {Liska}, M.; {Tchekhovskoy}, A.; {Markoff}, S.B.
\newblock {Accelerating AGN jets to parsec scales using general relativistic
MHD simulations}.
\newblock {\em \mnras} {\bf 2019}, {\em 490},~2200--2218.
\newblock  [\href{http://dx.doi.org/10.1093/mnras/stz2626}{CrossRef}]

\bibitem[{Vlahakis} and {K{\"o}nigl}(2003)]{VK2003}
{Vlahakis}, N.; {K{\"o}nigl}, A.
\newblock {Relativistic Magnetohydrodynamics with Application to Gamma-Ray
Burst Outflows. I. Theory and Semianalytic Trans-Alfv{\'e}nic Solutions}.
\newblock {\em \apj} {\bf 2003}, {\em 596},~1080--1103.
\newblock  [\href{http://dx.doi.org/10.1086/378226}{CrossRef}]

\bibitem[{Biretta} \em{et~al.}(1995){Biretta}, {Zhou}, and {Owen}]{Biretta1995}
{Biretta}, J.A.; {Zhou}, F.; {Owen}, F.N.
\newblock {Detection of Proper Motions in the M87 Jet}.
\newblock {\em \apj} {\bf 1995}, {\em 447},~582.
\newblock  [\href{http://dx.doi.org/10.1086/175901}{CrossRef}]

\bibitem[{Giroletti} \em{et~al.}(2012){Giroletti}, {Hada}, {Giovannini},
{Casadio}, {Beilicke}, {Cesarini}, {Cheung}, {Doi}, {Krawczynski}, {Kino},
{Lee}, and {Nagai}]{Giroletti2012}
{Giroletti}, M.; {Hada}, K.; {Giovannini}, G.; {Casadio}, C.; {Beilicke}, M.;
{Cesarini}, A.; {Cheung}, C.C.; {Doi}, A.; {Krawczynski}, H.; {Kino}, M.;
et~al.
\newblock {The kinematic of HST-1 in the jet of M 87}.
\newblock {\em \aap} {\bf 2012}, {\em 538},~L10.
\newblock  [\href{http://dx.doi.org/10.1051/0004-6361/201218794}{CrossRef}]

\bibitem[{Meyer} \em{et~al.}(2013){Meyer}, {Sparks}, {Biretta}, {Anderson},
{Sohn}, {van der Marel}, {Norman}, and {Nakamura}]{Meyer2013}
{Meyer}, E.T.; {Sparks}, W.B.; {Biretta}, J.A.; {Anderson}, J.; {Sohn}, S.T.;
{van der Marel}, R.P.; {Norman}, C.; {Nakamura}, M.
\newblock {Optical Proper Motion Measurements of the M87 Jet: New Results from
the Hubble Space Telescope}.
\newblock {\em \apjl} {\bf 2013}, {\em 774},~L21.
\newblock  [\href{http://dx.doi.org/10.1088/2041-8205/774/2/L21}{CrossRef}]

\bibitem[{Event Horizon Telescope Collaboration} \em{et~al.}(2021){Event
Horizon Telescope Collaboration}, {Akiyama}, {Algaba}, {Alberdi}, {Alef},
{Anantua}, {Asada}, {Azulay}, {Baczko}, {Ball}, {Balokovi{\'c}}, {Barrett},
{Benson}, {Bintley}, {Blackburn}, {Blundell}, {Boland}, {Bouman}, {Bower},
{Boyce}, {Bremer}, {Brinkerink}, {Brissenden}, {Britzen}, {Broderick},
{Broguiere}, {Bronzwaer}, {Byun}, {Carlstrom}, {Chael}, {Chan}, {Chatterjee},
{Chatterjee}, {Chen}, {Chen}, {Chesler}, {Cho}, {Christian}, {Conway},
{Cordes}, {Crawford}, {Crew}, {Cruz-Osorio}, {Cui}, {Davelaar}, {De
Laurentis}, {Deane}, {Dempsey}, {Desvignes}, {Dexter}, {Doeleman}, {Eatough},
{Falcke}, {Farah}, {Fish}, {Fomalont}, {Ford}, {Fraga-Encinas}, {Friberg},
{Fromm}, {Fuentes}, {Galison}, {Gammie}, {Garc{\'\i}a}, {Gelles}, {Gentaz},
{Georgiev}, {Goddi}, {Gold}, {G{\'o}mez}, {G{\'o}mez-Ruiz}, {Gu}, {Gurwell},
{Hada}, {Haggard}, {Hecht}, {Hesper}, {Himwich}, {Ho}, {Ho}, {Honma},
{Huang}, {Huang}, {Hughes}, {Ikeda}, {Inoue}, {Issaoun}, {James}, {Jannuzi},
{Janssen}, {Jeter}, {Jiang}, {Jimenez-Rosales}, {Johnson}, {Jorstad}, {Jung},
{Karami}, {Karuppusamy}, {Kawashima}, {Keating}, {Kettenis}, {Kim}, {Kim},
{Kim}, {Kim}, {Kino}, {Koay}, {Kofuji}, {Koch}, {Koyama}, {Kramer}, {Kramer},
{Krichbaum}, {Kuo}, {Lauer}, {Lee}, {Levis}, {Li}, {Li}, {Lindqvist}, {Lico},
{Lindahl}, {Liu}, {Liu}, {Liuzzo}, {Lo}, {Lobanov}, {Loinard}, {Lonsdale},
{Lu}, {MacDonald}, {Mao}, {Marchili}, {Markoff}, {Marrone}, {Marscher},
{Mart{\'\i}-Vidal}, {Matsushita}, {Matthews}, {Medeiros}, {Menten}, {Mizuno},
{Mizuno}, {Moran}, {Moriyama}, {Moscibrodzka}, {M{\"u}ller}, {Musoke}, {Mus
Mej{\'\i}as}, {Michalik}, {Nadolski}, {Nagai}, {Nagar}, {Nakamura},
{Narayan}, {Narayanan}, {Natarajan}, {Nathanail}, {Neilsen}, {Neri}, {Ni},
{Noutsos}, {Nowak}, {Okino}, {Olivares}, {Ortiz-Le{\'o}n}, {Oyama},
{{\"O}zel}, {Palumbo}, {Park}, {Patel}, {Pen}, {Pesce}, {Pi{\'e}tu},
{Plambeck}, {PopStefanija}, {Porth}, {P{\"o}tzl}, {Prather},
{Preciado-L{\'o}pez}, {Psaltis}, {Pu}, {Ramakrishnan}, {Rao}, {Rawlings},
{Raymond}, {Rezzolla}, {Ricarte}, {Ripperda}, {Roelofs}, {Rogers}, {Ros},
{Rose}, {Roshanineshat}, {Rottmann}, {Roy}, {Ruszczyk}, {Rygl},
{S{\'a}nchez}, {S{\'a}nchez-Arguelles}, {Sasada}, {Savolainen}, {Schloerb},
{Schuster}, {Shao}, {Shen}, {Small}, {Sohn}, {SooHoo}, {Sun}, {Tazaki},
{Tetarenko}, {Tiede}, {Tilanus}, {Titus}, {Toma}, {Torne}, {Trent},
{Traianou}, {Trippe}, {van Bemmel}, {van Langevelde}, {van Rossum}, {Wagner},
{Ward-Thompson}, {Wardle}, {Weintroub}, {Wex}, {Wharton}, {Wielgus}, {Wong},
{Wu}, {Yoon}, {Young}, {Young}, {Younsi}, {Yuan}, {Yuan}, {Zensus}, {Zhao},
and {Zhao}]{EHT2021b}
{Event Horizon Telescope Collaboration}.; {Akiyama}, K.; {Algaba}, J.C.;
{Alberdi}, A.; {Alef}, W.; {Anantua}, R.; {Asada}, K.; {Azulay}, R.;
{Baczko}, A.K.; {Ball}, D.;  et~al.
\newblock {First M87 Event Horizon Telescope Results. VIII. Magnetic Field
Structure near The Event Horizon}.
\newblock {\em \apjl} {\bf 2021}, {\em 910},~L13.
\newblock  [\href{http://dx.doi.org/10.3847/2041-8213/abe4de}{CrossRef}]

\bibitem[{Narayan} \em{et~al.}(2021){Narayan}, {Palumbo}, {Johnson}, {Gelles},
{Himwich}, {Chang}, {Ricarte}, {Dexter}, {Gammie}, {Chael}, {Event Horizon
Telescope Collaboration}, {Akiyama}, {Alberdi}, {Alef}, {Algaba}, {Anantua},
{Asada}, {Azulay}, {Baczko}, {Ball}, {Balokovi{\'c}}, {Barrett}, {Benson},
{Bintley}, {Blackburn}, {Blundell}, {Boland}, {Bouman}, {Bower}, {Boyce},
{Bremer}, {Brinkerink}, {Brissenden}, {Britzen}, {Broderick}, {Broguiere},
{Bronzwaer}, {Byun}, {Carlstrom}, {Chan}, {Chatterjee}, {Chatterjee}, {Chen},
{Chen}, {Chesler}, {Cho}, {Christian}, {Conway}, {Cordes}, {Crawford},
{Crew}, {Cruz-Osorio}, {Cui}, {Davelaar}, {De Laurentis}, {Deane}, {Dempsey},
{Desvignes}, {Doeleman}, {Eatough}, {Falcke}, {Farah}, {Fish}, {Fomalont},
{Ford}, {Fraga-Encinas}, {Friberg}, {Fromm}, {Fuentes}, {Galison},
{Garc{\'\i}a}, {Gentaz}, {Georgiev}, {Goddi}, {Gold}, {G{\'o}mez},
{G{\'o}mez-Ruiz}, {Gu}, {Gurwell}, {Hada}, {Haggard}, {Hecht}, {Hesper},
{Ho}, {Ho}, {Honma}, {Huang}, {Huang}, {Hughes}, {Ikeda}, {Inoue}, {Issaoun},
{James}, {Jannuzi}, {Janssen}, {Jeter}, {Jiang}, {Jimenez-Rosales},
{Jorstad}, {Jung}, {Karami}, {Karuppusamy}, {Kawashima}, {Keating},
{Kettenis}, {Kim}, {Kim}, {Kim}, {Kim}, {Kino}, {Koay}, {Kofuji}, {Koch},
{Koyama}, {Kramer}, {Kramer}, {Krichbaum}, {Kuo}, {Lauer}, {Lee}, {Levis},
{Li}, {Li}, {Lindqvist}, {Lico}, {Lindahl}, {Liu}, {Liu}, {Liuzzo}, {Lo},
{Lobanov}, {Loinard}, {Lonsdale}, {Lu}, {MacDonald}, {Mao}, {Marchili},
{Markoff}, {Marrone}, {Marscher}, {Mart{\'\i}-Vidal}, {Matsushita},
{Matthews}, {Medeiros}, {Menten}, {Mizuno}, {Mizuno}, {Moran}, {Moriyama},
{Moscibrodzka}, {M{\"u}ller}, {Musoke}, {Mej{\'\i}as}, {Nagai}, {Nagar},
{Nakamura}, {Narayanan}, {Natarajan}, {Nathanail}, {Neilsen}, {Neri}, {Ni},
{Noutsos}, {Nowak}, {Okino}, {Olivares}, {Ortiz-Le{\'o}n}, {Oyama},
{{\"O}zel}, {Park}, {Patel}, {Pen}, {Pesce}, {Pi{\'e}tu}, {Plambeck},
{PopStefanija}, {Porth}, {P{\"o}tzl}, {Prather}, {Preciado-L{\'o}pez},
{Psaltis}, {Pu}, {Ramakrishnan}, {Rao}, {Rawlings}, {Raymond}, {Rezzolla},
{Ripperda}, {Roelofs}, {Rogers}, {Ros}, {Rose}, {Roshanineshat}, {Rottmann},
{Roy}, {Ruszczyk}, {Rygl}, {S{\'a}nchez}, {S{\'a}nchez-Arguelles}, {Sasada},
{Savolainen}, {Schloerb}, {Schuster}, {Shao}, {Shen}, {Small}, {Sohn},
{SooHoo}, {Sun}, {Tazaki}, {Tetarenko}, {Tiede}, {Tilanus}, {Titus}, {Toma},
{Torne}, {Trent}, {Traianou}, {Trippe}, {van Bemmel}, {van Langevelde}, {van
Rossum}, {Wagner}, {Ward-Thompson}, {Wardle}, {Weintroub}, {Wex}, {Wharton},
{Wielgus}, {Wong}, {Wu}, {Yoon}, {Young}, {Young}, {Younsi}, {Yuan}, {Yuan},
{Zensus}, {Zhao}, and {Zhao}]{Narayan2021}
{Narayan}, R.; {Palumbo}, D.C.M.; {Johnson}, M.D.; {Gelles}, Z.; {Himwich}, E.;
{Chang}, D.O.; {Ricarte}, A.; {Dexter}, J.; {Gammie}, C.F.; {Chael}, A.A.;
et~al.
\newblock {The Polarized Image of a Synchrotron-emitting Ring of Gas Orbiting a
Black Hole}.
\newblock {\em \apj} {\bf 2021}, {\em 912},~35.
\newblock  [\href{http://dx.doi.org/10.3847/1538-4357/abf117}{CrossRef}]

\bibitem[{Narayan} \em{et~al.}(2012){Narayan}, {Sadowski}, {Penna}, and
{Kulkarni}]{Narayan2012}
{Narayan}, R.; {Sadowski}, A.; {Penna}, R.F.; {Kulkarni}, A.K.
\newblock {GRMHD simulations of magnetized advection-dominated accretion on a
non-spinning black hole: Role of outflows}.
\newblock {\em \mnras} {\bf 2012}, {\em 426},~3241--3259.
\newblock  [\href{http://dx.doi.org/10.1111/j.1365-2966.2012.22002.x}{CrossRef}]

\bibitem[{Ricarte} \em{et~al.}(2021){Ricarte}, {Qiu}, and
{Narayan}]{Ricarte2021}
{Ricarte}, A.; {Qiu}, R.; {Narayan}, R.
\newblock {Black hole magnetic fields and their imprint on circular
polarization images}.
\newblock {\em \mnras} {\bf 2021}, {\em 505},~523--539.
\newblock  [\href{http://dx.doi.org/10.1093/mnras/stab1289}{CrossRef}]

\bibitem[{Bisnovatyi-Kogan} and {Ruzmaikin}(1974)]{BR1974}
{Bisnovatyi-Kogan}, G.S.; {Ruzmaikin}, A.A.
\newblock {The Accretion of Matter by a Collapsing Star in the Presence of a
Magnetic Field}.
\newblock {\em \apss} {\bf 1974}, {\em 28},~45--59.
\newblock  [\href{http://dx.doi.org/10.1007/BF00642237}{CrossRef}]

\bibitem[{Igumenshchev} \em{et~al.}(2003){Igumenshchev}, {Narayan}, and
{Abramowicz}]{Igumenshchev2003}
{Igumenshchev}, I.V.; {Narayan}, R.; {Abramowicz}, M.A.
\newblock {Three-dimensional Magnetohydrodynamic Simulations of Radiatively
Inefficient Accretion Flows}.
\newblock {\em \apj} {\bf 2003}, {\em 592},~1042--1059.
\newblock  [\href{http://dx.doi.org/10.1086/375769}{CrossRef}]

\bibitem[{Tchekhovskoy} \em{et~al.}(2012){Tchekhovskoy}, {McKinney}, and
{Narayan}]{Tchekhovskoy2012}
{Tchekhovskoy}, A.; {McKinney}, J.C.; {Narayan}, R.
\newblock {General Relativistic Modeling of Magnetized Jets from Accreting
Black Holes}.
\newblock \emph{J. Phys. Conf. Ser.}  \textbf{2012}, \emph{372}, 012040.
\newblock  [\href{http://dx.doi.org/10.1088/1742-6596/372/1/012040}{CrossRef}]

\bibitem[{Bicknell} and {Begelman}(1996)]{BB1996}
{Bicknell}, G.V.; {Begelman}, M.C.
\newblock {Understanding the Kiloparsec-Scale Structure of M87}.
\newblock {\em \apj} {\bf 1996}, {\em 467},~597.
\newblock  [\href{http://dx.doi.org/10.1086/177636}{CrossRef}]

\bibitem[{Owen} \em{et~al.}(2000){Owen}, {Eilek}, and {Kassim}]{Owen2000}
{Owen}, F.N.; {Eilek}, J.A.; {Kassim}, N.E.
\newblock {M87 at 90 Centimeters: A Different Picture}.
\newblock {\em \apj} {\bf 2000}, {\em 543},~611--619.
\newblock  [\href{http://dx.doi.org/10.1086/317151}{CrossRef}]

\bibitem[{Allen} \em{et~al.}(2006){Allen}, {Dunn}, {Fabian}, {Taylor}, and
{Reynolds}]{Allen2006}
{Allen}, S.W.; {Dunn}, R.J.H.; {Fabian}, A.C.; {Taylor}, G.B.; {Reynolds}, C.S.
\newblock {The relation between accretion rate and jet power in X-ray luminous
elliptical galaxies}.
\newblock {\em \mnras} {\bf 2006}, {\em 372},~21--30.
\newblock  [\href{http://dx.doi.org/10.1111/j.1365-2966.2006.10778.x}{CrossRef}]

\bibitem[{Rafferty} \em{et~al.}(2006){Rafferty}, {McNamara}, {Nulsen}, and
{Wise}]{Rafferty2006}
{Rafferty}, D.A.; {McNamara}, B.R.; {Nulsen}, P.E.J.; {Wise}, M.W.
\newblock {The Feedback-regulated Growth of Black Holes and Bulges through Gas
Accretion and Starbursts in Cluster Central Dominant Galaxies}.
\newblock {\em \apj} {\bf 2006}, {\em 652},~216--231.
\newblock  [\href{http://dx.doi.org/10.1086/507672}{CrossRef}]

\bibitem[{Stawarz} \em{et~al.}(2006){Stawarz}, {Aharonian}, {Kataoka},
{Ostrowski}, {Siemiginowska}, and {Sikora}]{Stawarz2006}
{Stawarz}, {\L}.; {Aharonian}, F.; {Kataoka}, J.; {Ostrowski}, M.;
{Siemiginowska}, A.; {Sikora}, M.
\newblock {Dynamics and high-energy emission of the flaring HST-1 knot in the M
87 jet}.
\newblock {\em \mnras} {\bf 2006}, {\em 370},~981--992.
\newblock  [\href{http://dx.doi.org/10.1111/j.1365-2966.2006.10525.x}{CrossRef}]

\bibitem[{Broderick} and {Loeb}(2009)]{BL2009}
{Broderick}, A.E.; {Loeb}, A.
\newblock {Imaging the Black Hole Silhouette of M87: Implications for Jet
Formation and Black Hole Spin}.
\newblock {\em \apj} {\bf 2009}, {\em 697},~1164--1179.
\newblock  [\href{http://dx.doi.org/10.1088/0004-637X/697/2/1164}{CrossRef}]

\bibitem[{Broderick} \em{et~al.}(2015){Broderick}, {Narayan}, {Kormendy},
{Perlman}, {Rieke}, and {Doeleman}]{Broderick2015}
{Broderick}, A.E.; {Narayan}, R.; {Kormendy}, J.; {Perlman}, E.S.; {Rieke},
M.J.; {Doeleman}, S.S.
\newblock {The Event Horizon of M87}.
\newblock {\em \apj} {\bf 2015}, {\em 805},~179.
\newblock  [\href{http://dx.doi.org/10.1088/0004-637X/805/2/179}{CrossRef}]

\bibitem[{Strauss} \em{et~al.}(1992){Strauss}, {Huchra}, {Davis}, {Yahil},
{Fisher}, and {Tonry}]{Strauss1992}
{Strauss}, M.A.; {Huchra}, J.P.; {Davis}, M.; {Yahil}, A.; {Fisher}, K.B.;
{Tonry}, J.
\newblock {A Redshift Survey of IRAS Galaxies. VII. The Infrared and Redshift
Data for the 1.936 Jansky Sample}.
\newblock {\em \apjs} {\bf 1992}, {\em 83},~29.
\newblock  [\href{http://dx.doi.org/10.1086/191730}{CrossRef}]

\bibitem[{Scharw{\"a}chter} \em{et~al.}(2013){Scharw{\"a}chter}, {McGregor},
{Dopita}, and {Beck}]{Scharwachter2013}
{Scharw{\"a}chter}, J.; {McGregor}, P.J.; {Dopita}, M.A.; {Beck}, T.L.
\newblock {Kinematics and excitation of the molecular hydrogen accretion disc
in NGC 1275}.
\newblock {\em \mnras} {\bf 2013}, {\em 429},~2315--2332.
\newblock  [\href{http://dx.doi.org/10.1093/mnras/sts502}{CrossRef}]

\bibitem[{Giovannini} \em{et~al.}(2018){Giovannini}, {Savolainen}, {Orienti},
{Nakamura}, {Nagai}, {Kino}, {Giroletti}, {Hada}, {Bruni}, {Kovalev},
{Anderson}, {D'Ammando}, {Hodgson}, {Honma}, {Krichbaum}, {Lee}, {Lico},
{Lisakov}, {Lobanov}, {Petrov}, {Sohn}, {Sokolovsky}, {Voitsik}, {Zensus},
and {Tingay}]{Giovannini2018}
{Giovannini}, G.; {Savolainen}, T.; {Orienti}, M.; {Nakamura}, M.; {Nagai}, H.;
{Kino}, M.; {Giroletti}, M.; {Hada}, K.; {Bruni}, G.; {Kovalev}, Y.Y.;
et~al.
\newblock {A wide and collimated radio jet in 3C84 on the scale of a few
hundred gravitational radii}.
\newblock {\em Nat. Astron.} {\bf 2018}, {\em 2},~472--477.
\newblock  [\href{http://dx.doi.org/10.1038/s41550-018-0431-2}{CrossRef}]

\bibitem[{Walker} \em{et~al.}(2000){Walker}, {Dhawan}, {Romney}, {Kellermann},
and {Vermeulen}]{Walker2000}
{Walker}, R.C.; {Dhawan}, V.; {Romney}, J.D.; {Kellermann}, K.I.; {Vermeulen},
R.C.
\newblock {VLBA Absorption Imaging of Ionized Gas Associated with the Accretion
Disk in NGC 1275}.
\newblock {\em \apj} {\bf 2000}, {\em 530},~233--244.
\newblock  [\href{http://dx.doi.org/10.1086/308372}{CrossRef}]

\bibitem[{Kino} \em{et~al.}(2021){Kino}, {Niinuma}, {Kawakatu}, {Nagai},
{Giovannini}, {Orienti}, {Wajima}, {D'Ammando}, {Hada}, {Giroletti}, and
{Gurwell}]{Kino2021}
{Kino}, M.; {Niinuma}, K.; {Kawakatu}, N.; {Nagai}, H.; {Giovannini}, G.;
{Orienti}, M.; {Wajima}, K.; {D'Ammando}, F.; {Hada}, K.; {Giroletti}, M.;
et~al.
\newblock {Morphological Transition of the Compact Radio Lobe in 3C 84 via the
Strong Jet-Cloud Collision}.
\newblock {\em \apjl} {\bf 2021}, {\em 920},~L24.
\newblock  [\href{http://dx.doi.org/10.3847/2041-8213/ac24fa}{CrossRef}]

\bibitem[{Savolainen} \em{et~al.}(2021){Savolainen}, {Giovannini}, {Kovalev},
{Perucho}, {Anderson}, {Bruni}, {Edwards}, {Fuentes}, {Giroletti},
{G{\'o}mez}, {Hada}, {Lee}, {Lisakov}, {Lobanov}, {L{\'o}pez-Miralles},
{Orienti}, {Petrov}, {Plavin}, {Sohn}, {Sokolovsky}, {Voitsik}, and
{Zensus}]{Savolainen2021}
{Savolainen}, T.; {Giovannini}, G.; {Kovalev}, Y.Y.; {Perucho}, M.; {Anderson},
J.M.; {Bruni}, G.; {Edwards}, P.G.; {Fuentes}, A.; {Giroletti}, M.;
{G{\'o}mez}, J.L.;  et~al.
\newblock {RadioAstron discovers a mini-cocoon around the restarted
parsec-scale jet in 3C 84}.
\newblock {\em arXiv} {\bf 2021}, arXiv:2111.04481.

\bibitem[{Taylor} \em{et~al.}(2006){Taylor}, {Gugliucci}, {Fabian}, {Sanders},
{Gentile}, and {Allen}]{Taylor2006}
{Taylor}, G.B.; {Gugliucci}, N.E.; {Fabian}, A.C.; {Sanders}, J.S.; {Gentile},
G.; {Allen}, S.W.
\newblock {Magnetic fields in the centre of the Perseus cluster}.
\newblock {\em \mnras} {\bf 2006}, {\em 368},~1500--1506.
\newblock  [\href{http://dx.doi.org/10.1111/j.1365-2966.2006.10244.x}{CrossRef}]

\bibitem[{Plambeck} \em{et~al.}(2014){Plambeck}, {Bower}, {Rao}, {Marrone},
{Jorstad}, {Marscher}, {Doeleman}, {Fish}, and {Johnson}]{Plambeck2014}
{Plambeck}, R.L.; {Bower}, G.C.; {Rao}, R.; {Marrone}, D.P.; {Jorstad}, S.G.;
{Marscher}, A.P.; {Doeleman}, S.S.; {Fish}, V.L.; {Johnson}, M.D.
\newblock {Probing the Parsec-scale Accretion Flow of 3C 84 with Millimeter
Wavelength Polarimetry}.
\newblock {\em \apj} {\bf 2014}, {\em 797},~66.
\newblock  [\href{http://dx.doi.org/10.1088/0004-637X/797/1/66}{CrossRef}]

\bibitem[{Bower} \em{et~al.}(2003){Bower}, {Wright}, {Falcke}, and
{Backer}]{Bower2003}
{Bower}, G.C.; {Wright}, M.C.H.; {Falcke}, H.; {Backer}, D.C.
\newblock {Interferometric Detection of Linear Polarization from Sagittarius A*
at 230 GHz}.
\newblock {\em \apj} {\bf 2003}, {\em 588},~331--337.
\newblock  [\href{http://dx.doi.org/10.1086/373989}{CrossRef}]

\bibitem[{Marrone} \em{et~al.}(2007){Marrone}, {Moran}, {Zhao}, and
{Rao}]{Marrone2007}
{Marrone}, D.P.; {Moran}, J.M.; {Zhao}, J.H.; {Rao}, R.
\newblock {An Unambiguous Detection of Faraday Rotation in Sagittarius A*}.
\newblock {\em \apjl} {\bf 2007}, {\em 654},~L57--L60.
\newblock  [\href{http://dx.doi.org/10.1086/510850}{CrossRef}]

\bibitem[{Nagai} \em{et~al.}(2017){Nagai}, {Fujita}, {Nakamura}, {Orienti},
{Kino}, {Asada}, and {Giovannini}]{Nagai2017}
{Nagai}, H.; {Fujita}, Y.; {Nakamura}, M.; {Orienti}, M.; {Kino}, M.; {Asada},
K.; {Giovannini}, G.
\newblock {Enhanced Polarized Emission from the One-parsec-scale Hotspot of 3C
84 as a Result of the Interaction with the Clumpy Ambient Medium}.
\newblock {\em \apj} {\bf 2017}, {\em 849},~52.
\newblock  [\href{http://dx.doi.org/10.3847/1538-4357/aa8e43}{CrossRef}]

\bibitem[{Kim} \em{et~al.}(2019){Kim}, {Krichbaum}, {Marscher}, {Jorstad},
{Agudo}, {Thum}, {Hodgson}, {MacDonald}, {Ros}, {Lu}, {Bremer}, {de Vicente},
{Lindqvist}, {Trippe}, and {Zensus}]{Kim2019}
{Kim}, J.Y.; {Krichbaum}, T.P.; {Marscher}, A.P.; {Jorstad}, S.G.; {Agudo}, I.;
{Thum}, C.; {Hodgson}, J.A.; {MacDonald}, N.R.; {Ros}, E.; {Lu}, R.S.;
et~al.
\newblock {Spatially resolved origin of millimeter-wave linear polarization in
the nuclear region of 3C 84}.
\newblock {\em \aap} {\bf 2019}, {\em 622},~A196.
\newblock  [\href{http://dx.doi.org/10.1051/0004-6361/201832920}{CrossRef}]

\bibitem[{Owen} \em{et~al.}(1990){Owen}, {Eilek}, and {Keel}]{Owen1990}
{Owen}, F.N.; {Eilek}, J.A.; {Keel}, W.C.
\newblock {Detection of Large Faraday Rotation in the Inner 2 Kiloparsecs of
M87}.
\newblock {\em \apj} {\bf 1990}, {\em 362},~449.
\newblock  [\href{http://dx.doi.org/10.1086/169282}{CrossRef}]

\bibitem[{Pasetto} \em{et~al.}(2018){Pasetto}, {Carrasco-Gonz{\'a}lez},
{O'Sullivan}, {Basu}, {Bruni}, {Kraus}, {Curiel}, and {Mack}]{Pasetto2018}
{Pasetto}, A.; {Carrasco-Gonz{\'a}lez}, C.; {O'Sullivan}, S.; {Basu}, A.;
{Bruni}, G.; {Kraus}, A.; {Curiel}, S.; {Mack}, K.H.
\newblock {Broadband radio spectro-polarimetric observations of
high-Faraday-rotation-measure AGN}.
\newblock {\em \aap} {\bf 2018}, {\em 613},~A74.
\newblock  [\href{http://dx.doi.org/10.1051/0004-6361/201731804}{CrossRef}]

\bibitem[{Paltani} \em{et~al.}(1998){Paltani}, {Courvoisier}, and
{Walter}]{Paltani1998}
{Paltani}, S.; {Courvoisier}, T.J.L.; {Walter}, R.
\newblock {The blue-bump of 3C 273}.
\newblock {\em \aap} {\bf 1998}, {\em 340},~47--61.

\bibitem[{Meyer} \em{et~al.}(2016){Meyer}, {Sparks}, {Georganopoulos},
{Anderson}, {van der Marel}, {Biretta}, {Sohn}, {Chiaberge}, {Perlman}, and
{Norman}]{Meyer2016}
{Meyer}, E.T.; {Sparks}, W.B.; {Georganopoulos}, M.; {Anderson}, J.; {van der
Marel}, R.; {Biretta}, J.; {Sohn}, S.T.; {Chiaberge}, M.; {Perlman}, E.;
{Norman}, C.
\newblock {An HST Proper-motion Study of the Large-scale Jet of 3C273}.
\newblock {\em \apj} {\bf 2016}, {\em 818},~195.
\newblock  [\href{http://dx.doi.org/10.3847/0004-637X/818/2/195}{CrossRef}]

\bibitem[{Lobanov} and {Zensus}(2001)]{LZ2001}
{Lobanov}, A.P.; {Zensus}, J.A.
\newblock {A Cosmic Double Helix in the Archetypical Quasar 3C273}.
\newblock {\em Science} {\bf 2001}, {\em 294},~128--131.
\newblock  [\href{http://dx.doi.org/10.1126/science.1063239}{CrossRef}]

\bibitem[{Okino} \em{et~al.}(2021){Okino}, {Akiyama}, {Asada}, {G{\'o}mez},
{Hada}, {Honma}, {Krichbaum}, {Kino}, {Nagai}, {Nakamura}, {Bach},
{Blackburn}, {Bouman}, {Chael}, {Crew}, {Doeleman}, {Fish}, {Gabuzda},
{Goddi}, {Issaoun}, {Johnson}, {Jorstad}, {Koyama}, {Lonsdale},
{Mart{\'\i}-Vidal}, {Matthews}, {Mizuno}, {Moriyama}, {Pu}, {Ros},
{Savolainen}, {Tazaki}, {Wagner}, {Wielgus}, and {Zensus}]{Okino2021}
{Okino}, H.; {Akiyama}, K.; {Asada}, K.; {G{\'o}mez}, J.L.; {Hada}, K.;
{Honma}, M.; {Krichbaum}, T.P.; {Kino}, M.; {Nagai}, H.; {Nakamura}, M.;
et~al.
\newblock {Collimation of the relativistic jet in the quasar 3C 273}.
\newblock {\em arXiv} {\bf 2021}, arXiv:2112.12233.

\bibitem[{Attridge} \em{et~al.}(2005){Attridge}, {Wardle}, and
{Homan}]{Attridge2005}
{Attridge}, J.M.; {Wardle}, J.F.C.; {Homan}, D.C.
\newblock {Concurrent 43 and 86 GHz Very Long Baseline Polarimetry of 3C 273}.
\newblock {\em \apjl} {\bf 2005}, {\em 633},~L85--L88.
\newblock  [\href{http://dx.doi.org/10.1086/498392}{CrossRef}]

\bibitem[{Hada} \em{et~al.}(2016){Hada}, {Kino}, {Doi}, {Nagai}, {Honma},
{Akiyama}, {Tazaki}, {Lico}, {Giroletti}, {Giovannini}, {Orienti}, and
{Hagiwara}]{Hada2016}
{Hada}, K.; {Kino}, M.; {Doi}, A.; {Nagai}, H.; {Honma}, M.; {Akiyama}, K.;
{Tazaki}, F.; {Lico}, R.; {Giroletti}, M.; {Giovannini}, G.;  et~al.
\newblock {High-sensitivity 86 GHz (3.5 mm) VLBI Observations of M87: Deep
Imaging of the Jet Base at a Resolution of 10 Schwarzschild Radii}.
\newblock {\em \apj} {\bf 2016}, {\em 817},~131.
\newblock  [\href{http://dx.doi.org/10.3847/0004-637X/817/2/131}{CrossRef}]

\bibitem[{Hovatta} \em{et~al.}(2019){Hovatta}, {O'Sullivan},
{Mart{\'\i}-Vidal}, {Savolainen}, and {Tchekhovskoy}]{Hovatta2019}
{Hovatta}, T.; {O'Sullivan}, S.; {Mart{\'\i}-Vidal}, I.; {Savolainen}, T.;
{Tchekhovskoy}, A.
\newblock {Magnetic field at a jet base: Extreme Faraday rotation in 3C 273
revealed by ALMA}.
\newblock {\em \aap} {\bf 2019}, {\em 623},~A111.
\newblock  [\href{http://dx.doi.org/10.1051/0004-6361/201832587}{CrossRef}]

\bibitem[{Asada} \em{et~al.}(2002){Asada}, {Inoue}, {Uchida}, {Kameno},
{Fujisawa}, {Iguchi}, and {Mutoh}]{Asada2002}
{Asada}, K.; {Inoue}, M.; {Uchida}, Y.; {Kameno}, S.; {Fujisawa}, K.; {Iguchi},
S.; {Mutoh}, M.
\newblock {A Helical Magnetic Field in the Jet of 3C 273}.
\newblock {\em \pasj} {\bf 2002}, {\em 54},~L39--L43.
\newblock  [\href{http://dx.doi.org/10.1093/pasj/54.3.L39}{CrossRef}]

\bibitem[{Asada} \em{et~al.}(2008){Asada}, {Inoue}, {Kameno}, and
{Nagai}]{Asada2008}
{Asada}, K.; {Inoue}, M.; {Kameno}, S.; {Nagai}, H.
\newblock {Time Variation of the Rotation Measure Gradient in the 3C 273 Jet}.
\newblock {\em \apj} {\bf 2008}, {\em 675},~79--82.
\newblock  [\href{http://dx.doi.org/10.1086/524000}{CrossRef}]

\bibitem[{Zavala} and {Taylor}(2005)]{ZT2005}
{Zavala}, R.T.; {Taylor}, G.B.
\newblock {Faraday Rotation Measure Gradients from a Helical Magnetic Field in
3C 273}.
\newblock {\em \apjl} {\bf 2005}, {\em 626},~L73--L76.
\newblock  [\href{http://dx.doi.org/10.1086/431901}{CrossRef}]

\bibitem[{Hovatta} \em{et~al.}(2012){Hovatta}, {Lister}, {Aller}, {Aller},
{Homan}, {Kovalev}, {Pushkarev}, and {Savolainen}]{Hovatta2012}
{Hovatta}, T.; {Lister}, M.L.; {Aller}, M.F.; {Aller}, H.D.; {Homan}, D.C.;
{Kovalev}, Y.Y.; {Pushkarev}, A.B.; {Savolainen}, T.
\newblock {MOJAVE: Monitoring of Jets in Active Galactic Nuclei with VLBA
Experiments. VIII. Faraday Rotation in Parsec-scale AGN Jets}.
\newblock {\em \aj} {\bf 2012}, {\em 144},~105.
\newblock  [\href{http://dx.doi.org/10.1088/0004-6256/144/4/105}{CrossRef}]


\bibitem[{Lisakov} \em{et~al.}(2021){Lisakov}, {Kravchenko}, {Pushkarev},
{Kovalev}, {Savolainen}, and {Lister}]{Lisakov2021}
{Lisakov}, M.M.; {Kravchenko}, E.V.; {Pushkarev}, A.B.; {Kovalev}, Y.Y.;
{Savolainen}, T.K.; {Lister}, M.L.
\newblock {An Oversized Magnetic Sheath Wrapping around the Parsec-scale Jet in
3C 273}.
\newblock {\em \apj} {\bf 2021}, {\em 910},~35.
\newblock  [\href{http://dx.doi.org/10.3847/1538-4357/abe1bd}{CrossRef}]

\bibitem[{Thompson} \em{et~al.}(2017){Thompson}, {Moran}, and
{Swenson}]{Thompson2017}
{Thompson}, A.R.; {Moran}, J.M.; {Swenson}, G.W., Jr.
\newblock {\em {Interferometry and Synthesis in Radio Astronomy}}, 3rd ed.; Springer Nature: Cham, Switzerland,
2017.
\newblock  [\href{http://dx.doi.org/10.1007/978-3-319-44431-4}{CrossRef}]

\bibitem[{Leppanen} \em{et~al.}(1995){Leppanen}, {Zensus}, and
{Diamond}]{Leppanen1995}
{Leppanen}, K.J.; {Zensus}, J.A.; {Diamond}, P.J.
\newblock {Linear Polarization Imaging with Very Long Baseline Interferometry
at High Frequencies}.
\newblock {\em \aj} {\bf 1995}, {\em 110},~2479.
\newblock  [\href{http://dx.doi.org/10.1086/117706}{CrossRef}]

\bibitem[{Greisen}(2003)]{Greisen2003}
{Greisen}, E.W. {AIPS, the VLA, and the VLBA}.
\newblock In {\em Information Handling in Astronomy---Historical Vistas};
{Heck}, A., Ed.; {Astrophysics and Space Science
Library}; Spring: Cham, Switzerland, 2003; Volume 285, p. 109.
\newblock  [\href{http://dx.doi.org/10.1007/0-306-48080-8_7}{CrossRef}]

\bibitem[{Park} \em{et~al.}(2021){Park}, {Byun}, {Asada}, and {Yun}]{Park2021a}
{Park}, J.; {Byun}, D.Y.; {Asada}, K.; {Yun}, Y.
\newblock {GPCAL: A Generalized Calibration Pipeline for Instrumental
Polarization in VLBI Data}.
\newblock {\em \apj} {\bf 2021}, {\em 906},~85.
\newblock  [\href{http://dx.doi.org/10.3847/1538-4357/abcc6e}{CrossRef}]

\bibitem[{Mart{\'\i}-Vidal} \em{et~al.}(2021){Mart{\'\i}-Vidal}, {Mus},
{Janssen}, {de Vicente}, and {Gonz{\'a}lez}]{Martividal2021}
{Mart{\'\i}-Vidal}, I.; {Mus}, A.; {Janssen}, M.; {de Vicente}, P.;
{Gonz{\'a}lez}, J.
\newblock {Polarization calibration techniques for the new-generation VLBI}.
\newblock {\em \aap} {\bf 2021}, {\em 646},~A52.
\newblock  [\href{http://dx.doi.org/10.1051/0004-6361/202039527}{CrossRef}]

\bibitem[{Chael} \em{et~al.}(2016){Chael}, {Johnson}, {Narayan}, {Doeleman},
{Wardle}, and {Bouman}]{Chael2016}
{Chael}, A.A.; {Johnson}, M.D.; {Narayan}, R.; {Doeleman}, S.S.; {Wardle},
J.F.C.; {Bouman}, K.L.
\newblock {High-resolution Linear Polarimetric Imaging for the Event Horizon
Telescope}.
\newblock {\em \apj} {\bf 2016}, {\em 829},~11.
\newblock  [\href{http://dx.doi.org/10.3847/0004-637X/829/1/11}{CrossRef}]

\bibitem[{Chael} \em{et~al.}(2018){Chael}, {Johnson}, {Bouman}, {Blackburn},
{Akiyama}, and {Narayan}]{Chael2018}
{Chael}, A.A.; {Johnson}, M.D.; {Bouman}, K.L.; {Blackburn}, L.L.; {Akiyama},
K.; {Narayan}, R.
\newblock {Interferometric Imaging Directly with Closure Phases and Closure
Amplitudes}.
\newblock {\em \apj} {\bf 2018}, {\em 857},~23.
\newblock  [\href{http://dx.doi.org/10.3847/1538-4357/aab6a8}{CrossRef}]

\bibitem[{Pesce}(2021)]{Pesce2021}
{Pesce}, D.W.
\newblock {A D-term Modeling Code (DMC) for Simultaneous Calibration and
Full-Stokes Imaging of Very Long Baseline Interferometric Data}.
\newblock {\em \aj} {\bf 2021}, {\em 161},~178.
\newblock  [\href{http://dx.doi.org/10.3847/1538-3881/abe3f8}{CrossRef}]

\bibitem[{Broderick} \em{et~al.}(2020){Broderick}, {Gold}, {Karami},
{Preciado-L{\'o}pez}, {Tiede}, {Pu}, {Akiyama}, {Alberdi}, {Alef}, {Asada},
{Azulay}, {Baczko}, {Balokovi{\'c}}, {Barrett}, {Bintley}, {Blackburn},
{Boland}, {Bouman}, {Bower}, {Bremer}, {Brinkerink}, {Brissenden}, {Britzen},
{Broguiere}, {Bronzwaer}, {Byun}, {Carlstrom}, {Chael}, {Chatterjee},
{Chatterjee}, {Chen}, {Chen}, {Cho}, {Conway}, {Cordes}, {Crew}, {Cui},
{Davelaar}, {De Laurentis}, {Deane}, {Dempsey}, {Desvignes}, {Doeleman},
{Eatough}, {Falcke}, {Fish}, {Fomalont}, {Fraga-Encinas}, {Friberg}, {Fromm},
{Galison}, {Gammie}, {Garc{\'\i}a}, {Gentaz}, {Georgiev}, {Goddi},
{G{\'o}mez}, {Gu}, {Gurwell}, {Hada}, {Hecht}, {Hesper}, {Ho}, {Ho}, {Honma},
{Huang}, {Huang}, {Hughes}, {Inoue}, {Issaoun}, {James}, {Janssen}, {Jeter},
{Jiang}, {Jim{\'e}nez-Rosales}, {Johnson}, {Jorstad}, {Jung}, {Karuppusamy},
{Kawashima}, {Keating}, {Kettenis}, {Kim}, {Kim}, {Kino}, {Koay}, {Koch},
{Koyama}, {Kramer}, {Kramer}, {Krichbaum}, {Kuo}, {Lee}, {Li}, {Li},
{Lindqvist}, {Lico}, {Liu}, {Liuzzo}, {Lo}, {Lobanov}, {Loinard}, {Lonsdale},
{Lu}, {MacDonald}, {Mao}, {Marscher}, {Mart{\'\i}-Vidal}, {Matsushita},
{Matthews}, {Menten}, {Mizuno}, {Mizuno}, {Moran}, {Moriyama},
{Moscibrodzka}, {M{\"u}ller}, {Nagai}, {Nagar}, {Nakamura}, {Narayan},
{Narayanan}, {Natarajan}, {Neri}, {Ni}, {Noutsos}, {Okino}, {Olivares},
{Ortiz-Le{\'o}n}, {Oyama}, {Palumbo}, {Park}, {Pen}, {Pesce}, {Pi{\'e}tu},
{Plambeck}, {PopStefanija}, {Porth}, {Prather}, {Ramakrishnan}, {Rao},
{Rawlings}, {Raymond}, {Rezzolla}, {Ripperda}, {Roelofs}, {Rogers}, {Ros},
{Rose}, {Rottmann}, {Ruszczyk}, {Ryan}, {Rygl}, {S{\'a}nchez},
{S{\'a}nchez-Arguelles}, {Sasada}, {Savolainen}, {Schloerb}, {Schuster},
{Shao}, {Shen}, {Small}, {Sohn}, {SooHoo}, {Tazaki}, {Tilanus}, {Titus},
{Toma}, {Torne}, {Traianou}, {Trippe}, {Tsuda}, {van Bemmel}, {van
Langevelde}, {van Rossum}, {Wagner}, {Wardle}, {Weintroub}, {Wex}, {Wharton},
{Wielgus}, {Wong}, {Wu}, {Yoon}, {Young}, {Young}, {Younsi}, {Yuan}, {Yuan},
{Zensus}, {Zhao}, {Zhao}, {Zhu}, and {Event Horizon Telescope
Collaboration}]{Broderick2020}
{Broderick}, A.E.; {Gold}, R.; {Karami}, M.; {Preciado-L{\'o}pez}, J.A.;
{Tiede}, P.; {Pu}, H.Y.; {Akiyama}, K.; {Alberdi}, A.; {Alef}, W.; {Asada},
K.;  et~al.
\newblock {THEMIS: A Parameter Estimation Framework for the Event Horizon
Telescope}.
\newblock {\em \apj} {\bf 2020}, {\em 897},~139.
\newblock  [\href{http://dx.doi.org/10.3847/1538-4357/ab91a4}{CrossRef}]

\bibitem[{Park} \em{et~al.}(2021){Park}, {Asada}, {Nakamura}, {Kino}, {Pu},
{Hada}, {Kravchenko}, and {Giroletti}]{Park2021c}
{Park}, J.; {Asada}, K.; {Nakamura}, M.; {Kino}, M.; {Pu}, H.Y.; {Hada}, K.;
{Kravchenko}, E.V.; {Giroletti}, M.
\newblock {A Revised View of the Linear Polarization in the Subparsec Core of
M87 at 7 mm}.
\newblock {\em \apj} {\bf 2021}, {\em 922},~180.
\newblock  [\href{http://dx.doi.org/10.3847/1538-4357/ac26bf}{CrossRef}]

\bibitem[{Kravchenko} \em{et~al.}(2020){Kravchenko}, {Giroletti}, {Hada},
{Meier}, {Nakamura}, {Park}, and {Walker}]{Kravchenko2020}
{Kravchenko}, E.; {Giroletti}, M.; {Hada}, K.; {Meier}, D.L.; {Nakamura}, M.;
{Park}, J.; {Walker}, R.C.
\newblock {Linear polarization in the nucleus of M87 at 7 mm and 1.3 cm}.
\newblock {\em \aap} {\bf 2020}, {\em 637},~L6.
\newblock  [\href{http://dx.doi.org/10.1051/0004-6361/201937315}{CrossRef}]

\bibitem[{Broderick} and {Pesce}(2020)]{BP2020}
{Broderick}, A.E.; {Pesce}, D.W.
\newblock {Closure Traces: Novel Calibration-insensitive Quantities for Radio
Astronomy}.
\newblock {\em \apj} {\bf 2020}, {\em 904},~126.
\newblock  [\href{http://dx.doi.org/10.3847/1538-4357/abbd9d}{CrossRef}]

\bibitem[{Jones}(1988)]{Jones1988}
{Jones}, T.W.
\newblock {Polarization as a Probe of magnetic Fields and Plasma Properties of
Compact Radio Sources: Simulation of Relativistic Jets}.
\newblock {\em \apj} {\bf 1988}, {\em 332},~678.
\newblock  [\href{http://dx.doi.org/10.1086/166685}{CrossRef}]

\bibitem[{Gilbert} and {Conway}(1970)]{GilbertConway1970}
{Gilbert}, J.A.; {Conway}, R.G.
\newblock \textls[-25]{{Circular Polarization of Quasars at {\ensuremath{\lambda}}49 Cm}.}
\newblock {\em \nat} {\bf 1970}, {\em 227},~585--586.
\newblock  [\href{http://dx.doi.org/10.1038/227585b0}{CrossRef}]

\bibitem[{Weiler} and {de Pater}(1983)]{WeilerdePater1983}
{Weiler}, K.W.; {de Pater}, I.
\newblock {A catalog of high accuracy polarization measurements.}
\newblock {\em \apjs} {\bf 1983}, {\em 52},~293--327.
\newblock  [\href{http://dx.doi.org/10.1086/190869}{CrossRef}]

\bibitem[{Komesaroff} \em{et~al.}(1984){Komesaroff}, {Roberts}, {Milne},
{Rayner}, and {Cooke}]{Komesaroff1984}
{Komesaroff}, M.M.; {Roberts}, J.A.; {Milne}, D.K.; {Rayner}, P.T.; {Cooke},
D.J.
\newblock {Circular and linear polarization variations of compact radio
sources.}
\newblock {\em \mnras} {\bf 1984}, {\em 208},~409--425.
\newblock  [\href{http://dx.doi.org/10.1093/mnras/208.2.409}{CrossRef}]

\end{thebibliography}
\end{document}